\documentclass[aps, prd, twocolumn, nofootinbib, floatfix, superscriptaddress]{revtex4-1}

\usepackage{comment}
\usepackage{graphicx}
\usepackage{dcolumn}
\usepackage{bm}
\usepackage{subfigure}
\usepackage{epsfig}
\usepackage{amsmath}
\usepackage{amsfonts}
\usepackage{graphicx}
\usepackage{amssymb}
\usepackage{nccmath}
\usepackage{amssymb,amsmath,amsfonts}
\usepackage{xfrac}
\usepackage{color}
\usepackage{tikz}
\usepackage[T1]{fontenc}
\usepackage[utf8]{inputenc}
\usepackage{ulem}

\usepackage{tcolorbox}
\usepackage{hyperref}
\usepackage{booktabs}
\hypersetup{colorlinks=true, citecolor=red, linkcolor=blue, urlcolor=blue}

\definecolor{rossos}{cmyk}{0,1,1,0.55}
\definecolor{bluscuro}{rgb}{0.15, 0.2, .85}
\definecolor{bluchiaro}{cmyk}{1,.3,0.,0.1}

\let\oldsqrt\sqrt
\def\sqrt{\mathpalette\DHLhksqrt}
\def\DHLhksqrt#1#2{%
\setbox0=\hbox{$#1\oldsqrt{#2\,}$}\dimen0=\ht0
\advance\dimen0-0.2\ht0
\setbox2=\hbox{\vrule height\ht0 depth -\dimen0}%
{\box0\lower0.4pt\box2}}

\newcommand{\sss}[1]{{\scriptscriptstyle{#1}}}
\newcommand{\boldmathsymbol}[1]{{\ensuremath{\boldsymbol{#1}}}}

\newcommand{\uPl}{\mathrm{Pl}}

\newcommand{\usssPl}{\sss{\uPl}}

\newcommand{\ud}{\mathrm{d}}

\newcommand{\calH}{\mathcal{H}}

\newcommand{\Mp}{M_\usssPl}

\newcommand{\beq}{\begin{equation}}
\newcommand{\eeq}{\end{equation}}
\newcommand{\bea}{\begin{equation}\begin{aligned}}
\newcommand{\eea}{\end{aligned}\end{equation}}

\newlength{\wsingfig}
\newlength{\wdblefig}
\newlength{\wquadfig}
\newlength{\wtriplefig}
\newcommand{\Eq}[1]{Eq.~(\ref{#1})}

\newcommand{\Fig}[1]{Fig.~{\ref{#1}}}

\newcommand{\Sec}[1]{Sec.~\ref{#1}}

\newcommand{\be}{\begin{equation}}

\begin{document}
\rightline{KCL-PH-TH/2024-11}
\title{Revisiting string-inspired running-vacuum models under the lens of light primordial black holes}


\author{Theodoros~Papanikolaou}
\email{t.papanikolaou@ssmeridionale.it}
\affiliation{Scuola Superiore Meridionale, Largo S. Marcellino 10, 80138, Napoli, Italy}
\affiliation{Istituto Nazionale di Fisica Nucleare (INFN), Sezione di Napoli, Via Cinthia 21, 80126 Napoli, Italy}
\affiliation{National Observatory of Athens, Lofos Nymfon, 11852 Athens, 
Greece}

\author{Charalampos~Tzerefos}
\email{chtzeref@phys.uoa.gr}
\affiliation{Department of Physics, National \& Kapodistrian University of Athens, Zografou Campus GR 157 73, Athens, Greece}
\affiliation{National Observatory of Athens, Lofos Nymfon, 11852 Athens, 
Greece}

\author{Spyros~Basilakos}
\email{svasil@academyofathens.gr}
\affiliation{National Observatory of Athens, Lofos Nymfon, 11852 Athens, 
Greece}
\affiliation{Academy of Athens, Research Center for Astronomy and Applied Mathematics, Soranou Efesiou 4, 11527, Athens, Greece}
\affiliation{School of Sciences, European University Cyprus, Diogenes Street, Engomi, 1516 Nicosia, Cyprus}

\author{Emmanuel~N.~Saridakis}
\email{msaridak@noa.gr}
\affiliation{National Observatory of Athens, Lofos Nymfon, 11852 Athens, 
Greece}
\affiliation{CAS Key Laboratory for Researches in Galaxies and Cosmology, 
Department of Astronomy, University of Science and Technology of China, Hefei, 
Anhui 230026, P.R. China}
\affiliation{Departamento de Matem\'{a}ticas, Universidad Cat\'{o}lica del Norte, Avda.
Angamos 0610, Casilla 1280 Antofagasta, Chile}

\author{Nick~E.~Mavromatos}
\email{mavroman@mail.ntua.gr}
\affiliation{Physics Division, School of Applied Mathematical and Physical Sciences, National Technical University of Athens, Zografou Campus, Athens 157 80, Greece}
\affiliation{Theoretical Particle Physics and Cosmology group, Department of Physics, King's College London, London WC2R 2LS, UK}

 
\begin{abstract}
Light primordial black holes (PBHs) with masses $M_\mathrm{PBH}<10^9\mathrm{g}$ can interestingly dominate the Universe's energy budget and give rise to early matter-dominated (eMD) eras before Big Bang Nucleosyntesis (BBN). During this eMD era, one is met with an abundant production of induced gravitational waves (GWs) serving as a portal to constrain the underlying theory of gravity. In this work, we study this type of induced GWs  within the context of string-inspired running-vaccuum models (StRVMs), which, when expanded around de Sitter backgrounds, include logarithmic corrections of the space-time curvature. In particular,  we discuss in detail the effects of StRVMs on the source as well as on the propagation of these PBH-induced GWs. Remarkably, under the assumption that the logarithmic terms represent quantum gravity corrections in the PBH era, we show that 
GW overproduction can be avoided if one assumes 
a coefficient of these logarithmic corrections that is much larger than the square of the reduced Planck mass. The latter cannot characterise quantum gravity corrections, though, prompting the need for revision of the quantisation of StRVMs in different than de Sitter backgrounds, such as those characterising PBH-driven eMD eras. This non trivial result suggests the importance of light PBHs as probes of new physics.
\end{abstract}

\maketitle
\section{Introduction}

Running-Vacuum Models (RVM) (see, e.g. \cite{rvm1,rvm2,rvmde} and references therein), constitute a framework for describing the entire evolution of the Universe~\cite{Lima:2013dmf},
from inflation till the present era, including its thermodynamical aspects~\cite{Lima:2015mca,rvm3}.
The RVM framework is 
consistent with the plethora of the available cosmological data~\cite{rvmdata,rvmdata2,tsiapi}, including Big-Bang-Nucleosynthesis (BBN) data~\cite{rvmbbn}. Some RVM variants, are also capable of alleviating~\cite{rvmtenss8,rvmtens} the currently observed tensions in the cosmological data~\cite{tensions,tensions2}, namely 
the Hubble $H_0$ and the $\sigma_8$ tensions, associated with discrepancies of the $\Lambda$CDM-based best fits with the theoretically predicted values of the Hubble parameter $H_0$ today and the modern-era growth structures of the Universe, respectively.\footnote{Of course, these tensions may 
admit either more conventional astrophysical explanations, or even be artefacts of the current statistics in the data, thus  potentially disappearing in the future~\cite{freedman}.} 

The RVM framework is characterised by a vacuum energy density which (on account of general covariance) is a function of even powers of the Hubble parameter, $H^{2n}, \, n \in \mathbb Z^+$.
Microscopic frameworks which can result in RVM cosmologies are string theory (the so-called Stringy RVM (StRVM)~\cite{bms,bms2,bms3,ms1,ms2}), or quantum field theory of massive fields (of various spins) in  expanding-universe backgrounds~\cite{rvmqft1,rvmqft2,rvmqft3,rvmqft4}. However, such massive-quantum-matter fields lead also to logarithmic ${\rm ln}H$ corrections to the vacuum energy density. Such logarithmic corrections, but of the form $H^{2n}{\rm ln}H$, can also arise from 
graviton loops in quantum gravity models in (approximately) de-Sitter backgrounds, which may characterise both early~\cite{bmssugra,ms1,ms2} and late phases~\cite{Gomez-Valent:2023hov} of the StRVM. In the context of StRVM, such quantum-gravity-induced logarithmic corrections are compatible with the late-epoch Universe phenomenology, but they may also contribute~\cite{Gomez-Valent:2023hov} to the alleviation of the cosmic tensions~\cite{tensions,tensions2}, like the aforementioned variants of the conventional RVM. 
Remarkably, the order of magnitude of these StRVM logarithmic corrections,  as determined by the late Universe phenomenology~\cite{rvmdata,Gomez-Valent:2023hov}, is also compatible with BBN data~\cite{rvmbbn}, under the dark-energy interpretation of the models. 

On the other hand, primordial black holes (PBHs), firstly proposed in the early `70s~\cite{1967SvA....10..602Z, Carr:1974nx,1975ApJ...201....1C,1979A&A....80..104N}, and typically forming out of the collapse of enhanced cosmological perturbations, are currently attracting considerable attention, since they can address a number of issues of modern cosmology. In particular, they can 
potentially account for a part or all of the dark matter content of the Universe~\cite{Chapline:1975ojl}, offering in parallel an explanation for     
the large-scale structure formation through Poisson fluctuations~\cite{Meszaros:1975ef,Afshordi:2003zb}. Furthermore, they can consitute viable candidates for the origin of supermassive black holes residing in the galactic centres~\cite{1984MNRAS.206..315C, Bean:2002kx}, explaining as well some of the black-hole merging events recently 
detected by the LIGO/VIRGO collaboration~\cite{Sasaki:2016jop,Franciolini:2021tla}. Interestingly enough, PBHs can also account naturally for the procees of baryogenesis in the early Universe~\cite{Baumann:2007yr, Hook:2014mla, Carr:2019hud} as well as for the generation of cosmic magnetic fields~\cite{Safarzadeh:2017mdy,Papanikolaou:2023nkx}. For comprehensive reviews on PBHs the interested reader is referred to the refs.~\cite{Carr:2020gox,Escriva:2022duf} .

A particularly interesting kind of PBHs are the ultra-light ones, with masses lighter than $10^9 \mathrm{g}$, which have evaporated before BBN. Although light PBHs cannot constitute dark matter candidates, nonetheless, their existence addresses a plethora of cosmological issues. In particular, these ultra-light PBHs, naturally produced in the early Universe~\cite{Garcia-Bellido:1996mdl,Martin:2020fgl,Heydari:2021qsr,Banerjee:2022xft,Briaud:2023eae,Basilakos:2023jvp,del-Corral:2023apl,Heydari:2023xts}, can trigger early matter dominated eras (eDM)~\cite{Garcia-Bellido:1996mdl, Hidalgo:2011fj, Suyama:2014vga, Zagorac:2019ekv} before BBN and reheat the Universe through their evaporation~\cite{Lennon:2017tqq}. Interestingly enough, during this eMD era driven by PBHs, induced GWs~\cite{Papanikolaou:2020qtd,Domenech:2020ssp,Papanikolaou:2022chm,Domenech:2023jve} can be abundantly produced potentially being detectable by forthcoming GW observatories~\cite{Kozaczuk:2021wcl} and serving as new portal to constrain the underlying gravity theory~\cite{Papanikolaou:2021uhe, Tzerefos:2023mpe}. Notably, recent works~\cite{Basilakos:2023xof,Bhaumik:2023wmw} claim that light PBHs can explain as well the recently released Pulsar Timing Array (PTA) GW data. Moreover, ultra-light PBHs can account for the Hubble tension through the injection of light dark radiation degrees of freedom~\cite{Hooper:2019gtx,Nesseris:2019fwr,Lunardini:2019zob} 
to the primordial plasma (for a review see \cite{Papanikolaou:2023oxq}) while at the same time they can naturally produce the baryon asymmetry through CP violating out-of-equilibrium decays of their Hawking evaporation products~\cite{Barrow:1990he,Hamada:2016jnq,Bhaumik:2022pil}.

In this work, we study the production of GWs induced by  PBH energy density fluctuations. By studying the effect of quantum(-graviton) logarithmic corrections of StRVMs on the source and the propagation of these induced GWs, we find that fixing the parameters of the StRVM logarithmic corrections to take on values in the appropriate range to address cosmological tensions~\cite{Gomez-Valent:2023hov} leads to an incurable overproduction of GWs. By requiring, thus, the avoidance of such a GW overproduction issue, we impose lower bound constraints on the coefficient of the StRVM curvature logarithmic correction, being unacceptably large in terms of the reduced Planck mass squared.

This unnaturally large value of the curvature logarithmic correction coefficient present in PBH-dominated eras, sheds light on subtleties involved in imposing such GW overproduction constraints, which actually depend on the microscopic details of StRVM. In particular, during eMD eras, like the one considered here, such logarithmic corrections, having been derived only for de-Sitter backgrounds in the context of one-loop quantum (super)gravities, may be significantly suppressed, thus 
alleviating the inconsistency between StRVM and the scenario of light PBHs early-cosmological-epoch dominance.
The upshot of our work here is to highlight that the quantisation of StRVMs in such non-de Sitter space-time backgrounds, being related to specific microscopically realisations of the stringy running vacuum framework, needs to be revisited before definite conclusions are reached. 

The structure of the paper is the following: 
In \Sec{StRVM} we review the basic features of the gravitational effective theories stemming from the StRVM explaining also carefully the assumptions and theoretical considerations that underlie the logarithmic-curvature RVM corrections to the General Relativity (GR) terms.  
Then, in \Sec{sec:PBH_gas} we recap the physics of a population of ultra-light PBHs deriving at the end the power spectrum of the PBH energy density perturbations during a PBH-driven eMD era. Subsequently, in \Sec{sec:SIGW}, we present the basics of the GWs induced by PBH energy density fluctuations within the framework of StRVMs while in \Sec{sec:constraints} we set constraints on the coefficient $c_2$ of the StRVM curvature logarithmic correction. Finally, in \Sec{sec:conclusions}, we carefully assess the various interpretations of the derived constraints. Some technical aspects of our approach are given in the
Appendices \ref{app:Anisotropic_Stress} and \ref{app:Green_function_terms}. 

\section{Stringy running vacuum overview}\label{StRVM}

As discussed in the context of 
dynamically-broken one-loop N=1 (3+1)-dimensional supergravity effective field theories, in local de Sitter spacetime backgrounds~\cite{Alexandre:2013iva,Alexandre:2013nqa,Alexandre:2014lla}, 
characterised by a positive cosmological constant $\Lambda >0$, integrating out massless gravitons, following the pioneering work of 
\cite{Fradkin:1983mq},
leads to an effective one-loop action involving terms that depend on $\Lambda^n \log \Lambda$, $n=1,2$, in addition to $\Lambda$-dependent terms.
Taking into account that $R \propto  \Lambda$ is the de-Sitter background spacetime scalar curvature, there was made the suggestion in \cite{Alexandre:2013iva,Alexandre:2013nqa,Alexandre:2014lla}
that $R^{2l}, R^{2l} \log R$, $l=1,2$ corrections 
appear in one-loop corrected effective actions in such backgrounds, after massless graviton path-integration. 

This has been used in \cite{bmssugra,ms1} to suggest that such corrections may characterise the quantum-gravity StRVM effective actions describing the early and late eras of the Universe, which are 
characterised by de Sitter phases, such as inflation or current dark-energy-dominance epoch. 

Specific interest arises in the r\^ole of such corrections in the late Universe, where the data point out to a dark-energy dominance epoch, which is almost de Sitter. In such late de-Sitter-like eras, the dominant quantum-gravity corrections in the (3+1)-dimensional effective gravitational action of the StRVM may be parametrised as~\cite{Gomez-Valent:2023hov}:\footnote{We follow the conventions the convention
for the signature of the metric $(-, +, +, +)$, and the definitions of the Riemann Curvature tensor
$R^\lambda_{\,\,\,\,\mu \nu \sigma} = \partial_\nu \, \Gamma^\lambda_{\,\,\mu\sigma} + \Gamma^\rho_{\,\, \mu\sigma} \, \Gamma^\lambda_{\,\, \rho\nu} - (\nu \leftrightarrow \sigma)$, the Ricci tensor $R_{\mu\nu} = R^\lambda_{\,\,\,\,\mu \lambda \nu}$, and the Ricci scalar $R = R_{\mu\nu}g^{\mu\nu}$.} 
\beq\label{eq:action}
S=\int d^4x\,\sqrt{-g}\,\left
\{c_0+R\left[c_1+c_2\log\left(\frac{R}{R_0}\right)\right]+\mathcal{L}_{m} \right\} \,.
\eeq
In the above expression, $\mathcal L_m$ represents the matter action, and $R_0 =12H_0^2$ denotes the current-era scalar curvature of the expanding universe, with $H_0$ the present-era Hubble parameter.
The coefficient 
\begin{align}\label{c1def}
c_1= \frac{1}{16\pi\rm G} = \frac{M^2_{\rm Pl}}{2} + \tilde c_1 >0\,,
\end{align}
determines an effective (3+1)-dimensional gravitational constant G, including weak quantum-gravity corrections $\tilde c_1$, such that $\tilde c_1/M_{\rm Pl}^2 \ll 1$,
where $M_{\rm Pl}=\frac{1}{\sqrt{G_N}}$ (with $G_N$ the conventional (3+1)-dimensional Newton's constant) is the reduced Planck mass
$M_{\rm Pl} = 2.4 \cdot 10^{18}$~GeV. Throughout this work we work in units where $\hbar=c=1$. 

The constant $c_0$ parametrises the cosmological-constant dominance in the current-epoch cosmological data~\cite{Planck:2018jri}. 
For our phenomenological analysis we can take it to be:
\begin{align}\label{coval}
c_0 \simeq \frac{7}{3} \rho_m^{(0)} =   \frac{7}{3} \, H^2_0\, M^2_\mathrm{pl},
\end{align}
where $\rho_m^{(0)}$ is the present-day matter density.  
However, strictly speaking in the context of StRVM, $c_0$ is actually not a constant, but a slow-varying quantity parametrising dark energy in the modern era~\cite{bms}. We should stress at this point that, from the point of view of string theories, a de Sitter phase should only be metastable, in view of obstructions raised by both, perturbative (S-matrix~\cite{scat1,scat2}) and non-perturbative string (swampland~\cite{swamp1,swamp2,swamp3,swamp5,swamp4}) 
arguments. Thus, within the StRVM framework~\cite{bms,ms1,ms2,Gomez-Valent:2023hov}, we envisage a quitenssence-type situation in which, in the asymptotically far future, $c_0$ will eventually disappear, as required by consistency with the underlying string theory. This feature is of course in agreement with the current-data phenomenology of the StRVM.

It should be stressed that, as a result of the nature of the coefficient $c_2$, which is associated with quantum-gravity corrections, it must be that:
\begin{align}\label{c2c1}
    |c_2| \ll c_1,    .
\end{align}
with $c_2$ having the same units as $c_1$, that is $\Mp^2$. The late-era phenomenology of the model \eqref{eq:action}
can be most conveniently studied by means of the following two parameters~\cite{Gomez-Valent:2023hov}:
\begin{align}\label{parameters}
d \equiv c_1 + c_2, \qquad \epsilon  \equiv \frac{c_2}{c_1 + c_2}\,.
\end{align}
In \cite{Gomez-Valent:2023hov} 
it was demonstrated that the model 
can be easily made consistent with the current-era phenomenology, which is shared with conventional forms of RVM, upon restricting the ranges of $d,\epsilon$ (and thus $c_2$), in accordance with the condition \eqref{c2c1}. 
Indeed, since the quantity $d$ given in \eqref{parameters} essentially defines in the present era the ratio 
\begin{align}\label{dG}
d=\rm G_{\rm eff}/G_N\,
\end{align}
of an effective gravitational constant G$_{\rm eff}$ over the standard value of the Newton's constant G, we observe that for $d < 1$ ($d > 1$) we shall have a slower (faster) expansion of the Universe, and hence for $d \ne 1$ the temperature CMB spectra will be affected analogously. Moreover, as argued in \cite{Gomez-Valent:2023hov}, a $d < 1$ can alleviate the Hubble $H_0$ tension~\cite{tensions,tensions2}.  

The detailed analysis of \cite{Gomez-Valent:2023hov} has also demonstrated that consistency with the CMB and growth of structure data~\cite{Planck:2018jri} requires the condition
\begin{align}\label{eps}
\big|\epsilon \big|  \lesssim \mathcal O(10^{-7})\,.
\end{align}
To understand this restriction, one should take into account~\cite{Gomez-Valent:2023hov} that negative (positive) values of the parameter $\epsilon$ lead to a suppression (enhancement) of the amount of structures in the universe. Thus, by imposing the restriction \eqref{eps}, one can avoid changing drastically the matter fraction $\Omega_m$ in the current epoch, thus maintaining consistency with the data~\cite{Planck:2018jri}.

Moreover, as shown in \cite{Gomez-Valent:2023hov},  one can alleviate the observed tensions, both the $H_0$ and the growth of structure $\sigma_8$, by choosing the following values of the parameters:\footnote{It is interesting, and simultaneously quite curious, to remark that this value of $\epsilon <0$ corresponds to quantum-gravity induced logarithmic corrections in the dynamical broken N=1 supergravity model of \cite{Alexandre:2013nqa,Alexandre:2014lla,bmssugra} with a (sub-Planckian) supersymmetry breaking scale of order $\sqrt{f} \simeq 10^{-5/4} M_{\rm Pl} \sim 10^{17}~\rm GeV$, which appears quite consistent with the entire cosmological history of the StRVM as outlined in \cite{ms1}.}
\begin{align}\label{eps2}
0. 9 \lesssim d \lesssim  0.95\,, \qquad \big|\epsilon \big| \sim \mathcal O(10^{-7})\,,
\end{align}
and in this respect the StRVM \eqref{eq:action}
behaves as some variants of the conventional RVMI\cite{rvmtens}, specifically RVMII, which can alleviate both tensions simultaneously, by allowing for a mild cosmic time ($t$) dependence of the effective gravitational constant $G_{\rm eff}(t)$. In the context of the stringy RVM, such a dependence is replaced by the incorporation of the quantum-gravity logarithmic-curvature  corrections to the gravitational coupling.

In strong gravity backgrounds, such as the ones characterised by PBH domination, the quantum gravity corrections might be expected to be significant, making more demanding  a detailed embedding of the theory into an UltraViolet (UV) complete quantum gravity framework, such as string theory. 
At this point, we mention that string-inspired corrections to the General Relativity actions, such as higher-curvature corrections, e.g. quadratic Gauss Bonnet (GB) terms coupled to scalar dilaton fields, are known to make the finding de-Sitter-type vacuum solutions to  the dynamical equations para-metrically harder than easier~\cite{Cunillera:2021fbc}. This is in agreement with the Swampland criteria~\cite{swamp1,swamp2,swamp3,swamp4} of quantum gravity and theory generic incompatibility with de Sitter vacua.

In the context of the StRVM~\cite{bms,bms2,bms3,ms1,ms2}, dilatons have been assumed constant, in which case the non trivial quadratic curvature corrections come from the gravitational Chern-Simons (CS) term, coupled to the axion fields $b(x)$ of the string gravitational multiplet (gravitational axions), $\int d^4x \sqrt{-g} \frac{1}{2} b(x)\, \varepsilon^{\mu\nu\rho\sigma} R_{\sigma\rho}^{\quad\,\,\alpha\beta}\, R_{\mu\nu\alpha\beta} $, but also from higher than quadratic, non anomalous, curvature corrections. The latter are also expected, in analogy with the GB case, to affect the existence of proper de Sitter vacua. Such higher curvature terms are not important at the current cosmological era, for which the quantum corrections are strongly suppressed, and thus \eqref{eq:action} suffices to describe the gravitational dynamics, but they may become important in strong-gravity regimes, such as the aforementioned PBH dominance era. 

On the other hand, the anomalous CS terms of the StRVM can condense at early epochs, due to the presence of chiral GWs, leading to a \textit{metastable} condensate characterised by non trivial imaginary parts~\cite{Dorlis:2024yqw,Mavromatos:2024pho}. This, in turn, can lead to inflation of RVM type`\cite{bms,ms1}, with a duration determined by those imaginary parts, which can be in the phenomenologically correct ballpark (i.e. inflation with 50-60 e-foldings). We mention for completeness that the StRVM model is characterised by a pre-RVM inflationary phase, during which the gravitational axion $b$-field dominates the cosmic fluid with a stiff equation of state. It is at the end of this phase that chiral GW (which are produced in various scenarios~\cite{ms1,ms2}) can lead to CS condensates. In such eras, the CS anomaly terms alone suffice to describe the passage from the stiff era to a metastable de Sitter one, leading to RVM inflation.  
The metastability is notably compatible with the swampland criteria. It is towards the end of this RVM inflation, and subsequent eras, that one might have a dominance phase of PBHs, 
due to enhanced production~\cite{Mavromatos:2022yql},  
during which one should expect stringy corrections to affect significantly the RVM de Sitter vacua, and thus the form of the logarithmic-curvature  
correction terms in the pertinent effedctive action, which we stress once again is expected to be different from \eqref{eq:action}.

Although \eqref{eq:action} describes well the current de-Sitter-like era, an interesting question arises as to the potential dependence of the magnitude of the coefficient $c_2$ on the cosmic time, in other words on the specific cosmological era under consideration.  Actually, as already mentioned in the previous section, the form of the graviton-loop induced corrections  depends on the spacetime backgrounds about which one expands the theory, assuming weak graviton corrections to such backgrounds. 

Of course, the full answer to such questions would require the development of a complete theory of quantum gravity, or at least embedding the model into detailed, phenomenologically-realistic string theory models, which at present is not available. Nonetheless, 
it is the point of this article to stress that at least the first of the above questions, namely the cosmic time dependence of the magnitude of $c_2$, could be partly settled phenomenologically,  at least under some conditions that we shall specify below. The key to this lies on examining the effective StRVM gravitational theory \eqref{eq:action} under the lens of PBHs, in particular light PBHs. We stress at this point that, during the light-PBH-production era, of interest to us, the background spacetime is far from de Sitter, hence the considerations of \cite{Alexandre:2013iva,Alexandre:2014lla,Alexandre:2013nqa,Fradkin:1983mq,ms1} leading to the form \eqref{eq:action} are not strictly valid. We therefore first need to find a more appropriate parametrisation of the quantum-graviton corrected effective action for such an era.

In the next and the following sections we embark onto such a task. The conclusions, as we shall see, appear quite interesting, and in some cases, rather decisive, thus reinforcing the r\^ole of PBHs as interesting probes of new gravitational physics models.

\section{The primordial black hole gas}\label{sec:PBH_gas}
\subsection{Early matter dominated eras driven by primordial black holes}
We consider here a population (``gas") of PBHs forming in the RD era after inflation due to the collapse of enhanced cosmological perturbations~\cite{1967SvA....10..602Z,1975ApJ...201....1C}. For simplicity, we consider that all PBHs of our population share the same mass $M_\mathrm{PBH}$~\cite{Germani:2018jgr,MoradinezhadDizgah:2019wjf}. This can be achieved in general by a sharply-peaked primordial curvature power spectrum on small scales. At the end, we meet the formation of a PBH with a mass of the order of the cosmological horizon mass being recast as
\begin{equation}\label{eq:PBH_mass}
    M_{\rm PBH,f}=\frac{4\pi\gamma M_{\rm pl}^2}{H_{\rm f}}\,,
\end{equation}   
where $H_\mathrm{f}$ stands for the Hubble parameter at the time of PBH formation and $\gamma\simeq 0.2$ is the fraction of the horizon mass collapsing to PBHs during a RD era~\cite{Musco:2008hv}.

Regarding the dynamical evolution of the PBH abundance $\Omega_\mathrm{PBH}$, one should account for the fact that  $\Omega_\mathrm{PBH}$ will grow as $\Omega_\mathrm{PBH}\propto \rho_\mathrm{PBH}/\rho_\mathrm{r}\propto a^{-3}/a^{-4}\propto a$ since we have a matter component (PBHs in our case) evolving in a RD background. At the end, if the initial PBH abundance at PBH formation time $\Omega_\mathrm{PBH,f}$ is large enough, PBHs will dominate at some point the Universe's energy budget triggering an eMD era with the scale factor at PBH domination time being recast $a_\mathrm{d} = a_\mathrm{f}/\Omega_\mathrm{PBH,f}$, where $a_\mathrm{f}$ is the scale factor at PBH formation time. Note here that these transient eMDs should occur before BBN so as that Hawking radiated products do not ``disturb" the production of light elements in the early Universe, thus imposing that PBHs should evaporate before BBN.

If one now tracks back the dynamical evolution of the radiation energy density from today up to PBH formation era and takes into account the intervening transient eMD era driven by PBHs before BBN, they can show that $a_\mathrm{f}$ can be recast as~\cite{Domenech:2020ssp}
\beq
a_\mathrm{f} = \left(\frac{A\Mp^2}{2\pi\gamma\Omega^2_\mathrm{PBH,f}M^2_\mathrm{PBH}}\right)^{1/6}\Omega^{1/4}_{r,0}\sqrt{\frac{H_0M_\mathrm{PBH}}{4\pi\gamma\Mp^2}},
\eeq
where $H_0\simeq 70\mathrm{km/s/Mpc}$ is the Hubble parameter today, $\Omega_{r,0} = 4\times 10^{-5}$ is the present-day abundance of radiation and $A= 3.8\times g_\mathrm{eff}/960$, with $g_\mathrm{eff}\simeq 100$~\cite{Kolb:1990vq} being the effective number of relativistic degrees of freedom present at the epoch of PBH formation. At the end, the scale factor during the aforementioned eMD era will read as $a = a_\mathrm{d}(\eta/\eta_\mathrm{d})^2$, where $\eta$ is the conformal time defined as $\mathrm{d}t = a\mathrm{d}\eta$,  while the conformal Hubble parameter defined as $\mathcal{H} \equiv a^\prime/a$, where $\prime$ denotes derivation with respect to the conformal time, will be recast as $\mathcal{H} = 2/\eta$.

\subsection{The primordial black hole gravitational potential}
Having described above the dynamics of the PBH gas, let us discuss here the PBH gravitational potential $\Phi$ during the PBH-dominated era, which is associated with the PBH energy density perturbations themselves and will in fact constitute the source of our induced GW signal. PBHs form from the collapse of enhanced primordial curvature perturbations. Thus, the statistical properties of the primordial curvature perturbations will be inherited to PBH energy density fluctuations. Considering, then, Gaussian primordial curvature perturbations as imposed by Planck~\cite{Planck:2019kim}, distant curvature perturbations are uncorrelated, leading to a random spatial distribution of our PBH population~\cite{Desjacques:2018wuu, Ali-Haimoud:2018dau, MoradinezhadDizgah:2019wjf}. Consequently, PBH statistics follow the Poisson distribution, resulting in the following PBH matter power spectrum [see Sec. 2 and Appendix B of~\cite{Papanikolaou:2020qtd} for more details.]:
\begin{equation}\label{eq:ppoi}
    {\cal P}_{\delta_\mathrm{PBH}}(k)=\frac{2}{3\pi}\left(\frac{k}{k_{\rm UV}}\right)^3\,.
\end{equation}
In the above expression, we note the appearance of a UV-cutoff scale being related to the PBH mean separation scale and defined as $k_{\rm UV}\equiv (\gamma/\Omega_{\rm PBH,f})^{1/3}k_{\rm f}^{-1}$, where $k_\mathrm{f}$ is the typical PBH comoving scale crossing the horizon at PBH formation time. In particular, at distances  much larger than the mean PBH separation scale, namely $k^{-1}_\mathrm{UV}$, the PBH gas can be effectively treated as a pressureless fluid~\cite{Papanikolaou:2020qtd}. For scales smaller than $k^{-1}_\mathrm{UV}$ however, one enters the non-linear regime where ${\cal P}_{\delta_\mathrm{PBH}}(k)$ becomes larger than unity. In this regime, perturbation theory as well as the effective PBH pressureless fluid description break down. In the following, we consider scales where $k\leq k_\mathrm{UV}$.

Interestingly enough, the fact that PBHs can be viewed as discrete objects entails inhomogeneities in the PBH matter fluid, whereas the radiation background is homogeneous. Therefore, the initial PBH energy density perturbations being proportional to the PBH number density fluctuations can be regarded as isocurvature perturbations~\cite{Inman:2019wvr}, which are converted into adiabatic perturbations $\Phi$ deep in the PBH-driven eMD era. One can then  straightforwardly deduce~\cite{Papanikolaou:2020qtd,Domenech:2020ssp} that the power spectrum for the PBH gravitational potential 
$\Phi$ during the PBH-dominated era can be written in terms of the PBH matter power spectrum  as
\beq\label{eq:P_Phi_PBH_full}
\mathcal{P}_{\Phi}(k)=S^2_\Phi(k)\left(5+\frac{8}{9}\frac{k^2}{{k}^2_\mathrm{d}}\right)^{-2}\mathcal{P}_{\delta_\mathrm{PBH}}(k),
\eeq
where $\mathcal{P}_{\delta_\mathrm{PBH}}(k)$ is given by \Eq{eq:ppoi} and $k_\mathrm{d}$ is the comoving scale crossing the horizon at the onset of PBH-dominated era. $S_\Phi(k)$ is a suppression factor defined as $S_\Phi(k)\equiv \left({k}/{k_\mathrm{evap}}\right)^{-1/3}$, introduced here due to the fact that scales having a time variation which is larger than the PBH evaporation rate $\Gamma$, i.e. $k\gg \Gamma$, are effectively suppressed by the non-zero pressure of the radiation fluid~\cite{Inomata:2020lmk}. 

Accounting now for the effects of StRVMs, being seen as $f(R)$ gravity theories, on the aforementioned $\Phi$ power spectrum, one can show~\cite{Papanikolaou:2021uhe} that $\mathcal{P}_{\Phi}(k)$ will read as:
\begin{widetext}

\beq\label{eq:PowerSpectrum:Phi:PBHdom:f(R)}
\mathcal{P}_\Phi(k) = {\cal P}_{\delta_\mathrm{PBH}}(k)S^2_\Phi(k) \left[5+\frac{2}{3}\left(\frac{k}
{\mathcal{H}}\right)^2\frac{F}{\xi(a)} \left(\frac{1+ 3 \frac{k^2}{a^2} \frac{F_{,R}}{F}}{1+ 2 \frac{k^2}{a^2} \frac{F_{,R}}{F}}\right)\right]^{-2}.
\eeq
\end{widetext}
In the above expression, $\xi(a)$ is defined as
\beq\label{eq:xi_definition}
\xi(a)\equiv \frac{\delta_\mathrm{PBH}(a)}{\delta_\mathrm{PBH}(a_\mathrm{f})},
\eeq 
where ${\calH}$ is the conformal Hubble function and $\delta_\mathrm{PBH}(a)$ is the solution of
the sub-horizon Meszaros equation. Note that in the case of GR we have $F = 1$ and 
$\xi(a)\simeq \frac{3}{2}\frac{a}{a_\mathrm{d}}$ recovering the GR result.

\begin{figure}[h!]
\includegraphics[width=0.49\textwidth]{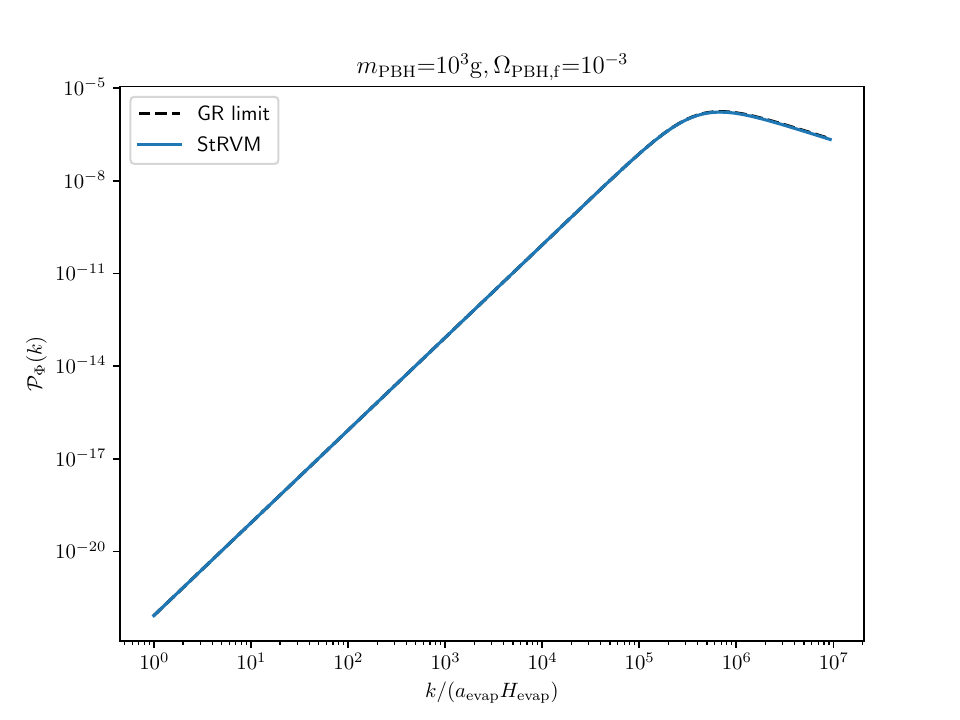}
\caption{\it{ The power spectrum of the PBH gravitational potential in GR (dashed black line) and in the StRVM framework (solid blue line) for $M_\mathrm{PBH} = 10^3\mathrm{g}$ and $\Omega_\mathrm{PBH,f}=10^{-3}$ and for $c_2/c_1 = 10^{-7}$.}}
\label{fig:P_Phi_StRVM}
\end{figure}

As one can see in \Fig{fig:P_Phi_StRVM}, where we plot $\mathcal{P}_\Phi(k)$ for $M_\mathrm{PBH} = 10^3\mathrm{g}$, $\Omega_\mathrm{PBH,f}=10^{-3}$ and $c_2/c_1 = 10^{-7}$, the deviations of the StRVM compared to GR is really small. The same conclusions are inferred by varying the PBH mass and the initial PBH abundance as long as $c_2/c_1<1$. One then can conclude that the effect of StRVMs at the level of the PBH gravitational potential power spectrum is negligible.

\subsection{The relevant PBH parameters}\label{sec:PBH_parameters}
Let us discuss here the relevant PBH parameters of the problem at hand. These are actually the PBH mass $M_\mathrm{PBH}$ and the initial PBH abundance at formation time $\Omega_\mathrm{PBH,f}$.
Regarding the PBH mass range, we need to stress that in our setup we focus on PBHs which  form after the end of inflation and evaporate before BBN time. This yields both, a lower and an upper bound on the PBH mass. In particular, on taking into account the current Planck upper bound on the tensor-to-scalar ratio for single-field slow-roll models of inflation, which gives  $\rho^{1/4}_\mathrm{inf}<10^{16}\mathrm{GeV}$ ~\cite{Planck:2018jri}, and that the PBH mass is given by \Eq{eq:PBH_mass}, by simply requiring that $\rho^{1/4}_\mathrm{f}\leq\rho^{1/4}_\mathrm{inf}<10^{16}\mathrm{GeV}$, one can show, using the Friedmann equation ($\rho = 3\Mp^2H^2$),  that $M_\mathrm{PBH}>1\mathrm{g}$. Then, by requiring that $\rho^{1/4}_\mathrm{evap}>\rho^{1/4}_\mathrm{BBN}\sim 1\mathrm{MeV}$ one arrives at $M_\mathrm{PBH}<10^9\mathrm{g}$. At the end, one has that~\cite{Papanikolaou:2020qtd}
\bea
\label{eq:domain:mPBHf}
10 \mathrm{g}< M_\mathrm{PBH}< 10^{9}\mathrm{g}.
\eea

Concerning now the range of $\Omega_\mathrm{PBH,f}$, one should require that PBHs evaporate after their PBH domination time, i.e. $t_\mathrm{evap}>t_\mathrm{d}$, in order to have eMD eras driven by PBHs. Thus, since $t_\mathrm{evap}\propto M^3_\mathrm{PBH}$~\cite{Hawking:1974rv}, $t_\mathrm{d}= 1/(2H_\mathrm{d})$ and $a_\mathrm{d}=a_\mathrm{f}/\Omega_\mathrm{PBH,f}$, one can straightforwardly show that 
\bea
\label{eq:Omega_PBH_f_min}
\Omega_\mathrm{PBH,f} > 7\times 10^{-14} \frac{10^4\mathrm{g}}{M_\mathrm{PBH}} .
\eea
Furthermore, one should account for the fact that GWs produced during an eMD era can contribute to the effective number of extra neutrino species $\Delta N_\mathrm{eff}$, which is tightly constrained by BBN and CMB observations as $\Delta N_\mathrm{eff}<0.3$~\cite{Planck:2018vyg}. This upper bound constraint on $\Delta N_\mathrm{eff}$ can then be translated to an upper bound on the GW amplitude  $\Omega_\mathrm{GW,0}h^2\leq 10^{-6}$~\cite{Smith:2006nka}. Employing this upper bound on the GW amplitude in the context of eMD eras driven by PBHs, one can derive, at least within GR, an upper bound on $\Omega_\mathrm{PBH,f}$ which reads as
\beq\label{eq:Omega_PBH_f_max}
\Omega_\mathrm{PBH,f}<10^{-6}\left(\frac{M_\mathrm{PBH}}{10^4\mathrm{g}}\right)^{-17/24}.
\eeq

Having derived here the relevant PBH parameter space, we are now well equipped to proceed to the study of light PBH-induced GWs within StRVMs, and thus derive constraints on the corresponding logarithmic quantum correction $c_2$ in that era.

\section{Scalar induced gravitational waves in the stringy running vaccum}
\label{sec:SIGW}
As already stated in the introduction, we shall consider a population of light PBHs, that is PBHs with masses smaller than $10^9\mathrm{g}$, which is present after the end of reheating. This population can dominate the energy budget of the Universe transiently after the end of inflation and evaporate before BBN. In the context of the StRVM~\cite{Basilakos:2019acj,ms1,ms2,Mavromatos:2022xdo} we will apply the analysis first presented in \cite{Papanikolaou:2021uhe} in which the gravitational wave signal induced by PBH energy density fluctuations is extracted in the framework of a generic $f(R)$ gravitational theory, since the RVM can easily be recast as an $f(R)$ theory as follows: 

\beq
f(R)= c_0+R\left(c_1+c_2\log\left(\frac{R}{R_0}\right)\right). \label{RVM}
\eeq

For our purposes, one of the most relevant features of $f(R)$ is the existence of an extra massive tensor polarization mode, the so-called ``scalaron" field \cite{Starobinsky:1980te} whose propagation equation reads as
\begin{equation}
    \Box F(R)= \frac{1}{3}\left[ 2f(R) - F(R)R + 8\pi G \, T^\mathrm{m} \right] \equiv \frac{dV}{dF}, \label{scalaron}
\end{equation}
where we have set $F \equiv \mathrm{d}f(R)/\mathrm{d}R$ and $T^\mathrm{m}$ is the trace of the energy-momentum tensor of the (total) matter content of the Universe. As we observe,  equation (\ref{scalaron}) is a wave equation for $\phi_{\mathrm{sc}} \equiv F(R)$ whose mass is given by $m_{\mathrm{sc}}^2 \equiv d^2 V/ dF^2$, reading as:
\begin{equation} \label{eq:scalmass}
    m_{\mathrm{sc}}^2 = \frac{1}{3}\left( \frac{F}{F_{,R}} - R \right),
\end{equation}
where $F_{,R} \equiv dF/dR = d^2f/dR^2$. In our running vacuum setup, from \eqref{RVM} we obtain that
\beq
   m_{\mathrm{sc}}^2= \frac{R}{3}\left(\frac{c_1}{c_2}+\log \left(\frac{R}{R_0}\right)\right) \label{mscal}
\eeq

\subsection{Tensor Perturbations}\label{subsec:tensor_perturbations}

We shall now study the tensor perturbations $h_{ij}$ induced by the gravitational potential $\Phi$. In particular, the perturbed metric in the Newtonian gauge, assuming as usual zero anisotropic stress and $\delta F/ F < 1$ [see \Fig{fig:k_evap}, \Fig{fig:k_d} and  \Fig{fig:k_UV} in Appendix \ref{app:Anisotropic_Stress}], is written as
\bea
\label{metric decomposition with tensor perturbations}
\mathrm{d}s^2 = a^2(\eta)\left\lbrace-(1+2\Phi)\mathrm{d}\eta^2  + \left[(1-2\Phi)\delta_{ij} + \frac{h_{ij}}{2}\right]\mathrm{d}x^i\mathrm{d}x^j\right\rbrace,
\eea
where we have multiplied by a factor $1/2$ the second order tensor perturbation as is standard in the literature. We remark that the contribution from the first-order tensor perturbations is not considered here since we concentrate on gravitational waves induced by scalar perturbations at second order.\footnote{It is worthy of pointing out that while tensor modes remain gauge invariant at first order, this isn't the case at second order \cite{Hwang_2017,Tomikawa:2019tvi,DeLuca:2019ufz,Yuan:2019fwv,Inomata:2019yww,Domenech:2020xin,Chang:2020tji}. This necessitates specifying the observational gauge for GWs. However, our study focuses on a GW backreaction problem, disregarding observational predictions. Specifically, if the energy density from induced GWs surpasses that of the background, perturbation theory is expected to fail regardless of the gauge. Thus, our findings are independent of the gauge choice.} Then, by Fourier transforming the tensor perturbations and taking into account the three polarization modes of the GWs in $f(R)$ gravity, namely the $(\times)$ and the $(+)$ as in GR and the scalaron one, denoted with $(\mathrm{sc})$, the equation of motion for the tensor modes $h_\boldmathsymbol{k}$ reads as
\beq
\label{Tensor Eq. of Motion}
h_\boldmathsymbol{k}^{s,\prime\prime} + 2\mathcal{H}h_\boldmathsymbol{k}^{s,\prime} + (k^{2}-\lambda m^2_\mathrm{sc}) h^s_\boldmathsymbol{k} = 4 S^s_\boldmathsymbol{k}\, ,
\eeq
where $\lambda=0$ when $s = (+), (\times)$ and $\lambda=1$ when $s=(\mathrm{sc})$. The scalaron mass term, $m^2_\mathrm{sc}$, is given by equation (\ref{mscal}) and the source term $S^s_\boldmathsymbol{k}$ can be written as 
\beq
\label{eq:Source:def}
\begin{split}
S^s_\boldmathsymbol{k} &  = \int\frac{\mathrm{d}^3 \boldmathsymbol{q}}{(2\pi)^{3/2}}e^s_{ij}(\boldmathsymbol{k})q_iq_j\Bigl[2\Phi_\boldmathsymbol{q}\Phi_\boldmathsymbol{k-q} \\ & + \frac{4}{3(1+w_\mathrm{tot})}(\mathcal{H}^{-1}\Phi_\boldmathsymbol{q} ^{\prime}+\Phi_\boldmathsymbol{q})(\mathcal{H}^{-1}\Phi_\boldmathsymbol{k-q} ^{\prime}+\Phi_\boldmathsymbol{k-q}) \Bigr],
\end{split}
\eeq
with $w_\mathrm{tot}$ being the effective parameter of state which arises from treating an $f(R)$ gravity theory as GR plus an effective curvature fluid (see e.g. \cite{Arjona:2018jhh} and \cite{Papanikolaou:2021uhe}) -its cumbersome explicit form is not important for our purposes. 

The polarization tensors  $e^{s}_{ij}(k)$ are defined as \cite{Capozziello:2011et}
\beq
e^{(+)}_{ij}(\boldmathsymbol{k}) = \frac{1}{\sqrt{2}}
\begin{pmatrix}
1 & 0 & 0\\
0 & -1 & 0 \\ 
0 & 0 & 0 
\end{pmatrix}, \quad
e^{(\times)}_{ij}(\boldmathsymbol{k}) = \frac{1}{\sqrt{2}}
\begin{pmatrix}
0 & 1 & 0\\
1 & 0 & 0 \\ 
0 & 0 & 0 
\end{pmatrix}, \nonumber
\eeq
\beq
e^{(\mathrm{sc})}_{ij}(\boldmathsymbol{k}) = \frac{1}{\sqrt{2}}
\begin{pmatrix}
0 & 0 & 0\\
0 & 0 & 0 \\ 
0 & 0 & 1 
\end{pmatrix}
,
\eeq
while the time evolution of the potential $\Phi$, considering only adiabatic perturbations, can be obtained from combining the first order perturbed field equations in $f(R)$ gravity, which yields \cite{Papanikolaou:2021uhe} 
\bea
\label{Bardeen potential 2}
\Phi_\boldmathsymbol{k}^{\prime\prime} + \frac{6(1+w_\mathrm{tot})}{1+3w_\mathrm{tot}}\frac{1}{\eta}\Phi_\boldmathsymbol{k}^{\prime} + w_\mathrm{tot}k^2\Phi_\boldmathsymbol{k} =0\, .
\eea
In the case of an MD era ($w_\mathrm{tot}\simeq 0$), the non-decaying solution of the above equation gives us a constant value of $\Phi$. Writing the solution of $\Phi$ as $\Phi_\boldmathsymbol{k}(\eta) = T_\Phi(\eta) \phi_\boldmathsymbol{k}$, where $\phi_\boldmathsymbol{k}$ is the value of the gravitational potential at some initial time (which here we consider it to be the time at which PBHs dominate the energy content of the Universe,  $x_\ud$) and $T_\Phi(\eta)$ is a transfer function one can normalise the MD transfer function to unity, i.e. $T_\Phi(\eta) = 1$. Consequently, \Eq{eq:Source:def} can be written in a more compact form as
\beq
\label{Source}
S^s_\boldmathsymbol{k}  =
\int\frac{\mathrm{d}^3 q}{(2\pi)^{3/2}}e^{s}(\boldmathsymbol{k},\boldmathsymbol{q})F(\boldmathsymbol{q},\boldmathsymbol{k-q},\eta)\phi_\boldmathsymbol{q}\phi_\boldmathsymbol{k-q}\, ,
\eeq
where
\bea
\label{F}
\!\!\!\!\!
F(\boldmathsymbol{q},\boldmathsymbol{k-q},\eta) & \equiv 2T_\Phi(q\eta)T_\Phi\left(|\boldmathsymbol{k}-\boldmathsymbol{q}|\eta\right) \\  & \kern-2em  + \frac{4}{3(1+w)}\left[\mathcal{H}^{-1}qT_\Phi^{\prime}(q\eta)+T_\Phi(q\eta)\right]
\\  & \kern-2em \cdot \left[\mathcal{H}^{-1}\vert\boldmathsymbol{k}-\boldmathsymbol{q}\vert T_\Phi^{\prime}\left(|\boldmathsymbol{k}-\boldmathsymbol{q}|\eta\right)+T_\Phi\left(|\boldmathsymbol{k}-\boldmathsymbol{q}|\eta\right)\right],
\eea
and $e^s_{ij}(\boldmathsymbol{k})q_iq_j \equiv e^s(\boldmathsymbol{k},\boldmathsymbol{q})$ can be written in terms of the spherical coordinates $(q,\theta,\varphi)$ of the vector $\bm{q}$ as 
\beq
e^s(\boldmathsymbol{k},\boldmathsymbol{q})=
\begin{cases}
\frac{1}{\sqrt{2}}q^2\sin^2\theta\cos 2\varphi \mathrm{\;for\;} s= (+)\\
\frac{1}{\sqrt{2}}q^2\sin^2\theta\sin 2\varphi  \mathrm{\;for\;} s= (\times) \\
\frac{1}{\sqrt{2}}q^2\cos^2\theta \mathrm{\;for\;} s= (\mathrm{sc})
\end{cases}
\, .
\eeq
At the end, \Eq{Tensor Eq. of Motion} can be solved by means of the Green's function formalism giving us a tensor perturbation $h^s_\boldmathsymbol{k}$ reading as
\bea
\label{tensor mode function}
a(\eta)h^s_\boldmathsymbol{k} (\eta)  = 4 \int^{\eta}_{\eta_\mathrm{d}}\mathrm{d}\bar{\eta}\,  G^s_\boldmathsymbol{k}(\eta,\bar{\eta})a(\bar{\eta})S^s_\boldmathsymbol{k}(\bar{\eta}),
\eea
where the Green's function  $G^s_{\bm{k}}(\eta,\bar{\eta})$ is actually the solution of the homogeneous equation being recast as
\beq\label{eq:Green_equation}
G_\boldmathsymbol{k}^{s,\prime\prime}(\eta,\bar{\eta})  + \left( k^{2} - \lambda m^2_\mathrm{sc} -\frac{a^{\prime\prime}}{a}\right)G^s_\boldmathsymbol{k}(\eta,\bar{\eta}) = \delta\left(\eta-\bar{\eta}\right) ,
\eeq
with the boundary conditions $\lim_{\eta\to \bar{\eta}}G^s_\boldmathsymbol{k}(\eta,\bar{\eta}) = 0$ and $ \lim_{\eta\to \bar{\eta}}G^{s,\prime}_\boldmathsymbol{k}(\eta,\bar{\eta})=1$.  

One then can extract the tensor power spectrum, $\mathcal{P}_{h}(\eta,k)$ for the different polarization modes, which is defined as follows:
\bea\label{tesnor power spectrum definition}
\langle h^r_{\boldmathsymbol{k}}(\eta)h^{s,*}_{\boldmathsymbol{k}^\prime}(\eta)\rangle \equiv \delta^{(3)}(\boldmathsymbol{k} - \boldmathsymbol{k}^\prime) \delta^{rs} \frac{2\pi^2}{k^3}\mathcal{P}^s_{h}(\eta,k),
\eea
where $s=(\times)$ or $(+)$ or $(\mathrm{sc})$.
At the end, after a lengthy but straightforward calculation, one gets $\mathcal{P}_{h}(\eta,k)$ reading as~\cite{Ananda:2006af,Baumann:2007zm,Kohri:2018awv,Espinosa:2018eve} 
\bea
\label{Tensor Power Spectrum}
\begin{split}   
\mathcal{P}^{(\times)\;\mathrm{or}\; (+)}_h(\eta,k) & = 4\int_{0}^{\infty} \mathrm{d}v\int_{|1-v|}^{1+v}\mathrm{d}u \left[ \frac{4v^2 - (1+v^2-u^2)^2}{4uv}\right]^{2}\\ & \times I^2(u,v,x)\mathcal{P}_\Phi(kv)\mathcal{P}_\Phi(ku),
\end{split}
\eea
whereas for the scalaron polarization one obtains that 
\bea
\begin{split} 
\label{Tensor Power Spectrum-scalaron}
\mathcal{P}^{(\mathrm{sc})}_h(\eta,k) & = 8\int_{0}^{\infty} \mathrm{d}v\int_{|1-v|}^{1+v}\mathrm{d}u \left[ \frac{ (1+v^2-u^2)^2}{4uv}\right]^{2} \\ & \times I^{2}(u,v,x)\mathcal{P}_\Phi(kv)\mathcal{P}_\Phi(ku).
\end{split} 
\eea
In Eqs.\eqref{Tensor Power Spectrum} and \eqref{Tensor Power Spectrum-scalaron} $u$ and $v$ are two auxiliary variables defined as $u \equiv |\boldmathsymbol{k} - \boldmathsymbol{q}|/k$ and $v \equiv q/k$, and the kernel function $I(u,v,x)$ is reads as
\bea
\label{I function}
I(u,v,x) = \int_{x_\mathrm{d}}^{x} \mathrm{d}\bar{x}\, \frac{a(\bar{x})}{a(x)}\, k\, G^s_{k}(x,\bar{x}) F_k(u,v,\bar{x}),
\eea
where $x=k\eta$. In a MD era as the one we consider here $T_\Phi = 1$, hence we obtain from \Eq{F} that $F=10/3$. 

At the end, one can straightforwardly show that the GW spectral abundance $\Omega_\mathrm{GW}(k,\eta)$ defined as the energy density contribution of GWs per logarithmic comoving scale, i.e. $\Omega_\mathrm{GW}(k,\eta)\equiv  \frac{1}{\rho_\mathrm{tot}}\frac{\mathrm{d}\rho_\mathrm{GW}(k,\eta)}{\mathrm{d}\ln k}$ can be recast as
\beq\label{eq:Omega_GW}
\Omega_\mathrm{GW}(k,\eta) = \frac{1}{96}\left(\frac{k}{\calH(\eta)}\right)^{2}\left[2\overline{\mathcal{P}^{(\times)}_h}(\eta,k)+\overline{\mathcal{P}^{(\mathrm{sc})}_h}(\eta,k)\right].
\eeq

\subsection{Redefining the stringy running vacuum parameters}\label{strvmparam}

Before proceeding with the calculation of the GW signal, let us make a brief discussion on the StRVM parameters, namely $c_0$ (\eqref{coval}), $c_1$ (\eqref{c1def}) and $c_2$. At this point, it is important to note a conceptual issue when using the effective action \eqref{eq:action} in the PBH- dominated era. Unfortunately, the derivation of this action holds only in de Sitter backgrounds,
at an one-loop approximation of weak quantum gravity~\cite{Fradkin:1983mq,Alexandre:2013iva,Alexandre:2013nqa,Alexandre:2014lla}. In the context of the StRVM~\cite{ms1}, which is based on superstring theory, such de-Sitter phases could occur: (i) either in the very early universe, shortly after the Big-Bang, which is assumed to be dominated by a phase of a dynamically broken (superstring-embeddable) supergravity~\cite{Alexandre:2013iva,Alexandre:2014lla}, characterised by an early (hill-top) inflationary era (approximately de Sitter)~\cite{Ellis:2013zsa,Alexandre:2014lla}, that precedes the RVM inflationary epoch, or (ii) at late eras, which are not supersymmetric, but during which the StRVM universe re-enters another approximately de-Sitter era. 

During the PBH-dominated era, the dominant spacetime background is far from de Sitter. Moreover, especially near light PBHs, which will introduce the most severe constraints of the model, perturbative quantum gravity is expected to cease to exist, and one enters possibly a strong gravity regime. In such a situation, the aforementioned derivations of \eqref{eq:action}~\cite{Alexandre:2013nqa,Alexandre:2014lla,bmssugra} are not valid, and thus the form 
\eqref{eq:action} does not give us the correct dynamics. It is not clear what modifications to the Einstein-Hilbert term one obtains in such a case, by including resummation (or actually non-perturbative string dynamics) effects. Even if the form \eqref{eq:action} were valid, the corresponding effective action should correctly 
represent an expansion about the spacetime background of the PBH-dominated era. Thus, 
the corresponding gravitational dynamics should be described by 
an action \eqref{eq:action} modified by the replacement of the quantity $R_0$ (which in late eras represents 
the current-era curvature scalar~\cite{Gomez-Valent:2023hov}) by 
\begin{align}\label{pbhcurv}
R_0 \, \rightarrow \, R_{\rm PBH}\,,
\end{align}
where $R_{\rm PBH}$ represents 
the curvature scalar during (light) PBH-domination era ({\it cf.} \eqref{rbh} below). In this setting, the coefficients 
$c_i$, $i=1,2$, in the current era appearing in \eqref{eq:action}  
should be replaced by the corresponding ones representing the quantum-gravity-corrected effective action at the epoch of primordial-PBH dominance:
\begin{align}\label{cipbh}
c_i \, \rightarrow \, c_i^{\rm {\small PBH}} \ne c_i\, \quad i=1,2\,. 
\end{align}
Under these assumptions, the gravitational model to be used in our phenomenology of the StRVM during the PBH domination will therefore be given by \eqref{eq:action} upon the replacements \eqref{pbhcurv} and \eqref{cipbh}, that is by:
\beq\label{eq:actionPBH}
\begin{split} 
S^{\rm PBH}= & \int d^4x\,\sqrt{-g}\,\Biggl\{c_0+R\Bigl[c_1^{\rm PBH} \\ & +c_2^{\rm PBH}\, \log\left(\frac{R}{R_{\rm PBH}}\right)\Bigr]+\mathcal{L}^{\rm PBH}_{m} \Biggr\} \,,
\end{split} 
\eeq
where $\mathcal{L}^{\rm PBH}_{m}$ denotes the corresponding matter Lagrangian during the PBH epoch. 

In the following section we shall make such an assumption \eqref{eq:actionPBH}, which is the most conservative, albeit, as explained above, not truly valid, scenario in a generic quantum gravity setting. Lacking a complete theory of quantum gravity, though, adopting such a scenario will allow us to impose much stronger constraints,   
in order to avoid GW overproduction during the era of PBH domination, on 
the coefficients $c_2^{\rm PBH}$ at that {\it era}, 
as compared to the corresponding values \eqref{eps2}, required for an alleviation of the current-era cosmological tensions~\cite{Gomez-Valent:2023hov}.

\subsection{The stringy running vacuum gravitational-wave propagator}
In this subsection we shall study GW propagation within the StRVM framework \eqref{eq:actionPBH}. In particular, we shall derive the modifications of the Green function \eqref{eq:Green_equation}, entering \Eq{tensor mode function}, which actually yields the GW propagator. To this end, we first note that for the polarisation modes $(\times)$ and $(+)$, one recovers the usual GR GW propagator, which, for a MD era, reads as: 
\beq
\begin{split} 
kG^{(\times)\;\mathrm{or}\;(+)}_k(x,\bar{x}) & = 
\frac{1}{x\bar{x}}\bigl[ (1+x\bar{x})\sin(x-\bar{x}) \\ & - (x-\bar{x})\cos(x-\bar{x})\bigr],
\end{split} 
\eeq
where $x \equiv k\eta$ and $\bar{x} \equiv k\bar{\eta}$.

We now proceed to derive the pertinent modifications of the scalaron GW propagator. For an eMD era, driven by PBHs, the scalaron mass \Eq{eq:scalmass}, $m^2_\mathrm{sc}$, is calculated to be:
\begin{widetext}
\beq\label{eq:m2_sc_full}
m^2_\mathrm{sc} = \frac{M^3_\mathrm{PBH}}{\eta^6 g_\mathrm{eff}H^3_0\Mp^4\Omega^{3/2}_{r,0}}\left[1286\frac{c_1^{\rm PBH}}{c_2^{\rm PBH}} + 2700 + 1286\ln\left(\frac{M^5_\mathrm{PBH}}{\eta^6\gamma^2 g_\mathrm{eff}H^3_0\Mp^8\Omega^4_\mathrm{PBH,f}\Omega^{3/2}_\mathrm{r,0}}\right)\right],
\eeq
\end{widetext}
where we have taken into account that the Ricci scalar during a PBH-dominated era is given by
\beq\label{rpbhdom}
R = \frac{3860M^3_\mathrm{PBH}}{\eta^6g_\mathrm{eff}H^3_0\Mp^4\Omega^{1/4}_{r,0}}\,,
\eeq
while the Ricci scalar at the onset of the PBH domination era, $R_\mathrm{PBH}$, can be recast as
\beq\label{rbh}
R_\mathrm{PBH}=\frac{474\gamma^2\Mp^4\Omega^4_\mathrm{PBH,f}}{M^2_\mathrm{PBH}}\,.
\eeq
For notational brevity in this and the following sections we shall use the notation $c_i$ for $c_i^{\rm PBH}$, $i=1,2$ that appear in \eqref{eq:actionPBH}. Since we exclusively concentrate here on the PBH-dominance epoch, there is no danger in bringing any confusion with such a notation. Substituting now in \Eq{eq:m2_sc_full} $g_\mathrm{eff} = 100$, being a typical order of magnitude of the number of the relativistic degrees of freedom at the time of PBH formation~\cite{Kolb:1990vq}, under the assumption that the standard model of particle physics describes the matter part $\mathcal L^{\rm PBH}_m$ of the effective action \eqref{eq:actionPBH},\footnote{We note though that, in general, in string-inspired models beyond the standard model of particle physics, this number can be much larger, up to about a 1000, in some phenomenologically  realistic string theories.} $\gamma \simeq 0.2$, being the fraction of the cosmological horizon collapsing into a PBH~\cite{Musco:2008hv}, $H_0 = 70\mathrm{km/s/Mpc}$, and $\Mp = 2.41\times 10^{18}\mathrm{GeV}$, we obtain: 
\beq\label{eq:m2_sc_vs_M_Omega_f_c1_c2}
m^2_\mathrm{sc} = \frac{3M^3_\mathrm{PBH}10^{60}}{\eta^6}\left[\frac{c_1}{c_2}+\ln\left(\frac{10^{-14}M^5_\mathrm{PBH}}{\eta^6\Omega^4_\mathrm{PBH,f}}\right) \right].
\eeq

\begin{figure*}[t!]
\centering
\includegraphics[width=0.49\textwidth]{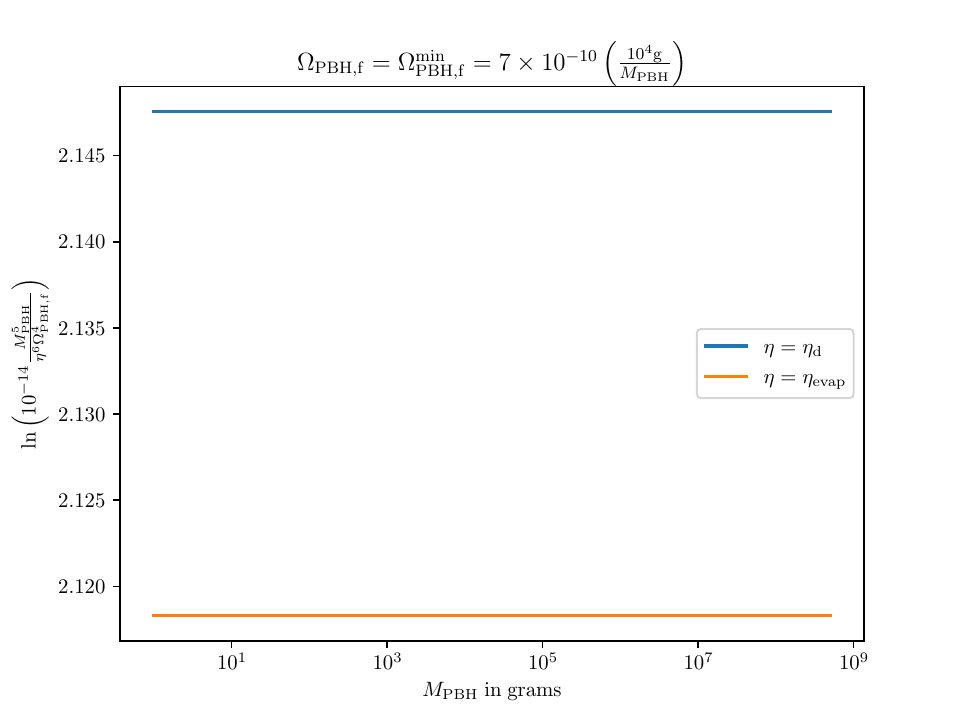}
\includegraphics[width=0.49\textwidth]{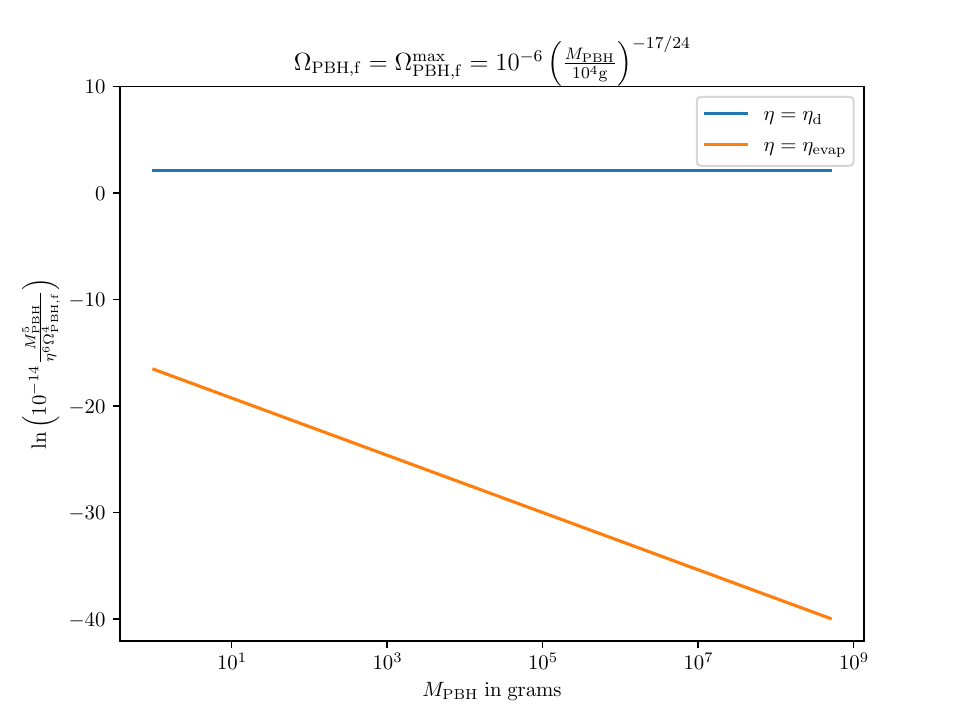}
\caption{\it{\underline{Upper Panel}: The logarithmic term $\ln\left(\frac{10^{-14}M^5_\mathrm{PBH}}{\eta^6\Omega^4_\mathrm{PBH,f}}\right)$ as a function of the PBH mass for the minimum value of the initial PBH abundance $\Omega^\mathrm{min}_\mathrm{PBH} = 7\times 10^{-10}\left(\frac{10^4\mathrm{g}}{M_\mathrm{PBH}}\right)$. \underline{Lower Panel}: The logarithmic term $\ln\left(\frac{10^{-14}M^5_\mathrm{PBH}}{\eta^6\Omega^4_\mathrm{PBH,f}}\right)$ as a function of the PBH mass for the maximum value of the initial PBH abundance $\Omega^\mathrm{max}_\mathrm{PBH} = 10^{-6}\left(\frac{M_\mathrm{PBH}}{10^4\mathrm{g}}\right)^{-17/24}$. }}
\label{fig:log_term}
\end{figure*}

Interestingly enough, for the relevant range of our parameter space $1\mathrm{g}<M_\mathrm{PBH}<10^9\mathrm{g}$, and $7\times 10^{-10}\frac{10^4\mathrm{g}}{M_\mathrm{PBH}}\leq\Omega_\mathrm{PBH,f}\leq 10^{-6}\left(\frac{M_\mathrm{PBH}}{10^4\mathrm{g}}\right)^{-17/24} $ [see discussion in \Sec{sec:PBH_parameters}], it was found numerically [see \Fig{fig:log_term}] that the  value of the logarithmic term $\ln\left(\frac{10^{-14}M^5_\mathrm{PBH}}{\eta^6\Omega^4_\mathrm{PBH,f}}\right)$ during the PBH-eMD era is below $10$. Thus since $c_1/c_2\gg 1$, in order for $c_2$ to be a quantum correction, the logarithmic term can be safely neglected, which leaves us with
\beq\label{scalmass2}
m^2_\mathrm{sc} (\eta,m_\mathrm{PBH}) \approx \frac{c_m c_1 m_{\mathrm{PBH}}^3}{3 c_2  \, \eta^6},
\eeq
where $c_m\equiv 3 \times 10^{60} \mathrm{GeV}^{-7}$.

One then can check numerically that the ``corrected" scalaron mass term \eqref{scalmass2} is the dominant one compared to the $ k^2$ and $a''/a$ ($= 2/\eta^2$ in a MD era) present in \Eq{eq:Green_equation} [see \Fig{fig:a_prime_prime_o_a_o_msc2} and \Fig{fig:k_2_o_msc2} of Appendix \ref{app:Green_function_terms}]. Thus, \Eq{eq:Green_equation} can be recast as
\beq
G_\boldmathsymbol{k}^{\mathrm{sc},\prime\prime}(\eta,\bar{\eta}) - \left( \frac{c_m c_1 m_{\mathrm{PBH}}^3}{3 c_2  \, \eta^6} \right)G^\mathrm{sc}_\boldmathsymbol{k}(\eta,\bar{\eta}) = \delta\left(\eta-\bar{\eta}\right),
\eeq
whose solution is
\begin{widetext}
\bea
kG^\mathrm{sc} (x , \bar{x}) & = \frac{\sqrt{x \bar{x}}}{2 \sqrt{2}} \pi  \Biggl[I_{-\frac{1}{4}}\left(\frac{\sqrt{c_1} \sqrt{c_m} M_{\mathrm{PBH}}^{3/2}k^2}{2 \sqrt{3} \sqrt{c_2} x^2}\right)+ I_{\frac{1}{4}}\left(\frac{\sqrt{c_1} \sqrt{c_m} M_{\mathrm{PBH}}^{3/2}k^2}{2 \sqrt{3} \sqrt{c_2} \bar{x}^2}\right) \\ & -I_{\frac{1}{4}}\left(\frac{\sqrt{c_1} \sqrt{c_m} M_{\mathrm{PBH}}^{3/2}k^2}{2 \sqrt{3} \sqrt{c_2} x^2}\right) I_{-\frac{1}{4}}\left(\frac{\sqrt{c_1} \sqrt{c_m} M_{\mathrm{PBH}}^{3/2}k^2}{2 \sqrt{3} \sqrt{c_2} \bar{x}^2}\right) \Biggr]
\eea
\end{widetext}
where $I_n$ is the modified Bessel function of the first kind. At the end, the kernel function for a MD era where $F = 10/3$ reads as
\begin{widetext}
\beq
\begin{split}
 I^{(\mathrm{sc})2}_\mathrm{MD}(x) & = \frac{25}{63504x^{12}}\Bigg\{ 56 _0F_1\left(\frac{3}{4};\frac{c_v^2}{4x^4}\right) \Biggl[x^3~_1F_2\left(-\frac{3}{4};\frac{1}{4},\frac{5}{4};\frac{c_v^2}{4x^4}\right)  -x^3_\mathrm{d}~_1F_2\left(-\frac{3}{4};\frac{1}{4},\frac{5}{4};\frac{c_v^2}{4x^4_\mathrm{d}}\right) \Biggr] + \frac{1}{x^4_\mathrm{d}}~ _0F_1\left(\frac{5}{4};\frac{c_v^2}{4x^4}\right) \\ & \times \Biggl[c_v^4x^4_\mathrm{d}~_2F_3\left(1,1;2,\frac{11}{4},3;\frac{c_v^2}{4x^4}\right) - c_v^4x^4~_2F_3\left(1,1;2,\frac{11}{4},3;\frac{c_v^2}{4x^4_\mathrm{d}}\right) + 14x^4x^4_\mathrm{d}\Bigl( -3x^4 + 3x^4_\mathrm{d} + 4 c_v^2 \ln \frac{x_\mathrm{d}}{x} \Bigr) \Biggr] \Biggr\}^2,
\end{split}
\eeq
\end{widetext}
where $c_v \equiv 10^{30}\mathrm{GeV}^{-7/2} \sqrt{\frac{c_1}{c_2}}M^{3/2}k^2$ and $_pF_q({a_1,..a_p};{b_1,...,b_q},z)$ is the generalised hypergeometric function.

\section{Constraining the stringy running vacuum logarithmic corrections}\label{sec:constraints}
Having discussed in previous sections the PBH gas and the associated scalar-induced GWs within the context of StRVMs, we now proceed to impose constraints on the parameter $c_2$ of the StRVM quantum logarithmic correction \eqref{eq:actionPBH},  using the aforementioned GW portal. 

\begin{figure}[h]
\includegraphics[width=0.49\textwidth]{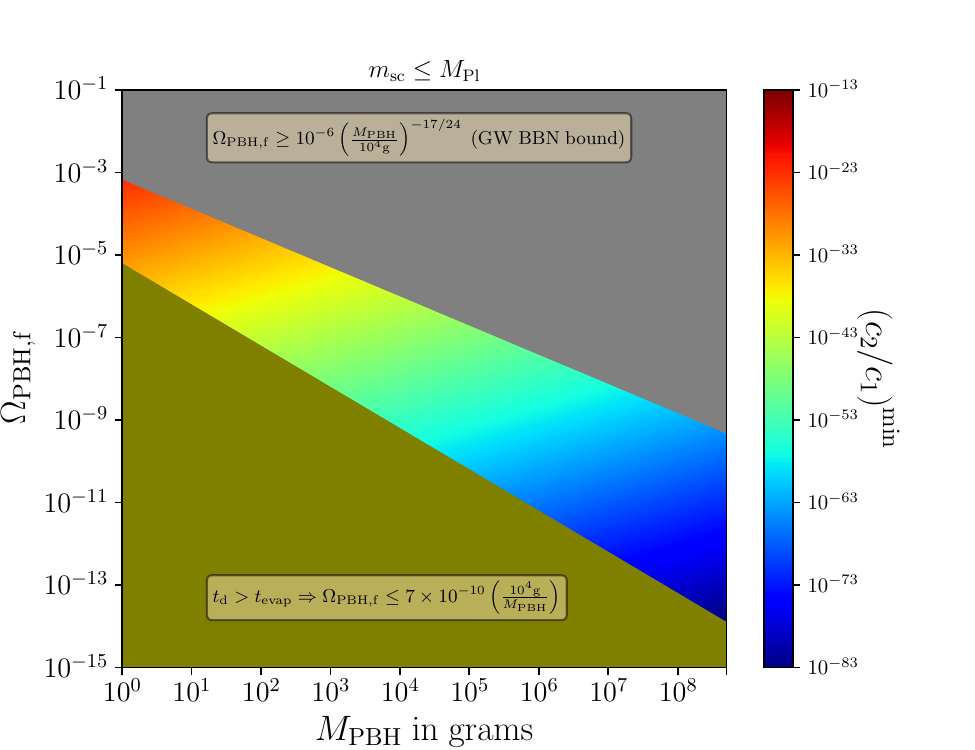}
\caption{\it{The lower bound constraints on $c_2/c_1$ (color bar axis) as a function of the PBH mass $M_\mathrm{PBH}$ ($x$-axis) and the initial PBH abundance at formation $\Omega_\mathrm{PBH,f}$ ($y$-axis) accounting for the fact that scalaron mass should be smaller than the Planck scale.}}
\label{fig:c_2_c_1_M_sc_constraints}
\end{figure}

Before doing so, it is worthwhile to clarify the status of the scalaron polarisation mode. Interestingly enough, if one views the scalaron as a new physical degree of freedom~\cite{Capozziello:2011et}, one should require that its mass be smaller than the UV cut-off of the theory, which in the context of StRVM can be naturally taken to be the string scale $M_\mathrm{s}$~\cite{Mavromatos:2022xdo}. Since $M_\mathrm{s}$ should be smaller than, or at most equal to, the Planck scale $\Mp$,  one obtains $m_\mathrm{sc}\leq \Mp$. The regime where $m_\mathrm{sc}\geq \Mp$ corresponds actually to the limit where $c_2\rightarrow 0$ (GR limit) where $m_\mathrm{sc}\rightarrow \infty$. In this large mass regime, the scalaron becomes in fact a ``heavy" mode, not contributing to the polarisation modes of GWs. Consequently, by the simple requirement that the scalaron mass should be smaller than the reduced Planck scale $\Mp$ at both, the onset of the PBH domination era ($\eta_\mathrm{d}$), and the PBH evaporation time ($\eta_\mathrm{evap}$), 
so as to avoid transplanckian modes, one can derive a lower limit on the ratio $c_2/c_1$ as a function of the PBH mass and their initial abundance at formation time reading as
\beq\label{c_2_o_c_1_constraints_M_Sc}
\frac{c_2}{c_1}\geq  10^{-46} \mathrm{max}\left[\left(\frac{\Omega_\mathrm{PBH}}{10^{-7}}\right)^4\left(\frac{10^4\mathrm{g}}{M_\mathrm{PBH}}\right)^2\, ,\, 10^{-9}\left(\frac{10^4\mathrm{g}}{M_\mathrm{PBH}}\right)^6\right],
\eeq
where the right term within the brackets of the right-hand side of \Eq{c_2_o_c_1_constraints_M_Sc} comes by requiring $m_\mathrm{sc}(\eta_\mathrm{evap},M_\mathrm{PBH})<\Mp$ while the left one originates by demanding that $m_\mathrm{sc}(\eta_\mathrm{d},M_\mathrm{PBH})<\Mp$. In \Fig{fig:c_2_c_1_M_sc_constraints}, we show this lower bound constraints on the ratio of $c_2/c_1$ as a function of $M_\mathrm{PBH}$ and $\Omega_\mathrm{PBH,f}$.

Focusing now on the GW amplitude on the PBH-driven eMD era, one can constrain $c_2$ just by requiring that the GW amplitude at the end of the PBH-dominated era, namely at the end of PBH evaporation, should be smaller than unity so as to avoid GW overproduction. Remarkably, $I^{(\mathrm{sc})2}_\mathrm{MD}(x_\mathrm{evap})$ diverges for values of $c^2_v/(4x^4) \gtrsim 1000 $ signalling a GW backreaction issue, since the spectral abundance of the induced GWs associated with the scalaron mode produced during the PBH-dominated era is proportional to $I^{(\mathrm{sc})2}_\mathrm{MD}(x)$, as it can be seen from \Eq{Tensor Power Spectrum-scalaron} and \Eq{eq:Omega_GW}, or in other words $\Omega^{(\mathrm{sc})}_\mathrm{GW}(\eta,k)\propto I^{(\mathrm{sc})2}_\mathrm{MD}(x)$ (see also~\cite{Maggiore_2000}). Therefore, in order to avoid such a GW overproduction issue, one should require that $c^2_v/(4x^4) \leq 1000$. Since $x_\mathrm{evap}>x_\mathrm{d}$, we require conservatively that $c^2_v/(4x^4_\mathrm{d}) \leq 1000$ leading to the following lower bound constraint on $c_2/c_1$:
\beq\label{eq:c_2_o_c_1_constraints_GW}
\frac{c_2}{c_1}\geq 2\times 10^{59}\left(\frac{10^4\mathrm{g}}{M_\mathrm{PBH}}\right)^{1/3} \Omega^{1/3}_\mathrm{PBH,f}\,.
\eeq

\begin{figure}[h]
\includegraphics[width=0.49\textwidth]{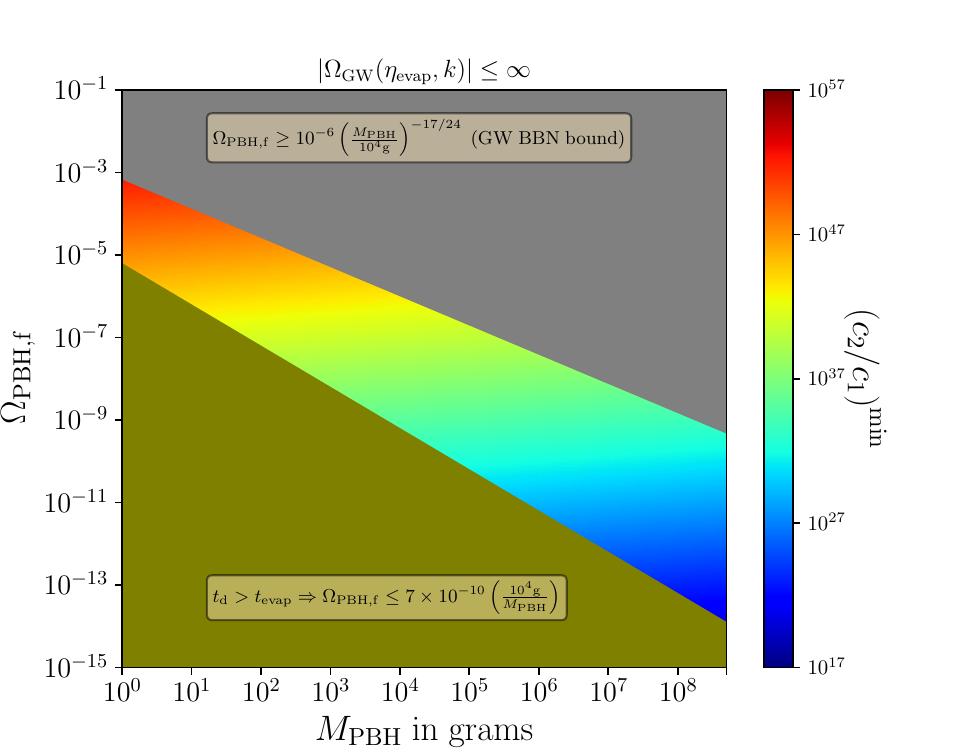}
\caption{\it{ The lower bound constraints on $c_2/c_1$ (color bar axis) as a function of the PBH mass $M_\mathrm{PBH}$ ($x$-axis) and the initial PBH abundance at formation $\Omega_\mathrm{PBH,f}$ ($y$-axis) accounting for GW overproduction during an eMD era driven by light PBHs.}}
\label{fig:c_2_o_c_1_constraints_GW}
\end{figure}

In \Fig{fig:c_2_o_c_1_constraints_GW} we show the $c_2/c_1$ GW lower bound constraints \eqref{eq:c_2_o_c_1_constraints_GW} as a function of the PBH mass $M_\mathrm{PBH}$ and the initial PBH abundance $\Omega_\mathrm{PBH,f}$. Notably, in the entirety of the relevant parameter space of our physical setup, the lower bound on $c_2/c_1$ is too large, many orders of magnitude larger than unity. As already mentioned, for consistency within a perturbative quantum gravity framework (see \Sec{strvmparam}), such a ratio has to be much smaller than one, which leads to constraints on the relevant PBH parameters, in particular on the ratio $\Omega_\mathrm{PBH,f}/M$~\footnote{We need to mention that our analysis does not hold only for the case of an eMD driven by light PBHs but it can be extended as well to any physical setups accepting eMD eras such as the ones driven by  massive scalar~\cite{Basilakos:2023jvp} or vector ~\cite{Dimopoulos:2006ms} fields as well as Q-balls~\cite{Kawasaki:2023rfx}. In the latter cases, the lower bound constraints on the ratio $c_2/c_1$ and the consistency of the StRVMs with such eMD eras would depend on the parameters of the physical setup at hand.}. Thus, as it can be seen from \Eq{eq:c_2_o_c_1_constraints_GW}, in order to get a lower bound on $c_2/c_1$ smaller than unity, one should either go to very large PBH masses, above $10^{9}\mathrm{g}$, or assume very small initial PBH abundances, many orders of magnitude below the lower bound on $\Omega_\mathrm{PBH,f}$ \eqref{eq:Omega_PBH_f_min}, required from the existence of a PBH-driven eMD era. 

Consequently, one can conclude that eMD eras driven by light PBHs are inconsistent with quantum logarithmic corrections within the StRVM framework. This is somehow justified based on theoretical grounds [see also discussion in \Sec{strvmparam}] since the logarithmic corrections present in \Eq{eq:action} were derived by quantising StRVMs within a de-Sitter background, which is not the case in MD era. Interestingly enough, if in the future we detect light PBHs with masses smaller than $10^9\mathrm{g}$, this will further necessitate  the recasting of the StRVM quantum logarithmic corrections.

At this point, it is important to stress that such stringy quantum corrections can affect PBHs themselves as well, e.g. the PBH lifetime. In particular, regarding the PBH evaporation process, there have been some recent works~\cite{Alexandre:2024nuo,Thoss:2024hsr,Dvali:2024hsb} suggesting that quantum effects such as memory burden puts the process of Hawking evaporation out of the self-similar semi-classical regime prolonging at the end the black hole evaporation time. In such a case, a prolonged PBH-dominated era will lead to enhanced GW amplitude thus to smaller $c_2/c_1$ lower bounds, potentially smaller to unity since the GW amplitude scales inversely with the ratio  $c_2/c_1$  as it can be seen by Fig. 4. However, since such quantum corrections on the Hawking evaporation time have been treated up to now only at the phenomenological level~\cite{Dvali:2020wft,Balaji:2024hpu}, one cannot give a definite answer whether or not higher than one lower bound constraints on the ratio $c_2/c_1$ can be avoided. To address this issue, one needs to perform a quite technical calculation extracting the PBH evaporation time within the framework of StRVMs, something which goes beyond the scope of the current work.
\section{Discussion}\label{sec:conclusions}

In this work we have studied GWs induced by the energy density perturbations of ultra-light PBHs ($M_\mathrm{PBH}<10^9\mathrm{g}$) within the framework of the stringy running-vacuum models (StRVMs). We accounted for the effects of employing StRVMs both, on the source and on the propagation of the induced GW signal. Regarding the effect on the source of the induced GW signal, namely the power spectrum of the PBH gravitational potential, $\mathcal{P}_\mathrm{\Phi}(k)$, we found a minor effect of StRVMs giving us a behaviour similar to that of GR. See \Fig{fig:P_Phi_StRVM}.

By treating the scalaron mode as a physical degree of freedom, we required its mass to be smaller than the UV cut-off of the theory which can saturate at the Planck scale. Interestingly enough, imposing such a requirement we found a lower bound constraint on $c_2$ as a function of $M_\mathrm{PBH}$ and $\Omega_\mathrm{PBH,f}$.

Focusing on the GW propagator, namely the Green function of the tensor perturbations, we found considerable GW overproduction for small values of the StRVM quantum correction $c_2$. In order to avoid such a GW overproduction issue, we found that $c_2$ should be unnaturally large, at least greater than $10^{17}$ in reduced Planck mass units square, something which is inconsistent with the quantum nature of the correction $c_2$, which was derived when expanding StRVMs around de Sitter backgrounds. 

There are thus two possible ways of interpreting these very large lower bounds on the value of the $c_2$ coefficient of the quantum corrections: (i) if we consider the scalaron field as a physical mode ($m_\mathrm{sc}\leq\Mp$), then we find that, the requirement of avoiding GW overproduction, leads to a lower limit of the quantity $c_2/c_1$ of more than $10^{17}$ [see Fig. 2]. This signals that this quantum correction should not be valid in the early Universe. One can then claim that in order to solve such an inconsistency, they should impose GR in such early cosmic times before BBN. As a consequence, $m_\mathrm{sc}>\Mp$, leading to $c_2/c_1$ at best smaller than $10^{-13}$ [see Fig. 1], $6$ orders of magnitude tighter than the $c_2/c_1$ constraint derived from late-Universe observations from the requirement of alleviating the observed cosmic tensions~\cite{Gomez-Valent:2023hov}, giving us $c_2/c_1 \sim 10^{-7}$.
(ii) On the other hand, if we do not consider the scalaron field as a physical mode, then the scalaron mass is unbounded from above. In such a case, one obtains only an extremely large lower bound constraint of $c_2/c_1$ [see \Fig{fig:c_2_o_c_1_constraints_GW}] in order to avoid GW overproduction. This points, therefore, to the fact that, either there are no light PBHs in case 
the dynamics of the physical early universe is described by stringy RVMs with actions of the form \Eq{eq:actionPBH}, or, if light PBHs are detected in the future, then stringy RVMs should be reconsidered due to GW overproduction.

We recapitulate by stressing that the aforementioned analysis does not rule out the StRVM, but dictates that the corrections to the Einstein gravity induced by quantum-graviton loops are necessarily very different during the light PBH-dominated era, 
as compared to those in the modern epoch, otherwise one would be faced with an unacceptably large GW amplitude. In particular, if such corrections were logarithmic in the scalar curvature, then their coefficients in the light-PBH-domination era should be much more enhanced than their counterparts in the current era in order to avoid GW overproduction. This non trivial result demonstrates the importance of PBHs as probes of new physics.

{\bf Acknowledgements} -- . 
TP acknowledges the support of the INFN Sezione di Napoli \textit{initiativa specifica} QGSKY. SB, TP, ENS and CT acknowledge the 
contribution of the LISA Cosmology Working Group. 
All authors acknowledge participation in the COST Action CA21136 ``Addressing observational tensions in cosmology with systematics and 
fundamental physics (CosmoVerse)''.
TP and CT acknowledge as well financial support from the Foundation for Education and European Culture in Greece and A.G. Leventis Foundation respectively.
The work of NEM is partially supported by the UK Science and Technology Facilities research Council (STFC) and UK Engineering and Physical Sciences Research Council (EPSRC) under the research grants  ST/X000753/1 and  EP/V002821/1, respectively. 

\section{Appendix A: The anisotropic stress}\label{app:Anisotropic_Stress}
\begin{figure*}
\centering
\includegraphics[width=0.49\textwidth]{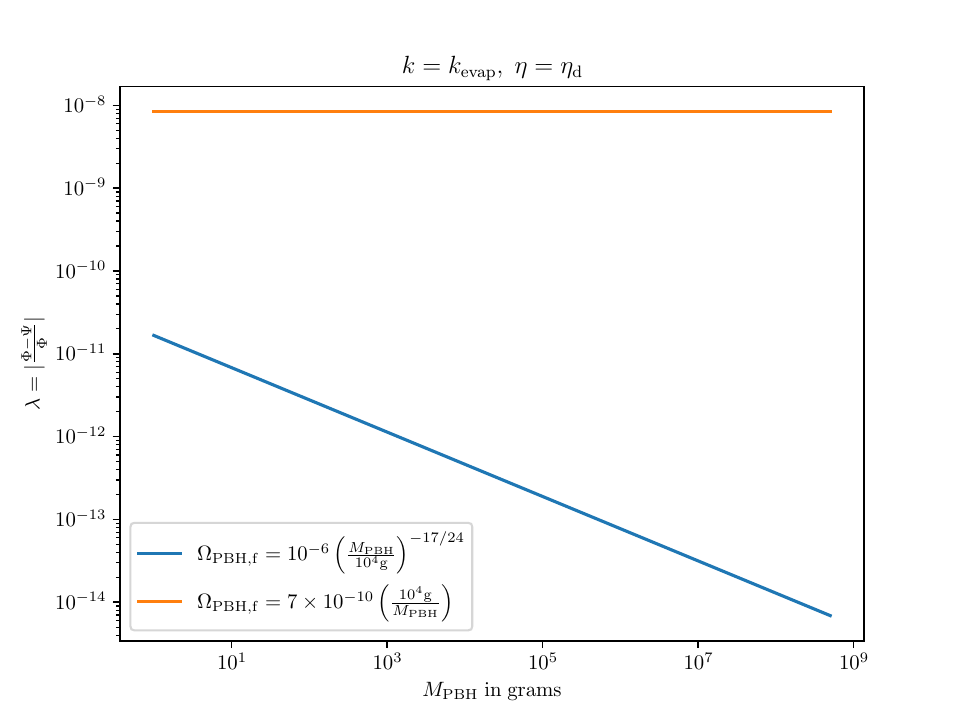}
\includegraphics[width=0.49\textwidth]{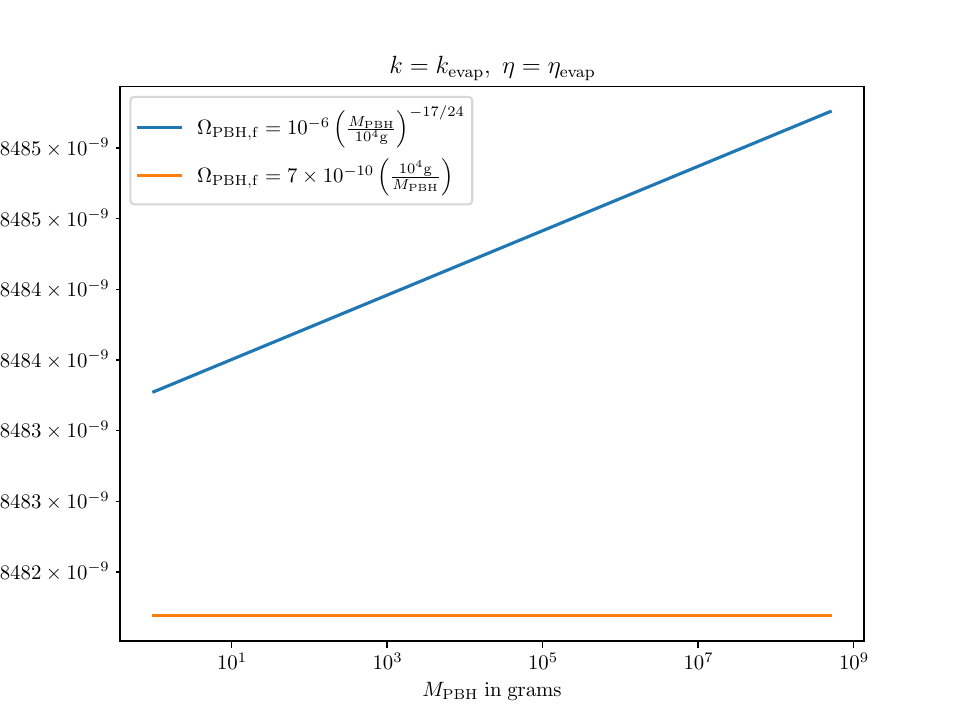}
\caption{\it{\underline{Left Panel}: The dimensionless parameter $\lambda$ for $k=k_\mathrm{evap}$ and $\eta =\eta_\mathrm{d}$ for different values of the PBH masses $M_\mathrm{PBH}$ and the initial PBH abundances $\Omega_\mathrm{PBH,f}$. 
\underline{Right Panel}: The dimensionless parameter $\lambda$ for $k=k_\mathrm{evap}$ and $\eta =\eta_\mathrm{evap}$ for different values of the PBH masses $M_\mathrm{PBH}$ and the initial PBH abundances $\Omega_\mathrm{PBH,f}$. }}
\label{fig:k_evap}
\end{figure*}

\begin{figure*}
\centering
\includegraphics[width=0.49\textwidth]{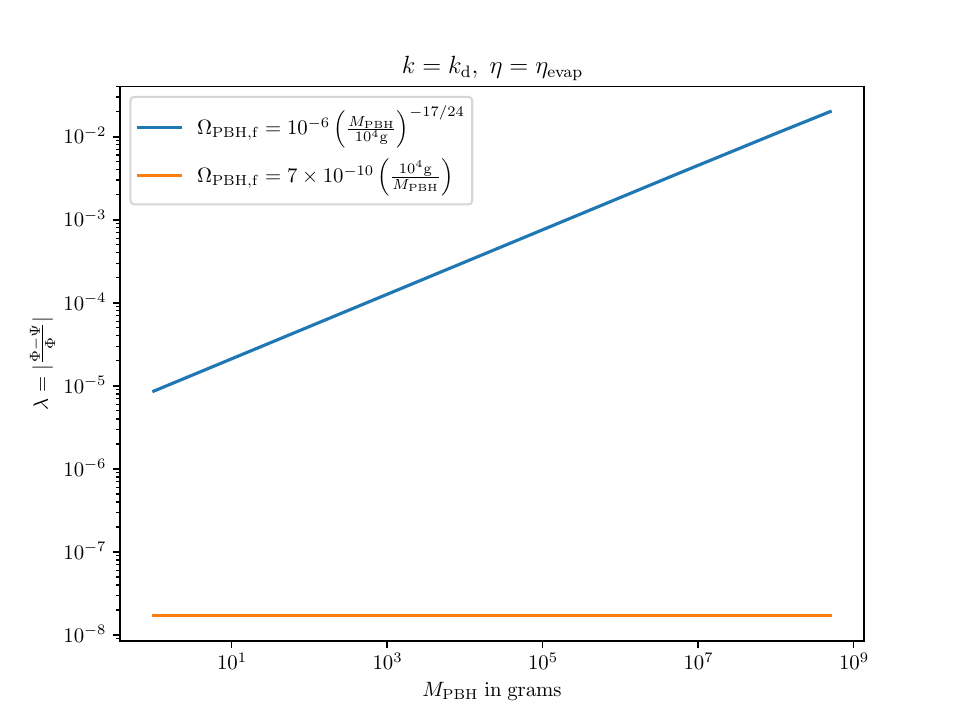}
\includegraphics[width=0.49\textwidth]{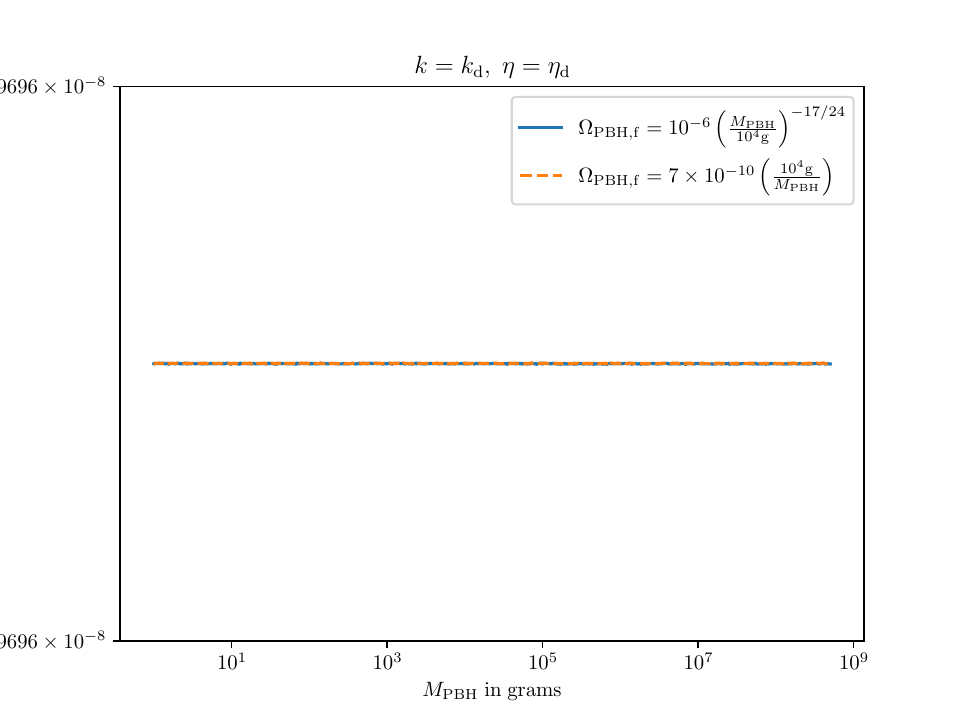}
\caption{\it{\underline{Left Panel}: The dimensionless parameter $\lambda$ for $k=k_\mathrm{d}$ and $\eta =\eta_\mathrm{evap}$ for different values of the PBH masses $M_\mathrm{PBH}$ and the initial PBH abundances $\Omega_\mathrm{PBH,f}$. \underline{Right Panel}: The dimensionless parameter $\lambda$ for $k=k_\mathrm{d}$ and $\eta =\eta_\mathrm{d}$ for different values of the PBH masses $M_\mathrm{PBH}$ and the initial PBH abundances $\Omega_\mathrm{PBH,f}$.}}
\label{fig:k_d}
\end{figure*}

\begin{figure*}
\centering
\includegraphics[width=0.49\textwidth]{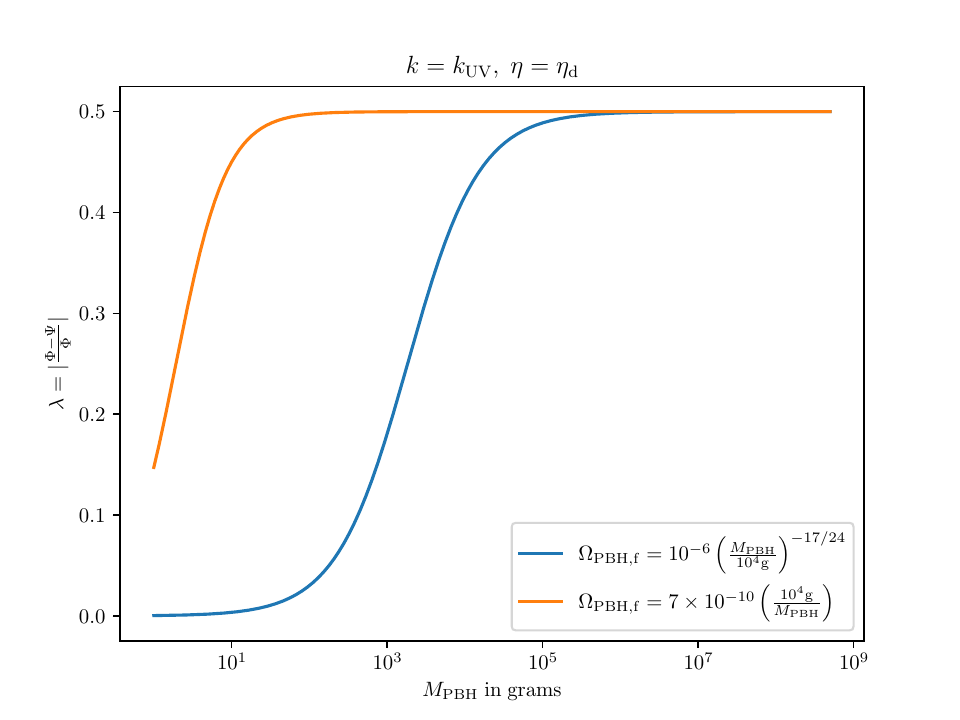}
\includegraphics[width=0.49\textwidth]{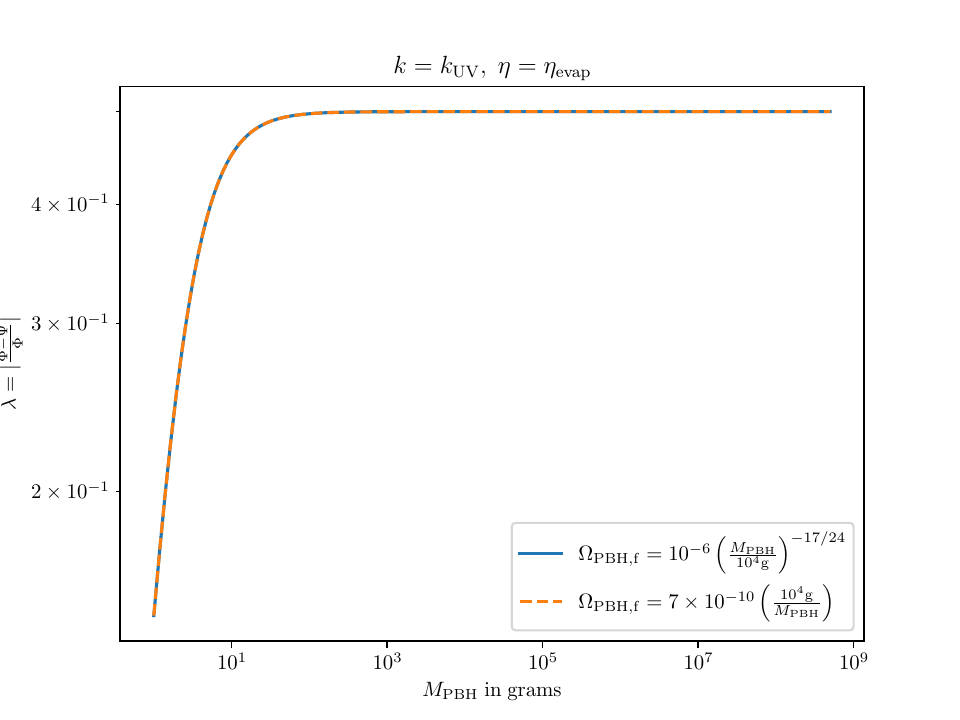}
\caption{\it{\underline{Left Panel}: The dimensionless parameter $\lambda$ for $k=k_\mathrm{UV}$ and $\eta =\eta_\mathrm{d}$ for different values of the PBH masses $M_\mathrm{PBH}$ and the initial PBH abundances $\Omega_\mathrm{PBH,f}$. \underline{Right Panel}: The dimensionless parameter $\lambda$ for $k=k_\mathrm{UV}$ and $\eta =\eta_\mathrm{evap}$ for different values of the PBH masses $M_\mathrm{PBH}$ and the initial PBH abundances $\Omega_\mathrm{PBH,f}$. }}
\label{fig:k_UV}
\end{figure*}

Before BBN, which is the period we focus on, there are no free streaming particles, namely neutrinos or photons, with the dominant matter species being in the form of PBHs. We can then safely consider that $\Pi_\mathrm{r}=\Pi_\mathrm{m}=0$ with the only source of anisotropic stress being the $f(R)$ gravity effective fluid, having a pure geometrical origin. In particular, following~\cite{Papanikolaou:2021uhe} one can show that the scalar metric perturbations $\Phi$ and $\Psi$ are related as
\beq\label{eq:Phi-Psi_geometrical}
\Phi - \Psi = \frac{\delta F}{F},
\eeq
with $\delta F = F_{,R}\delta R$ and $\delta R$, denoting the first order perturbation of the scalar curvature, which can be recast as~\cite{Tsujikawa:2007gd}
\beq\label{eq:delta_R}
\delta R = -2 \frac{k^2}{a^2}\frac{\Phi}{1 + 4\frac{k^2}{a^2}\frac{F_{,R}}{F}}.
\eeq
Defining thus the dimensionless quantity $\lambda$ as 
\beq
\lambda \equiv \frac{\Phi - \Psi}{\Phi},
\eeq
we can actually quantify the anisotropic stress of geometrical origin. In the case of $f(R)$ gravity, which category StRVMs belong to, plugging \Eq{eq:delta_R} into $\delta F = F_{,R}\delta R$ and then $\delta F$ into \Eq{eq:Phi-Psi_geometrical}, one obtains that
\beq\label{eq:gravitional_slip}
\lambda = \frac{-2\frac{k^2}{a^2}\frac{F_{,R}}{F}}{1 + 4\frac{k^2}{a^2}\frac{F_{,R}}{F}}.
\eeq
Below we plot this quantity for different values of the wave number $k$ within the range $[k_\mathrm{evap},k_\mathrm{UV}]$, and for different values of the PBH masses $M_\mathrm{PBH}$ within the range $[10\mathrm{g},10^9\mathrm{g}]$. With regards to the initial PBH abundance $\Omega_\mathrm{PBH,f}$, the latter is considered  within the range defined by \Eq{eq:Omega_PBH_f_min} and \Eq{eq:Omega_PBH_f_max}. As one may see from \Fig{fig:k_evap}, \Fig{fig:k_d} and \Fig{fig:k_UV},  the parameter $\lambda$ is much smaller than one, except for values of $k$ close to the UV cut-off where $\lambda$ is close to $0.5$, still remaining though less than unity. 
One then can legitimately neglect the anisotropic stress of geometrical origin and take $\Phi = \Psi$ (Newtonian gauge) in the derivation of the induced GW spectrum.

\section{Appendix B: The different terms in the Green equation}\label{app:Green_function_terms}
\begin{figure}
\centering
\includegraphics[width=0.49\textwidth]{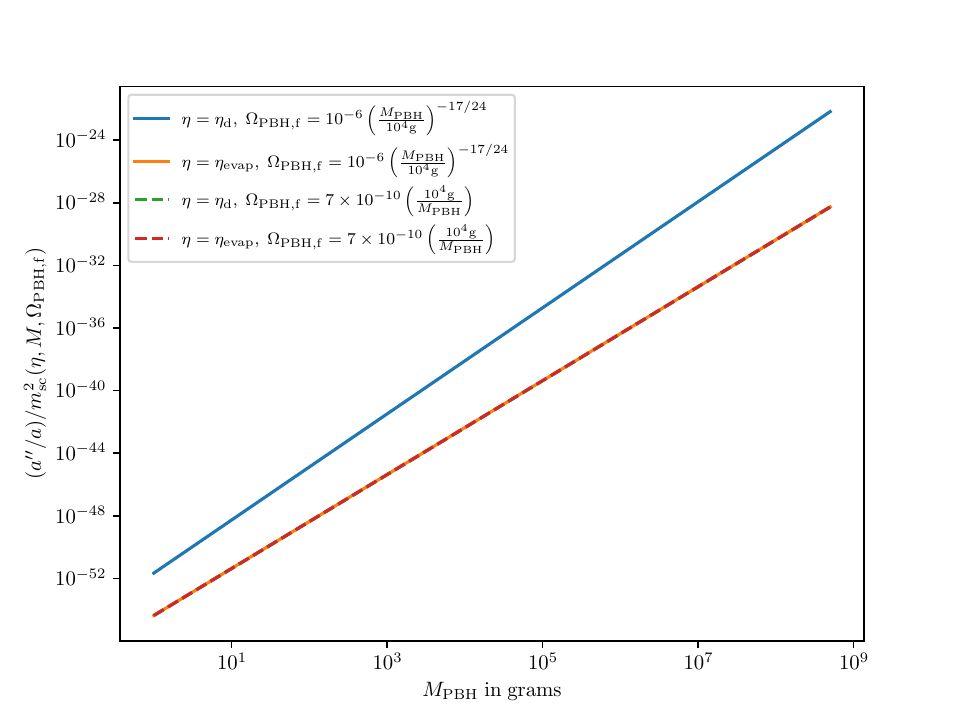}
\caption{\it{The ratio $(a^{\prime\prime}/a)/m^2_\mathrm{sc}$ as a function of the PBH mass for different values of the initial PBH abundances $\Omega_\mathrm{PBH,f}$.}}
\label{fig:a_prime_prime_o_a_o_msc2}
\end{figure}

\begin{figure*}
\centering
\includegraphics[width=0.49\textwidth]{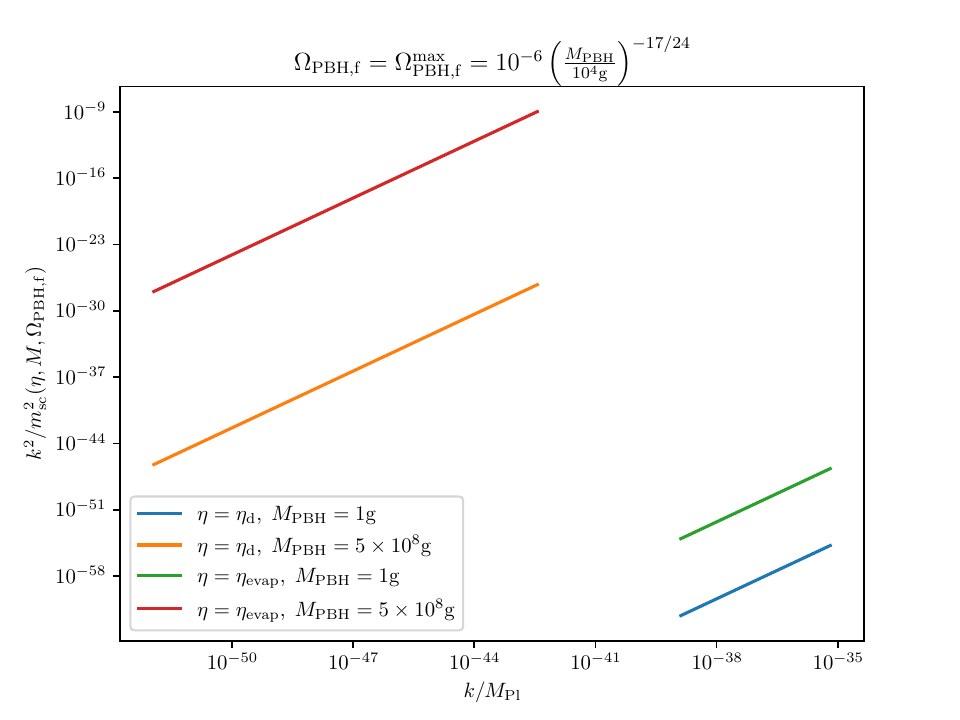}
\includegraphics[width=0.49\textwidth]{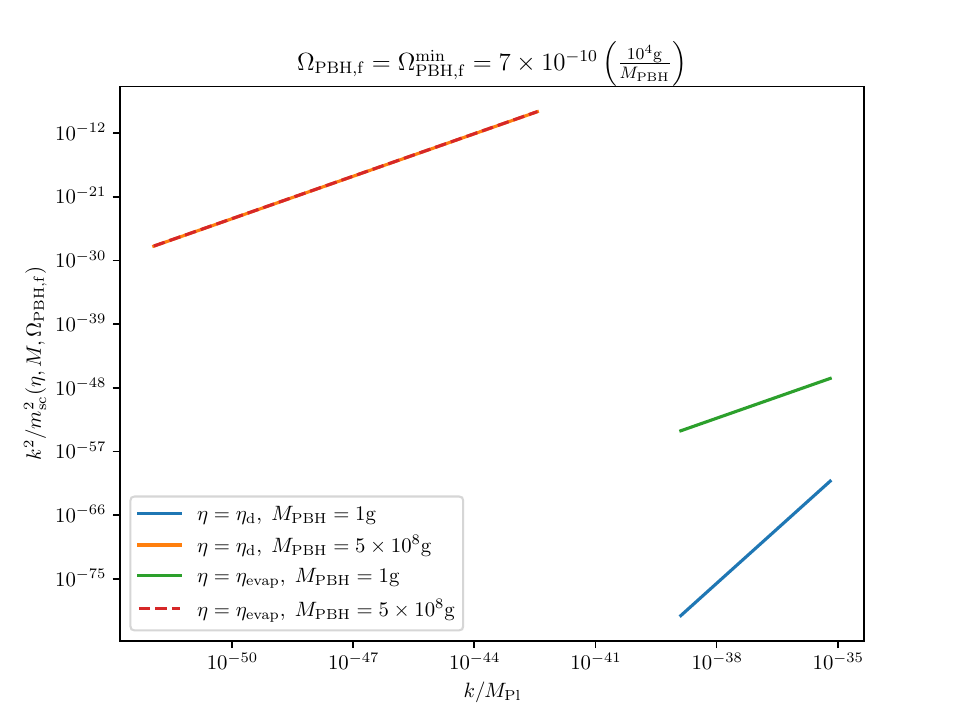}
\caption{\it{The ratio $k^2/m^2_\mathrm{sc}$ as a function of the comoving scale $k$ for different values of the PBH masses $M_\mathrm{PBH}$ and of the initial PBH abundances $\Omega_\mathrm{PBH,f}$.}}
\label{fig:k_2_o_msc2}
\end{figure*}

We compare below in \Fig{fig:a_prime_prime_o_a_o_msc2} and in  \Fig{fig:k_2_o_msc2} the three terms $k^2$, $a^{\prime\prime}/a$ and $m^2_\mathrm{sc}$ present in the 
equation \eqref{eq:Green_equation} governing the evolution of the Green function (GW propagator),  by plotting the ratios $k^2/m^2_\mathrm{sc}$ and $(a^{\prime\prime}/a)/m^2_\mathrm{sc}$ for different values of the PBH masses $M_\mathrm{PBH}$ within the range $[10\mathrm{g},10^9\mathrm{g}]$ and of the initial PBH abundance $\Omega_\mathrm{PBH,f}$ within the range defined by \Eq{eq:Omega_PBH_f_min} and \Eq{eq:Omega_PBH_f_max}. The comoving scales $k$ are varying between $k_\mathrm{evap}$ and $k_\mathrm{UV}$.

\bibliography{PBH}

\begin{thebibliography}{126}%
\makeatletter
\providecommand \@ifxundefined [1]{%
 \@ifx{#1\undefined}
}%
\providecommand \@ifnum [1]{%
 \ifnum #1\expandafter \@firstoftwo
 \else \expandafter \@secondoftwo
 \fi
}%
\providecommand \@ifx [1]{%
 \ifx #1\expandafter \@firstoftwo
 \else \expandafter \@secondoftwo
 \fi
}%
\providecommand \natexlab [1]{#1}%
\providecommand \enquote  [1]{``#1''}%
\providecommand \bibnamefont  [1]{#1}%
\providecommand \bibfnamefont [1]{#1}%
\providecommand \citenamefont [1]{#1}%
\providecommand \href@noop [0]{\@secondoftwo}%
\providecommand \href [0]{\begingroup \@sanitize@url \@href}%
\providecommand \@href[1]{\@@startlink{#1}\@@href}%
\providecommand \@@href[1]{\endgroup#1\@@endlink}%
\providecommand \@sanitize@url [0]{\catcode `\\12\catcode `\$12\catcode
  `\&12\catcode `\#12\catcode `\^12\catcode `\_12\catcode `\%12\relax}%
\providecommand \@@startlink[1]{}%
\providecommand \@@endlink[0]{}%
\providecommand \url  [0]{\begingroup\@sanitize@url \@url }%
\providecommand \@url [1]{\endgroup\@href {#1}{\urlprefix }}%
\providecommand \urlprefix  [0]{URL }%
\providecommand \Eprint [0]{\href }%
\providecommand \doibase [0]{http://dx.doi.org/}%
\providecommand \selectlanguage [0]{\@gobble}%
\providecommand \bibinfo  [0]{\@secondoftwo}%
\providecommand \bibfield  [0]{\@secondoftwo}%
\providecommand \translation [1]{[#1]}%
\providecommand \BibitemOpen [0]{}%
\providecommand \bibitemStop [0]{}%
\providecommand \bibitemNoStop [0]{.\EOS\space}%
\providecommand \EOS [0]{\spacefactor3000\relax}%
\providecommand \BibitemShut  [1]{\csname bibitem#1\endcsname}%
\let\auto@bib@innerbib\@empty
\bibitem [{\citenamefont {Sola~Peracaula}(2022)}]{rvm1}%
  \BibitemOpen
  \bibfield  {author} {\bibinfo {author} {\bibfnamefont {J.}~\bibnamefont
  {Sola~Peracaula}},\ }\href {\doibase 10.1098/rsta.2021.0182} {\bibfield
  {journal} {\bibinfo  {journal} {Phil. Trans. Roy. Soc. Lond. A}\ }\textbf
  {\bibinfo {volume} {380}},\ \bibinfo {pages} {20210182} (\bibinfo {year}
  {2022})},\ \Eprint {http://arxiv.org/abs/2203.13757} {arXiv:2203.13757
  [gr-qc]} \BibitemShut {NoStop}%
\bibitem [{\citenamefont {Sol\`a}\ and\ \citenamefont
  {G\'omez-Valent}(2015)}]{rvm2}%
  \BibitemOpen
  \bibfield  {author} {\bibinfo {author} {\bibfnamefont {J.}~\bibnamefont
  {Sol\`a}}\ and\ \bibinfo {author} {\bibfnamefont {A.}~\bibnamefont
  {G\'omez-Valent}},\ }\href {\doibase 10.1142/S0218271815410035} {\bibfield
  {journal} {\bibinfo  {journal} {Int. J. Mod. Phys. D}\ }\textbf {\bibinfo
  {volume} {24}},\ \bibinfo {pages} {1541003} (\bibinfo {year} {2015})},\
  \Eprint {http://arxiv.org/abs/1501.03832} {arXiv:1501.03832 [gr-qc]}
  \BibitemShut {NoStop}%
\bibitem [{\citenamefont {Sola}\ \emph {et~al.}(2017)\citenamefont {Sola},
  \citenamefont {Gomez-Valent},\ and\ \citenamefont {de~Cruz~P\'erez}}]{rvmde}%
  \BibitemOpen
  \bibfield  {author} {\bibinfo {author} {\bibfnamefont {J.}~\bibnamefont
  {Sola}}, \bibinfo {author} {\bibfnamefont {A.}~\bibnamefont {Gomez-Valent}},
  \ and\ \bibinfo {author} {\bibfnamefont {J.}~\bibnamefont
  {de~Cruz~P\'erez}},\ }\href {\doibase 10.1142/S0217732317500547} {\bibfield
  {journal} {\bibinfo  {journal} {Mod. Phys. Lett. A}\ }\textbf {\bibinfo
  {volume} {32}},\ \bibinfo {pages} {1750054} (\bibinfo {year} {2017})},\
  \Eprint {http://arxiv.org/abs/1610.08965} {arXiv:1610.08965 [astro-ph.CO]}
  \BibitemShut {NoStop}%
\bibitem [{\citenamefont {Lima}\ \emph {et~al.}(2013)\citenamefont {Lima},
  \citenamefont {Basilakos},\ and\ \citenamefont {Sola}}]{Lima:2013dmf}%
  \BibitemOpen
  \bibfield  {author} {\bibinfo {author} {\bibfnamefont {J.~A.~S.}\
  \bibnamefont {Lima}}, \bibinfo {author} {\bibfnamefont {S.}~\bibnamefont
  {Basilakos}}, \ and\ \bibinfo {author} {\bibfnamefont {J.}~\bibnamefont
  {Sola}},\ }\href {\doibase 10.1093/mnras/stt220} {\bibfield  {journal}
  {\bibinfo  {journal} {Mon. Not. Roy. Astron. Soc.}\ }\textbf {\bibinfo
  {volume} {431}},\ \bibinfo {pages} {923} (\bibinfo {year} {2013})},\ \Eprint
  {http://arxiv.org/abs/1209.2802} {arXiv:1209.2802 [gr-qc]} \BibitemShut
  {NoStop}%
\bibitem [{\citenamefont {Lima}\ \emph {et~al.}(2016)\citenamefont {Lima},
  \citenamefont {Basilakos},\ and\ \citenamefont {Sol\`a}}]{Lima:2015mca}%
  \BibitemOpen
  \bibfield  {author} {\bibinfo {author} {\bibfnamefont {J.~A.~S.}\
  \bibnamefont {Lima}}, \bibinfo {author} {\bibfnamefont {S.}~\bibnamefont
  {Basilakos}}, \ and\ \bibinfo {author} {\bibfnamefont {J.}~\bibnamefont
  {Sol\`a}},\ }\href {\doibase 10.1140/epjc/s10052-016-4060-6} {\bibfield
  {journal} {\bibinfo  {journal} {Eur. Phys. J. C}\ }\textbf {\bibinfo {volume}
  {76}},\ \bibinfo {pages} {228} (\bibinfo {year} {2016})},\ \Eprint
  {http://arxiv.org/abs/1509.00163} {arXiv:1509.00163 [gr-qc]} \BibitemShut
  {NoStop}%
\bibitem [{\citenamefont {Sol\`a~Peracaula}\ and\ \citenamefont
  {Yu}(2020)}]{rvm3}%
  \BibitemOpen
  \bibfield  {author} {\bibinfo {author} {\bibfnamefont {J.}~\bibnamefont
  {Sol\`a~Peracaula}}\ and\ \bibinfo {author} {\bibfnamefont {H.}~\bibnamefont
  {Yu}},\ }\href {\doibase 10.1007/s10714-020-2657-4} {\bibfield  {journal}
  {\bibinfo  {journal} {Gen. Rel. Grav.}\ }\textbf {\bibinfo {volume} {52}},\
  \bibinfo {pages} {17} (\bibinfo {year} {2020})},\ \Eprint
  {http://arxiv.org/abs/1910.01638} {arXiv:1910.01638 [gr-qc]} \BibitemShut
  {NoStop}%
\bibitem [{\citenamefont {Sol\`a}\ \emph {et~al.}(2017)\citenamefont {Sol\`a},
  \citenamefont {G\'omez-Valent},\ and\ \citenamefont
  {de~Cruz~P\'erez}}]{rvmdata}%
  \BibitemOpen
  \bibfield  {author} {\bibinfo {author} {\bibfnamefont {J.}~\bibnamefont
  {Sol\`a}}, \bibinfo {author} {\bibfnamefont {A.}~\bibnamefont
  {G\'omez-Valent}}, \ and\ \bibinfo {author} {\bibfnamefont {J.}~\bibnamefont
  {de~Cruz~P\'erez}},\ }\href {\doibase 10.3847/1538-4357/836/1/43} {\bibfield
  {journal} {\bibinfo  {journal} {Astrophys. J.}\ }\textbf {\bibinfo {volume}
  {836}},\ \bibinfo {pages} {43} (\bibinfo {year} {2017})},\ \Eprint
  {http://arxiv.org/abs/1602.02103} {arXiv:1602.02103 [astro-ph.CO]}
  \BibitemShut {NoStop}%
\bibitem [{\citenamefont {Sol\`a~Peracaula}\ \emph {et~al.}(2018)\citenamefont
  {Sol\`a~Peracaula}, \citenamefont {de~Cruz~P\'erez},\ and\ \citenamefont
  {G\'omez-Valent}}]{rvmdata2}%
  \BibitemOpen
  \bibfield  {author} {\bibinfo {author} {\bibfnamefont {J.}~\bibnamefont
  {Sol\`a~Peracaula}}, \bibinfo {author} {\bibfnamefont {J.}~\bibnamefont
  {de~Cruz~P\'erez}}, \ and\ \bibinfo {author} {\bibfnamefont {A.}~\bibnamefont
  {G\'omez-Valent}},\ }\href {\doibase 10.1209/0295-5075/121/39001} {\bibfield
  {journal} {\bibinfo  {journal} {EPL}\ }\textbf {\bibinfo {volume} {121}},\
  \bibinfo {pages} {39001} (\bibinfo {year} {2018})},\ \Eprint
  {http://arxiv.org/abs/1606.00450} {arXiv:1606.00450 [gr-qc]} \BibitemShut
  {NoStop}%
\bibitem [{\citenamefont {Papagiannopoulos}\ \emph {et~al.}(2020)\citenamefont
  {Papagiannopoulos}, \citenamefont {Tsiapi}, \citenamefont {Basilakos},\ and\
  \citenamefont {Paliathanasis}}]{tsiapi}%
  \BibitemOpen
  \bibfield  {author} {\bibinfo {author} {\bibfnamefont {G.}~\bibnamefont
  {Papagiannopoulos}}, \bibinfo {author} {\bibfnamefont {P.}~\bibnamefont
  {Tsiapi}}, \bibinfo {author} {\bibfnamefont {S.}~\bibnamefont {Basilakos}}, \
  and\ \bibinfo {author} {\bibfnamefont {A.}~\bibnamefont {Paliathanasis}},\
  }\href {\doibase 10.1140/epjc/s10052-019-7600-z} {\bibfield  {journal}
  {\bibinfo  {journal} {Eur. Phys. J. C}\ }\textbf {\bibinfo {volume} {80}},\
  \bibinfo {pages} {55} (\bibinfo {year} {2020})},\ \Eprint
  {http://arxiv.org/abs/1911.12431} {arXiv:1911.12431 [gr-qc]} \BibitemShut
  {NoStop}%
\bibitem [{\citenamefont {Asimakis}\ \emph {et~al.}(2022)\citenamefont
  {Asimakis}, \citenamefont {Basilakos}, \citenamefont {Mavromatos},\ and\
  \citenamefont {Saridakis}}]{rvmbbn}%
  \BibitemOpen
  \bibfield  {author} {\bibinfo {author} {\bibfnamefont {P.}~\bibnamefont
  {Asimakis}}, \bibinfo {author} {\bibfnamefont {S.}~\bibnamefont {Basilakos}},
  \bibinfo {author} {\bibfnamefont {N.~E.}\ \bibnamefont {Mavromatos}}, \ and\
  \bibinfo {author} {\bibfnamefont {E.~N.}\ \bibnamefont {Saridakis}},\ }\href
  {\doibase 10.1103/PhysRevD.105.084010} {\bibfield  {journal} {\bibinfo
  {journal} {Phys. Rev. D}\ }\textbf {\bibinfo {volume} {105}},\ \bibinfo
  {pages} {084010} (\bibinfo {year} {2022})},\ \Eprint
  {http://arxiv.org/abs/2112.10863} {arXiv:2112.10863 [gr-qc]} \BibitemShut
  {NoStop}%
\bibitem [{\citenamefont {Gomez-Valent}\ and\ \citenamefont
  {Sola}(2017)}]{rvmtenss8}%
  \BibitemOpen
  \bibfield  {author} {\bibinfo {author} {\bibfnamefont {A.}~\bibnamefont
  {Gomez-Valent}}\ and\ \bibinfo {author} {\bibfnamefont {J.}~\bibnamefont
  {Sola}},\ }\href {\doibase 10.1209/0295-5075/120/39001} {\bibfield  {journal}
  {\bibinfo  {journal} {EPL}\ }\textbf {\bibinfo {volume} {120}},\ \bibinfo
  {pages} {39001} (\bibinfo {year} {2017})},\ \Eprint
  {http://arxiv.org/abs/1711.00692} {arXiv:1711.00692 [astro-ph.CO]}
  \BibitemShut {NoStop}%
\bibitem [{\citenamefont {Sol\`a~Peracaula}\ \emph {et~al.}(2021)\citenamefont
  {Sol\`a~Peracaula}, \citenamefont {G\'omez-Valent}, \citenamefont
  {de~Cruz~Perez},\ and\ \citenamefont {Moreno-Pulido}}]{rvmtens}%
  \BibitemOpen
  \bibfield  {author} {\bibinfo {author} {\bibfnamefont {J.}~\bibnamefont
  {Sol\`a~Peracaula}}, \bibinfo {author} {\bibfnamefont {A.}~\bibnamefont
  {G\'omez-Valent}}, \bibinfo {author} {\bibfnamefont {J.}~\bibnamefont
  {de~Cruz~Perez}}, \ and\ \bibinfo {author} {\bibfnamefont {C.}~\bibnamefont
  {Moreno-Pulido}},\ }\href {\doibase 10.1209/0295-5075/134/19001} {\bibfield
  {journal} {\bibinfo  {journal} {EPL}\ }\textbf {\bibinfo {volume} {134}},\
  \bibinfo {pages} {19001} (\bibinfo {year} {2021})},\ \Eprint
  {http://arxiv.org/abs/2102.12758} {arXiv:2102.12758 [astro-ph.CO]}
  \BibitemShut {NoStop}%
\bibitem [{\citenamefont {Verde}\ \emph {et~al.}(2019)\citenamefont {Verde},
  \citenamefont {Treu},\ and\ \citenamefont {Riess}}]{tensions}%
  \BibitemOpen
  \bibfield  {author} {\bibinfo {author} {\bibfnamefont {L.}~\bibnamefont
  {Verde}}, \bibinfo {author} {\bibfnamefont {T.}~\bibnamefont {Treu}}, \ and\
  \bibinfo {author} {\bibfnamefont {A.~G.}\ \bibnamefont {Riess}},\ }\href
  {\doibase 10.1038/s41550-019-0902-0} {\bibfield  {journal} {\bibinfo
  {journal} {Nature Astron.}\ }\textbf {\bibinfo {volume} {3}},\ \bibinfo
  {pages} {891} (\bibinfo {year} {2019})},\ \Eprint
  {http://arxiv.org/abs/1907.10625} {arXiv:1907.10625 [astro-ph.CO]}
  \BibitemShut {NoStop}%
\bibitem [{\citenamefont {Abdalla}\ \emph {et~al.}(2022)\citenamefont {Abdalla}
  \emph {et~al.}}]{tensions2}%
  \BibitemOpen
  \bibfield  {author} {\bibinfo {author} {\bibfnamefont {E.}~\bibnamefont
  {Abdalla}} \emph {et~al.},\ }\href {\doibase 10.1016/j.jheap.2022.04.002}
  {\bibfield  {journal} {\bibinfo  {journal} {JHEAp}\ }\textbf {\bibinfo
  {volume} {34}},\ \bibinfo {pages} {49} (\bibinfo {year} {2022})},\ \Eprint
  {http://arxiv.org/abs/2203.06142} {arXiv:2203.06142 [astro-ph.CO]}
  \BibitemShut {NoStop}%
\bibitem [{\citenamefont {Freedman}(2017)}]{freedman}%
  \BibitemOpen
  \bibfield  {author} {\bibinfo {author} {\bibfnamefont {W.~L.}\ \bibnamefont
  {Freedman}},\ }\href {\doibase 10.1038/s41550-017-0121} {\bibfield  {journal}
  {\bibinfo  {journal} {Nature Astron.}\ }\textbf {\bibinfo {volume} {1}},\
  \bibinfo {pages} {0121} (\bibinfo {year} {2017})},\ \Eprint
  {http://arxiv.org/abs/1706.02739} {arXiv:1706.02739 [astro-ph.CO]}
  \BibitemShut {NoStop}%
\bibitem [{\citenamefont {Basilakos}\ \emph
  {et~al.}(2020{\natexlab{a}})\citenamefont {Basilakos}, \citenamefont
  {Mavromatos},\ and\ \citenamefont {Sol\`a~Peracaula}}]{bms}%
  \BibitemOpen
  \bibfield  {author} {\bibinfo {author} {\bibfnamefont {S.}~\bibnamefont
  {Basilakos}}, \bibinfo {author} {\bibfnamefont {N.~E.}\ \bibnamefont
  {Mavromatos}}, \ and\ \bibinfo {author} {\bibfnamefont {J.}~\bibnamefont
  {Sol\`a~Peracaula}},\ }\href {\doibase 10.1103/PhysRevD.101.045001}
  {\bibfield  {journal} {\bibinfo  {journal} {Phys. Rev. D}\ }\textbf {\bibinfo
  {volume} {101}},\ \bibinfo {pages} {045001} (\bibinfo {year}
  {2020}{\natexlab{a}})},\ \Eprint {http://arxiv.org/abs/1907.04890}
  {arXiv:1907.04890 [hep-ph]} \BibitemShut {NoStop}%
\bibitem [{\citenamefont {Basilakos}\ \emph
  {et~al.}(2020{\natexlab{b}})\citenamefont {Basilakos}, \citenamefont
  {Mavromatos},\ and\ \citenamefont {Sol\`a~Peracaula}}]{bms2}%
  \BibitemOpen
  \bibfield  {author} {\bibinfo {author} {\bibfnamefont {S.}~\bibnamefont
  {Basilakos}}, \bibinfo {author} {\bibfnamefont {N.~E.}\ \bibnamefont
  {Mavromatos}}, \ and\ \bibinfo {author} {\bibfnamefont {J.}~\bibnamefont
  {Sol\`a~Peracaula}},\ }\href {\doibase 10.1016/j.physletb.2020.135342}
  {\bibfield  {journal} {\bibinfo  {journal} {Phys. Lett. B}\ }\textbf
  {\bibinfo {volume} {803}},\ \bibinfo {pages} {135342} (\bibinfo {year}
  {2020}{\natexlab{b}})},\ \Eprint {http://arxiv.org/abs/2001.03465}
  {arXiv:2001.03465 [gr-qc]} \BibitemShut {NoStop}%
\bibitem [{\citenamefont {Basilakos}\ \emph {et~al.}(2019)\citenamefont
  {Basilakos}, \citenamefont {Mavromatos},\ and\ \citenamefont
  {Sol\`a~Peracaula}}]{bms3}%
  \BibitemOpen
  \bibfield  {author} {\bibinfo {author} {\bibfnamefont {S.}~\bibnamefont
  {Basilakos}}, \bibinfo {author} {\bibfnamefont {N.~E.}\ \bibnamefont
  {Mavromatos}}, \ and\ \bibinfo {author} {\bibfnamefont {J.}~\bibnamefont
  {Sol\`a~Peracaula}},\ }\href {\doibase 10.1142/S0218271819440024} {\bibfield
  {journal} {\bibinfo  {journal} {Int. J. Mod. Phys. D}\ }\textbf {\bibinfo
  {volume} {28}},\ \bibinfo {pages} {1944002} (\bibinfo {year} {2019})},\
  \Eprint {http://arxiv.org/abs/1905.04685} {arXiv:1905.04685 [hep-th]}
  \BibitemShut {NoStop}%
\bibitem [{\citenamefont {Mavromatos}\ and\ \citenamefont
  {Sol\`a~Peracaula}(2021{\natexlab{a}})}]{ms1}%
  \BibitemOpen
  \bibfield  {author} {\bibinfo {author} {\bibfnamefont {N.~E.}\ \bibnamefont
  {Mavromatos}}\ and\ \bibinfo {author} {\bibfnamefont {J.}~\bibnamefont
  {Sol\`a~Peracaula}},\ }\href {\doibase 10.1140/epjs/s11734-021-00197-8}
  {\bibfield  {journal} {\bibinfo  {journal} {Eur. Phys. J. ST}\ }\textbf
  {\bibinfo {volume} {230}},\ \bibinfo {pages} {2077} (\bibinfo {year}
  {2021}{\natexlab{a}})},\ \Eprint {http://arxiv.org/abs/2012.07971}
  {arXiv:2012.07971 [hep-ph]} \BibitemShut {NoStop}%
\bibitem [{\citenamefont {Mavromatos}\ and\ \citenamefont
  {Sol\`a~Peracaula}(2021{\natexlab{b}})}]{ms2}%
  \BibitemOpen
  \bibfield  {author} {\bibinfo {author} {\bibfnamefont {N.~E.}\ \bibnamefont
  {Mavromatos}}\ and\ \bibinfo {author} {\bibfnamefont {J.}~\bibnamefont
  {Sol\`a~Peracaula}},\ }\href {\doibase 10.1140/epjp/s13360-021-02149-6}
  {\bibfield  {journal} {\bibinfo  {journal} {Eur. Phys. J. Plus}\ }\textbf
  {\bibinfo {volume} {136}},\ \bibinfo {pages} {1152} (\bibinfo {year}
  {2021}{\natexlab{b}})},\ \Eprint {http://arxiv.org/abs/2105.02659}
  {arXiv:2105.02659 [hep-th]} \BibitemShut {NoStop}%
\bibitem [{\citenamefont {Moreno-Pulido}\ and\ \citenamefont
  {Sola}(2020)}]{rvmqft1}%
  \BibitemOpen
  \bibfield  {author} {\bibinfo {author} {\bibfnamefont {C.}~\bibnamefont
  {Moreno-Pulido}}\ and\ \bibinfo {author} {\bibfnamefont {J.}~\bibnamefont
  {Sola}},\ }\href {\doibase 10.1140/epjc/s10052-020-8238-6} {\bibfield
  {journal} {\bibinfo  {journal} {Eur. Phys. J. C}\ }\textbf {\bibinfo {volume}
  {80}},\ \bibinfo {pages} {692} (\bibinfo {year} {2020})},\ \Eprint
  {http://arxiv.org/abs/2005.03164} {arXiv:2005.03164 [gr-qc]} \BibitemShut
  {NoStop}%
\bibitem [{\citenamefont {Moreno-Pulido}\ and\ \citenamefont
  {Sola~Peracaula}(2022{\natexlab{a}})}]{rvmqft2}%
  \BibitemOpen
  \bibfield  {author} {\bibinfo {author} {\bibfnamefont {C.}~\bibnamefont
  {Moreno-Pulido}}\ and\ \bibinfo {author} {\bibfnamefont {J.}~\bibnamefont
  {Sola~Peracaula}},\ }\href {\doibase 10.1140/epjc/s10052-022-10484-w}
  {\bibfield  {journal} {\bibinfo  {journal} {Eur. Phys. J. C}\ }\textbf
  {\bibinfo {volume} {82}},\ \bibinfo {pages} {551} (\bibinfo {year}
  {2022}{\natexlab{a}})},\ \Eprint {http://arxiv.org/abs/2201.05827}
  {arXiv:2201.05827 [gr-qc]} \BibitemShut {NoStop}%
\bibitem [{\citenamefont {Moreno-Pulido}\ and\ \citenamefont
  {Sola~Peracaula}(2022{\natexlab{b}})}]{rvmqft3}%
  \BibitemOpen
  \bibfield  {author} {\bibinfo {author} {\bibfnamefont {C.}~\bibnamefont
  {Moreno-Pulido}}\ and\ \bibinfo {author} {\bibfnamefont {J.}~\bibnamefont
  {Sola~Peracaula}},\ }\href {\doibase 10.1140/epjc/s10052-022-11117-y}
  {\bibfield  {journal} {\bibinfo  {journal} {Eur. Phys. J. C}\ }\textbf
  {\bibinfo {volume} {82}},\ \bibinfo {pages} {1137} (\bibinfo {year}
  {2022}{\natexlab{b}})},\ \Eprint {http://arxiv.org/abs/2207.07111}
  {arXiv:2207.07111 [gr-qc]} \BibitemShut {NoStop}%
\bibitem [{\citenamefont {Moreno-Pulido}\ \emph {et~al.}(2023)\citenamefont
  {Moreno-Pulido}, \citenamefont {Sola~Peracaula},\ and\ \citenamefont
  {Cheraghchi}}]{rvmqft4}%
  \BibitemOpen
  \bibfield  {author} {\bibinfo {author} {\bibfnamefont {C.}~\bibnamefont
  {Moreno-Pulido}}, \bibinfo {author} {\bibfnamefont {J.}~\bibnamefont
  {Sola~Peracaula}}, \ and\ \bibinfo {author} {\bibfnamefont {S.}~\bibnamefont
  {Cheraghchi}},\ }\href {\doibase 10.1140/epjc/s10052-023-11772-9} {\bibfield
  {journal} {\bibinfo  {journal} {Eur. Phys. J. C}\ }\textbf {\bibinfo {volume}
  {83}},\ \bibinfo {pages} {637} (\bibinfo {year} {2023})},\ \Eprint
  {http://arxiv.org/abs/2301.05205} {arXiv:2301.05205 [gr-qc]} \BibitemShut
  {NoStop}%
\bibitem [{\citenamefont {Basilakos}\ \emph {et~al.}(2016)\citenamefont
  {Basilakos}, \citenamefont {Mavromatos},\ and\ \citenamefont
  {Sol\`a}}]{bmssugra}%
  \BibitemOpen
  \bibfield  {author} {\bibinfo {author} {\bibfnamefont {S.}~\bibnamefont
  {Basilakos}}, \bibinfo {author} {\bibfnamefont {N.~E.}\ \bibnamefont
  {Mavromatos}}, \ and\ \bibinfo {author} {\bibfnamefont {J.}~\bibnamefont
  {Sol\`a}},\ }\href {\doibase 10.3390/universe2030014} {\bibfield  {journal}
  {\bibinfo  {journal} {Universe}\ }\textbf {\bibinfo {volume} {2}},\ \bibinfo
  {pages} {14} (\bibinfo {year} {2016})},\ \Eprint
  {http://arxiv.org/abs/1505.04434} {arXiv:1505.04434 [gr-qc]} \BibitemShut
  {NoStop}%
\bibitem [{\citenamefont {G\'omez-Valent}\ \emph {et~al.}(2024)\citenamefont
  {G\'omez-Valent}, \citenamefont {Mavromatos},\ and\ \citenamefont
  {Sol\`a~Peracaula}}]{Gomez-Valent:2023hov}%
  \BibitemOpen
  \bibfield  {author} {\bibinfo {author} {\bibfnamefont {A.}~\bibnamefont
  {G\'omez-Valent}}, \bibinfo {author} {\bibfnamefont {N.~E.}\ \bibnamefont
  {Mavromatos}}, \ and\ \bibinfo {author} {\bibfnamefont {J.}~\bibnamefont
  {Sol\`a~Peracaula}},\ }\href {\doibase 10.1088/1361-6382/ad0fb8} {\bibfield
  {journal} {\bibinfo  {journal} {Class. Quant. Grav.}\ }\textbf {\bibinfo
  {volume} {41}},\ \bibinfo {pages} {015026} (\bibinfo {year} {2024})},\
  \Eprint {http://arxiv.org/abs/2305.15774} {arXiv:2305.15774 [gr-qc]}
  \BibitemShut {NoStop}%
\bibitem [{\citenamefont {{Zel'dovich}}\ and\ \citenamefont
  {{Novikov}}(1967)}]{1967SvA....10..602Z}%
  \BibitemOpen
  \bibfield  {author} {\bibinfo {author} {\bibfnamefont {Y.~B.}\ \bibnamefont
  {{Zel'dovich}}}\ and\ \bibinfo {author} {\bibfnamefont {I.~D.}\ \bibnamefont
  {{Novikov}}},\ }\href@noop {} {\bibfield  {journal} {\bibinfo  {journal}
  {Soviet Astronomy}\ }\textbf {\bibinfo {volume} {10}},\ \bibinfo {pages}
  {602} (\bibinfo {year} {1967})}\BibitemShut {NoStop}%
\bibitem [{\citenamefont {Carr}\ and\ \citenamefont
  {Hawking}(1974)}]{Carr:1974nx}%
  \BibitemOpen
  \bibfield  {author} {\bibinfo {author} {\bibfnamefont {B.~J.}\ \bibnamefont
  {Carr}}\ and\ \bibinfo {author} {\bibfnamefont {S.~W.}\ \bibnamefont
  {Hawking}},\ }\href@noop {} {\bibfield  {journal} {\bibinfo  {journal} {Mon.
  Not. Roy. Astron. Soc.}\ }\textbf {\bibinfo {volume} {168}},\ \bibinfo
  {pages} {399} (\bibinfo {year} {1974})}\BibitemShut {NoStop}%
\bibitem [{\citenamefont {{Carr}}(1975)}]{1975ApJ...201....1C}%
  \BibitemOpen
  \bibfield  {author} {\bibinfo {author} {\bibfnamefont {B.~J.}\ \bibnamefont
  {{Carr}}},\ }\href {\doibase 10.1086/153853} {\bibfield  {journal} {\bibinfo
  {journal} {ApJ}\ }\textbf {\bibinfo {volume} {201}},\ \bibinfo {pages} {1}
  (\bibinfo {year} {1975})}\BibitemShut {NoStop}%
\bibitem [{\citenamefont {{Novikov}}\ \emph {et~al.}(1979)\citenamefont
  {{Novikov}}, \citenamefont {{Polnarev}}, \citenamefont {{Starobinskii}},\
  and\ \citenamefont {{Zeldovich}}}]{1979A&A....80..104N}%
  \BibitemOpen
  \bibfield  {author} {\bibinfo {author} {\bibfnamefont {I.~D.}\ \bibnamefont
  {{Novikov}}}, \bibinfo {author} {\bibfnamefont {A.~G.}\ \bibnamefont
  {{Polnarev}}}, \bibinfo {author} {\bibfnamefont {A.~A.}\ \bibnamefont
  {{Starobinskii}}}, \ and\ \bibinfo {author} {\bibfnamefont {I.~B.}\
  \bibnamefont {{Zeldovich}}},\ }\href@noop {} {\bibfield  {journal} {\bibinfo
  {journal} {Astronomy and Astrophysics}\ }\textbf {\bibinfo {volume} {80}},\
  \bibinfo {pages} {104} (\bibinfo {year} {1979})}\BibitemShut {NoStop}%
\bibitem [{\citenamefont {Chapline}(1975)}]{Chapline:1975ojl}%
  \BibitemOpen
  \bibfield  {author} {\bibinfo {author} {\bibfnamefont {G.~F.}\ \bibnamefont
  {Chapline}},\ }\href {\doibase 10.1038/253251a0} {\bibfield  {journal}
  {\bibinfo  {journal} {Nature}\ }\textbf {\bibinfo {volume} {253}},\ \bibinfo
  {pages} {251} (\bibinfo {year} {1975})}\BibitemShut {NoStop}%
\bibitem [{\citenamefont {Meszaros}(1975)}]{Meszaros:1975ef}%
  \BibitemOpen
  \bibfield  {author} {\bibinfo {author} {\bibfnamefont {P.}~\bibnamefont
  {Meszaros}},\ }\href@noop {} {\bibfield  {journal} {\bibinfo  {journal}
  {Astron. Astrophys.}\ }\textbf {\bibinfo {volume} {38}},\ \bibinfo {pages}
  {5} (\bibinfo {year} {1975})}\BibitemShut {NoStop}%
\bibitem [{\citenamefont {Afshordi}\ \emph {et~al.}(2003)\citenamefont
  {Afshordi}, \citenamefont {McDonald},\ and\ \citenamefont
  {Spergel}}]{Afshordi:2003zb}%
  \BibitemOpen
  \bibfield  {author} {\bibinfo {author} {\bibfnamefont {N.}~\bibnamefont
  {Afshordi}}, \bibinfo {author} {\bibfnamefont {P.}~\bibnamefont {McDonald}},
  \ and\ \bibinfo {author} {\bibfnamefont {D.}~\bibnamefont {Spergel}},\ }\href
  {\doibase 10.1086/378763} {\bibfield  {journal} {\bibinfo  {journal}
  {Astrophys. J. Lett.}\ }\textbf {\bibinfo {volume} {594}},\ \bibinfo {pages}
  {L71} (\bibinfo {year} {2003})},\ \Eprint
  {http://arxiv.org/abs/astro-ph/0302035} {arXiv:astro-ph/0302035} \BibitemShut
  {NoStop}%
\bibitem [{\citenamefont {{Carr}}\ and\ \citenamefont
  {{Rees}}(1984)}]{1984MNRAS.206..315C}%
  \BibitemOpen
  \bibfield  {author} {\bibinfo {author} {\bibfnamefont {B.~J.}\ \bibnamefont
  {{Carr}}}\ and\ \bibinfo {author} {\bibfnamefont {M.~J.}\ \bibnamefont
  {{Rees}}},\ }\href {\doibase 10.1093/mnras/206.2.315} {\bibfield  {journal}
  {\bibinfo  {journal} {Monthly Notices of Royal Academy of Science}\ }\textbf
  {\bibinfo {volume} {206}},\ \bibinfo {pages} {315} (\bibinfo {year}
  {1984})}\BibitemShut {NoStop}%
\bibitem [{\citenamefont {Bean}\ and\ \citenamefont
  {Magueijo}(2002)}]{Bean:2002kx}%
  \BibitemOpen
  \bibfield  {author} {\bibinfo {author} {\bibfnamefont {R.}~\bibnamefont
  {Bean}}\ and\ \bibinfo {author} {\bibfnamefont {J.}~\bibnamefont
  {Magueijo}},\ }\href {\doibase 10.1103/PhysRevD.66.063505} {\bibfield
  {journal} {\bibinfo  {journal} {Phys. Rev. D}\ }\textbf {\bibinfo {volume}
  {66}},\ \bibinfo {pages} {063505} (\bibinfo {year} {2002})},\ \Eprint
  {http://arxiv.org/abs/astro-ph/0204486} {arXiv:astro-ph/0204486} \BibitemShut
  {NoStop}%
\bibitem [{\citenamefont {Sasaki}\ \emph {et~al.}(2016)\citenamefont {Sasaki},
  \citenamefont {Suyama}, \citenamefont {Tanaka},\ and\ \citenamefont
  {Yokoyama}}]{Sasaki:2016jop}%
  \BibitemOpen
  \bibfield  {author} {\bibinfo {author} {\bibfnamefont {M.}~\bibnamefont
  {Sasaki}}, \bibinfo {author} {\bibfnamefont {T.}~\bibnamefont {Suyama}},
  \bibinfo {author} {\bibfnamefont {T.}~\bibnamefont {Tanaka}}, \ and\ \bibinfo
  {author} {\bibfnamefont {S.}~\bibnamefont {Yokoyama}},\ }\href {\doibase
  10.1103/PhysRevLett.121.059901, 10.1103/PhysRevLett.117.061101} {\bibfield
  {journal} {\bibinfo  {journal} {Phys. Rev. Lett.}\ }\textbf {\bibinfo
  {volume} {117}},\ \bibinfo {pages} {061101} (\bibinfo {year} {2016})},\
  \bibinfo {note} {[erratum: Phys. Rev. Lett.121,no.5,059901(2018)]},\ \Eprint
  {http://arxiv.org/abs/1603.08338} {arXiv:1603.08338 [astro-ph.CO]}
  \BibitemShut {NoStop}%
\bibitem [{\citenamefont {Franciolini}\ \emph {et~al.}(2022)\citenamefont
  {Franciolini}, \citenamefont {Baibhav}, \citenamefont {De~Luca},
  \citenamefont {Ng}, \citenamefont {Wong}, \citenamefont {Berti},
  \citenamefont {Pani}, \citenamefont {Riotto},\ and\ \citenamefont
  {Vitale}}]{Franciolini:2021tla}%
  \BibitemOpen
  \bibfield  {author} {\bibinfo {author} {\bibfnamefont {G.}~\bibnamefont
  {Franciolini}}, \bibinfo {author} {\bibfnamefont {V.}~\bibnamefont
  {Baibhav}}, \bibinfo {author} {\bibfnamefont {V.}~\bibnamefont {De~Luca}},
  \bibinfo {author} {\bibfnamefont {K.~K.~Y.}\ \bibnamefont {Ng}}, \bibinfo
  {author} {\bibfnamefont {K.~W.~K.}\ \bibnamefont {Wong}}, \bibinfo {author}
  {\bibfnamefont {E.}~\bibnamefont {Berti}}, \bibinfo {author} {\bibfnamefont
  {P.}~\bibnamefont {Pani}}, \bibinfo {author} {\bibfnamefont {A.}~\bibnamefont
  {Riotto}}, \ and\ \bibinfo {author} {\bibfnamefont {S.}~\bibnamefont
  {Vitale}},\ }\href {\doibase 10.1103/PhysRevD.105.083526} {\bibfield
  {journal} {\bibinfo  {journal} {Phys. Rev. D}\ }\textbf {\bibinfo {volume}
  {105}},\ \bibinfo {pages} {083526} (\bibinfo {year} {2022})},\ \Eprint
  {http://arxiv.org/abs/2105.03349} {arXiv:2105.03349 [gr-qc]} \BibitemShut
  {NoStop}%
\bibitem [{\citenamefont {Baumann}\ \emph
  {et~al.}(2007{\natexlab{a}})\citenamefont {Baumann}, \citenamefont
  {Steinhardt},\ and\ \citenamefont {Turok}}]{Baumann:2007yr}%
  \BibitemOpen
  \bibfield  {author} {\bibinfo {author} {\bibfnamefont {D.}~\bibnamefont
  {Baumann}}, \bibinfo {author} {\bibfnamefont {P.~J.}\ \bibnamefont
  {Steinhardt}}, \ and\ \bibinfo {author} {\bibfnamefont {N.}~\bibnamefont
  {Turok}},\ }\href@noop {} {\  (\bibinfo {year} {2007}{\natexlab{a}})},\
  \Eprint {http://arxiv.org/abs/hep-th/0703250} {arXiv:hep-th/0703250}
  \BibitemShut {NoStop}%
\bibitem [{\citenamefont {Hook}(2014)}]{Hook:2014mla}%
  \BibitemOpen
  \bibfield  {author} {\bibinfo {author} {\bibfnamefont {A.}~\bibnamefont
  {Hook}},\ }\href {\doibase 10.1103/PhysRevD.90.083535} {\bibfield  {journal}
  {\bibinfo  {journal} {Phys. Rev. D}\ }\textbf {\bibinfo {volume} {90}},\
  \bibinfo {pages} {083535} (\bibinfo {year} {2014})},\ \Eprint
  {http://arxiv.org/abs/1404.0113} {arXiv:1404.0113 [hep-ph]} \BibitemShut
  {NoStop}%
\bibitem [{\citenamefont {Carr}\ \emph
  {et~al.}(2021{\natexlab{a}})\citenamefont {Carr}, \citenamefont {Clesse},\
  and\ \citenamefont {Garc\'\i{}a-Bellido}}]{Carr:2019hud}%
  \BibitemOpen
  \bibfield  {author} {\bibinfo {author} {\bibfnamefont {B.}~\bibnamefont
  {Carr}}, \bibinfo {author} {\bibfnamefont {S.}~\bibnamefont {Clesse}}, \ and\
  \bibinfo {author} {\bibfnamefont {J.}~\bibnamefont {Garc\'\i{}a-Bellido}},\
  }\href {\doibase 10.1093/mnras/staa3726} {\bibfield  {journal} {\bibinfo
  {journal} {Mon. Not. Roy. Astron. Soc.}\ }\textbf {\bibinfo {volume} {501}},\
  \bibinfo {pages} {1426} (\bibinfo {year} {2021}{\natexlab{a}})},\ \Eprint
  {http://arxiv.org/abs/1904.02129} {arXiv:1904.02129 [astro-ph.CO]}
  \BibitemShut {NoStop}%
\bibitem [{\citenamefont {Safarzadeh}\ \emph {et~al.}(2018)\citenamefont
  {Safarzadeh}, \citenamefont {Naoz}, \citenamefont {Sadowski}, \citenamefont
  {Sironi},\ and\ \citenamefont {Narayan}}]{Safarzadeh:2017mdy}%
  \BibitemOpen
  \bibfield  {author} {\bibinfo {author} {\bibfnamefont {M.}~\bibnamefont
  {Safarzadeh}}, \bibinfo {author} {\bibfnamefont {S.}~\bibnamefont {Naoz}},
  \bibinfo {author} {\bibfnamefont {A.}~\bibnamefont {Sadowski}}, \bibinfo
  {author} {\bibfnamefont {L.}~\bibnamefont {Sironi}}, \ and\ \bibinfo {author}
  {\bibfnamefont {R.}~\bibnamefont {Narayan}},\ }\href {\doibase
  10.1093/mnras/sty1486} {\bibfield  {journal} {\bibinfo  {journal} {Mon. Not.
  Roy. Astron. Soc.}\ }\textbf {\bibinfo {volume} {479}},\ \bibinfo {pages}
  {315} (\bibinfo {year} {2018})},\ \Eprint {http://arxiv.org/abs/1701.03800}
  {arXiv:1701.03800 [astro-ph.HE]} \BibitemShut {NoStop}%
\bibitem [{\citenamefont {Papanikolaou}\ and\ \citenamefont
  {Gourgouliatos}(2023)}]{Papanikolaou:2023nkx}%
  \BibitemOpen
  \bibfield  {author} {\bibinfo {author} {\bibfnamefont {T.}~\bibnamefont
  {Papanikolaou}}\ and\ \bibinfo {author} {\bibfnamefont {K.~N.}\ \bibnamefont
  {Gourgouliatos}},\ }\href {\doibase 10.1103/PhysRevD.107.103532} {\bibfield
  {journal} {\bibinfo  {journal} {Phys. Rev. D}\ }\textbf {\bibinfo {volume}
  {107}},\ \bibinfo {pages} {103532} (\bibinfo {year} {2023})},\ \Eprint
  {http://arxiv.org/abs/2301.10045} {arXiv:2301.10045 [astro-ph.CO]}
  \BibitemShut {NoStop}%
\bibitem [{\citenamefont {Carr}\ \emph
  {et~al.}(2021{\natexlab{b}})\citenamefont {Carr}, \citenamefont {Kohri},
  \citenamefont {Sendouda},\ and\ \citenamefont {Yokoyama}}]{Carr:2020gox}%
  \BibitemOpen
  \bibfield  {author} {\bibinfo {author} {\bibfnamefont {B.}~\bibnamefont
  {Carr}}, \bibinfo {author} {\bibfnamefont {K.}~\bibnamefont {Kohri}},
  \bibinfo {author} {\bibfnamefont {Y.}~\bibnamefont {Sendouda}}, \ and\
  \bibinfo {author} {\bibfnamefont {J.}~\bibnamefont {Yokoyama}},\ }\href
  {\doibase 10.1088/1361-6633/ac1e31} {\bibfield  {journal} {\bibinfo
  {journal} {Rept. Prog. Phys.}\ }\textbf {\bibinfo {volume} {84}},\ \bibinfo
  {pages} {116902} (\bibinfo {year} {2021}{\natexlab{b}})},\ \Eprint
  {http://arxiv.org/abs/2002.12778} {arXiv:2002.12778 [astro-ph.CO]}
  \BibitemShut {NoStop}%
\bibitem [{\citenamefont {Escriv\`a}\ \emph {et~al.}(2022)\citenamefont
  {Escriv\`a}, \citenamefont {Kuhnel},\ and\ \citenamefont
  {Tada}}]{Escriva:2022duf}%
  \BibitemOpen
  \bibfield  {author} {\bibinfo {author} {\bibfnamefont {A.}~\bibnamefont
  {Escriv\`a}}, \bibinfo {author} {\bibfnamefont {F.}~\bibnamefont {Kuhnel}}, \
  and\ \bibinfo {author} {\bibfnamefont {Y.}~\bibnamefont {Tada}},\ }\href@noop
  {} {\  (\bibinfo {year} {2022})},\ \Eprint {http://arxiv.org/abs/2211.05767}
  {arXiv:2211.05767 [astro-ph.CO]} \BibitemShut {NoStop}%
\bibitem [{\citenamefont {Garcia-Bellido}\ \emph {et~al.}(1996)\citenamefont
  {Garcia-Bellido}, \citenamefont {Linde},\ and\ \citenamefont
  {Wands}}]{Garcia-Bellido:1996mdl}%
  \BibitemOpen
  \bibfield  {author} {\bibinfo {author} {\bibfnamefont {J.}~\bibnamefont
  {Garcia-Bellido}}, \bibinfo {author} {\bibfnamefont {A.~D.}\ \bibnamefont
  {Linde}}, \ and\ \bibinfo {author} {\bibfnamefont {D.}~\bibnamefont
  {Wands}},\ }\href {\doibase 10.1103/PhysRevD.54.6040} {\bibfield  {journal}
  {\bibinfo  {journal} {Phys. Rev. D}\ }\textbf {\bibinfo {volume} {54}},\
  \bibinfo {pages} {6040} (\bibinfo {year} {1996})},\ \Eprint
  {http://arxiv.org/abs/astro-ph/9605094} {arXiv:astro-ph/9605094} \BibitemShut
  {NoStop}%
\bibitem [{\citenamefont {Martin}\ \emph {et~al.}(2020)\citenamefont {Martin},
  \citenamefont {Papanikolaou}, \citenamefont {Pinol},\ and\ \citenamefont
  {Vennin}}]{Martin:2020fgl}%
  \BibitemOpen
  \bibfield  {author} {\bibinfo {author} {\bibfnamefont {J.}~\bibnamefont
  {Martin}}, \bibinfo {author} {\bibfnamefont {T.}~\bibnamefont
  {Papanikolaou}}, \bibinfo {author} {\bibfnamefont {L.}~\bibnamefont {Pinol}},
  \ and\ \bibinfo {author} {\bibfnamefont {V.}~\bibnamefont {Vennin}},\ }\href
  {\doibase 10.1088/1475-7516/2020/05/003} {\bibfield  {journal} {\bibinfo
  {journal} {JCAP}\ }\textbf {\bibinfo {volume} {05}},\ \bibinfo {pages} {003}
  (\bibinfo {year} {2020})},\ \Eprint {http://arxiv.org/abs/2002.01820}
  {arXiv:2002.01820 [astro-ph.CO]} \BibitemShut {NoStop}%
\bibitem [{\citenamefont {Heydari}\ and\ \citenamefont
  {Karami}(2022)}]{Heydari:2021qsr}%
  \BibitemOpen
  \bibfield  {author} {\bibinfo {author} {\bibfnamefont {S.}~\bibnamefont
  {Heydari}}\ and\ \bibinfo {author} {\bibfnamefont {K.}~\bibnamefont
  {Karami}},\ }\href {\doibase 10.1088/1475-7516/2022/03/033} {\bibfield
  {journal} {\bibinfo  {journal} {JCAP}\ }\textbf {\bibinfo {volume} {03}},\
  \bibinfo {pages} {033} (\bibinfo {year} {2022})},\ \Eprint
  {http://arxiv.org/abs/2111.00494} {arXiv:2111.00494 [gr-qc]} \BibitemShut
  {NoStop}%
\bibitem [{\citenamefont {Banerjee}\ \emph {et~al.}(2022)\citenamefont
  {Banerjee}, \citenamefont {Papanikolaou},\ and\ \citenamefont
  {Saridakis}}]{Banerjee:2022xft}%
  \BibitemOpen
  \bibfield  {author} {\bibinfo {author} {\bibfnamefont {S.}~\bibnamefont
  {Banerjee}}, \bibinfo {author} {\bibfnamefont {T.}~\bibnamefont
  {Papanikolaou}}, \ and\ \bibinfo {author} {\bibfnamefont {E.~N.}\
  \bibnamefont {Saridakis}},\ }\href {\doibase 10.1103/PhysRevD.106.124012}
  {\bibfield  {journal} {\bibinfo  {journal} {Phys. Rev. D}\ }\textbf {\bibinfo
  {volume} {106}},\ \bibinfo {pages} {124012} (\bibinfo {year} {2022})},\
  \Eprint {http://arxiv.org/abs/2206.01150} {arXiv:2206.01150 [gr-qc]}
  \BibitemShut {NoStop}%
\bibitem [{\citenamefont {Briaud}\ and\ \citenamefont
  {Vennin}(2023)}]{Briaud:2023eae}%
  \BibitemOpen
  \bibfield  {author} {\bibinfo {author} {\bibfnamefont {V.}~\bibnamefont
  {Briaud}}\ and\ \bibinfo {author} {\bibfnamefont {V.}~\bibnamefont
  {Vennin}},\ }\href {\doibase 10.1088/1475-7516/2023/06/029} {\bibfield
  {journal} {\bibinfo  {journal} {JCAP}\ }\textbf {\bibinfo {volume} {06}},\
  \bibinfo {pages} {029} (\bibinfo {year} {2023})},\ \Eprint
  {http://arxiv.org/abs/2301.09336} {arXiv:2301.09336 [astro-ph.CO]}
  \BibitemShut {NoStop}%
\bibitem [{\citenamefont {Basilakos}\ \emph
  {et~al.}(2024{\natexlab{a}})\citenamefont {Basilakos}, \citenamefont
  {Nanopoulos}, \citenamefont {Papanikolaou}, \citenamefont {Saridakis},\ and\
  \citenamefont {Tzerefos}}]{Basilakos:2023jvp}%
  \BibitemOpen
  \bibfield  {author} {\bibinfo {author} {\bibfnamefont {S.}~\bibnamefont
  {Basilakos}}, \bibinfo {author} {\bibfnamefont {D.~V.}\ \bibnamefont
  {Nanopoulos}}, \bibinfo {author} {\bibfnamefont {T.}~\bibnamefont
  {Papanikolaou}}, \bibinfo {author} {\bibfnamefont {E.~N.}\ \bibnamefont
  {Saridakis}}, \ and\ \bibinfo {author} {\bibfnamefont {C.}~\bibnamefont
  {Tzerefos}},\ }\href {\doibase 10.1016/j.physletb.2024.138446} {\bibfield
  {journal} {\bibinfo  {journal} {Phys. Lett. B}\ }\textbf {\bibinfo {volume}
  {849}},\ \bibinfo {pages} {138446} (\bibinfo {year} {2024}{\natexlab{a}})},\
  \Eprint {http://arxiv.org/abs/2309.15820} {arXiv:2309.15820 [astro-ph.CO]}
  \BibitemShut {NoStop}%
\bibitem [{\citenamefont {del Corral}\ \emph {et~al.}(2023)\citenamefont {del
  Corral}, \citenamefont {Gondolo}, \citenamefont {Kumar},\ and\ \citenamefont
  {Marto}}]{del-Corral:2023apl}%
  \BibitemOpen
  \bibfield  {author} {\bibinfo {author} {\bibfnamefont {D.}~\bibnamefont {del
  Corral}}, \bibinfo {author} {\bibfnamefont {P.}~\bibnamefont {Gondolo}},
  \bibinfo {author} {\bibfnamefont {K.~S.}\ \bibnamefont {Kumar}}, \ and\
  \bibinfo {author} {\bibfnamefont {J.~a.}\ \bibnamefont {Marto}},\ }\href@noop
  {} {\  (\bibinfo {year} {2023})},\ \Eprint {http://arxiv.org/abs/2311.02754}
  {arXiv:2311.02754 [astro-ph.CO]} \BibitemShut {NoStop}%
\bibitem [{\citenamefont {Heydari}\ and\ \citenamefont
  {Karami}(2024)}]{Heydari:2023xts}%
  \BibitemOpen
  \bibfield  {author} {\bibinfo {author} {\bibfnamefont {S.}~\bibnamefont
  {Heydari}}\ and\ \bibinfo {author} {\bibfnamefont {K.}~\bibnamefont
  {Karami}},\ }\href {\doibase 10.1088/1475-7516/2024/02/047} {\bibfield
  {journal} {\bibinfo  {journal} {JCAP}\ }\textbf {\bibinfo {volume} {02}},\
  \bibinfo {pages} {047} (\bibinfo {year} {2024})},\ \Eprint
  {http://arxiv.org/abs/2309.01239} {arXiv:2309.01239 [astro-ph.CO]}
  \BibitemShut {NoStop}%
\bibitem [{\citenamefont {Hidalgo}\ \emph {et~al.}(2012)\citenamefont
  {Hidalgo}, \citenamefont {Urena-Lopez},\ and\ \citenamefont
  {Liddle}}]{Hidalgo:2011fj}%
  \BibitemOpen
  \bibfield  {author} {\bibinfo {author} {\bibfnamefont {J.~C.}\ \bibnamefont
  {Hidalgo}}, \bibinfo {author} {\bibfnamefont {L.~A.}\ \bibnamefont
  {Urena-Lopez}}, \ and\ \bibinfo {author} {\bibfnamefont {A.~R.}\ \bibnamefont
  {Liddle}},\ }\href {\doibase 10.1103/PhysRevD.85.044055} {\bibfield
  {journal} {\bibinfo  {journal} {Phys. Rev.}\ }\textbf {\bibinfo {volume}
  {D85}},\ \bibinfo {pages} {044055} (\bibinfo {year} {2012})},\ \Eprint
  {http://arxiv.org/abs/1107.5669} {arXiv:1107.5669 [astro-ph.CO]} \BibitemShut
  {NoStop}%
\bibitem [{\citenamefont {Suyama}\ \emph {et~al.}(2014)\citenamefont {Suyama},
  \citenamefont {Wu},\ and\ \citenamefont {Yokoyama}}]{Suyama:2014vga}%
  \BibitemOpen
  \bibfield  {author} {\bibinfo {author} {\bibfnamefont {T.}~\bibnamefont
  {Suyama}}, \bibinfo {author} {\bibfnamefont {Y.-P.}\ \bibnamefont {Wu}}, \
  and\ \bibinfo {author} {\bibfnamefont {J.}~\bibnamefont {Yokoyama}},\ }\href
  {\doibase 10.1103/PhysRevD.90.043514} {\bibfield  {journal} {\bibinfo
  {journal} {Phys. Rev.}\ }\textbf {\bibinfo {volume} {D90}},\ \bibinfo {pages}
  {043514} (\bibinfo {year} {2014})},\ \Eprint {http://arxiv.org/abs/1406.0249}
  {arXiv:1406.0249 [astro-ph.CO]} \BibitemShut {NoStop}%
\bibitem [{\citenamefont {Zagorac}\ \emph {et~al.}(2019)\citenamefont
  {Zagorac}, \citenamefont {Easther},\ and\ \citenamefont
  {Padmanabhan}}]{Zagorac:2019ekv}%
  \BibitemOpen
  \bibfield  {author} {\bibinfo {author} {\bibfnamefont {J.~L.}\ \bibnamefont
  {Zagorac}}, \bibinfo {author} {\bibfnamefont {R.}~\bibnamefont {Easther}}, \
  and\ \bibinfo {author} {\bibfnamefont {N.}~\bibnamefont {Padmanabhan}},\
  }\href {\doibase 10.1088/1475-7516/2019/06/052} {\bibfield  {journal}
  {\bibinfo  {journal} {JCAP}\ }\textbf {\bibinfo {volume} {1906}},\ \bibinfo
  {pages} {052} (\bibinfo {year} {2019})},\ \Eprint
  {http://arxiv.org/abs/1903.05053} {arXiv:1903.05053 [astro-ph.CO]}
  \BibitemShut {NoStop}%
\bibitem [{\citenamefont {Lennon}\ \emph {et~al.}(2018)\citenamefont {Lennon},
  \citenamefont {March-Russell}, \citenamefont {Petrossian-Byrne},\ and\
  \citenamefont {Tillim}}]{Lennon:2017tqq}%
  \BibitemOpen
  \bibfield  {author} {\bibinfo {author} {\bibfnamefont {O.}~\bibnamefont
  {Lennon}}, \bibinfo {author} {\bibfnamefont {J.}~\bibnamefont
  {March-Russell}}, \bibinfo {author} {\bibfnamefont {R.}~\bibnamefont
  {Petrossian-Byrne}}, \ and\ \bibinfo {author} {\bibfnamefont
  {H.}~\bibnamefont {Tillim}},\ }\href {\doibase 10.1088/1475-7516/2018/04/009}
  {\bibfield  {journal} {\bibinfo  {journal} {JCAP}\ }\textbf {\bibinfo
  {volume} {04}},\ \bibinfo {pages} {009} (\bibinfo {year} {2018})},\ \Eprint
  {http://arxiv.org/abs/1712.07664} {arXiv:1712.07664 [hep-ph]} \BibitemShut
  {NoStop}%
\bibitem [{\citenamefont {Papanikolaou}\ \emph {et~al.}(2021)\citenamefont
  {Papanikolaou}, \citenamefont {Vennin},\ and\ \citenamefont
  {Langlois}}]{Papanikolaou:2020qtd}%
  \BibitemOpen
  \bibfield  {author} {\bibinfo {author} {\bibfnamefont {T.}~\bibnamefont
  {Papanikolaou}}, \bibinfo {author} {\bibfnamefont {V.}~\bibnamefont
  {Vennin}}, \ and\ \bibinfo {author} {\bibfnamefont {D.}~\bibnamefont
  {Langlois}},\ }\href {\doibase 10.1088/1475-7516/2021/03/053} {\bibfield
  {journal} {\bibinfo  {journal} {JCAP}\ }\textbf {\bibinfo {volume} {03}},\
  \bibinfo {pages} {053} (\bibinfo {year} {2021})},\ \Eprint
  {http://arxiv.org/abs/2010.11573} {arXiv:2010.11573 [astro-ph.CO]}
  \BibitemShut {NoStop}%
\bibitem [{\citenamefont {Dom\`enech}\ \emph {et~al.}(2021)\citenamefont
  {Dom\`enech}, \citenamefont {Lin},\ and\ \citenamefont
  {Sasaki}}]{Domenech:2020ssp}%
  \BibitemOpen
  \bibfield  {author} {\bibinfo {author} {\bibfnamefont {G.}~\bibnamefont
  {Dom\`enech}}, \bibinfo {author} {\bibfnamefont {C.}~\bibnamefont {Lin}}, \
  and\ \bibinfo {author} {\bibfnamefont {M.}~\bibnamefont {Sasaki}},\ }\href
  {\doibase 10.1088/1475-7516/2021/04/062} {\bibfield  {journal} {\bibinfo
  {journal} {JCAP}\ }\textbf {\bibinfo {volume} {04}},\ \bibinfo {pages} {062}
  (\bibinfo {year} {2021})},\ \Eprint {http://arxiv.org/abs/2012.08151}
  {arXiv:2012.08151 [gr-qc]} \BibitemShut {NoStop}%
\bibitem [{\citenamefont {Papanikolaou}(2022)}]{Papanikolaou:2022chm}%
  \BibitemOpen
  \bibfield  {author} {\bibinfo {author} {\bibfnamefont {T.}~\bibnamefont
  {Papanikolaou}},\ }\href {\doibase 10.1088/1475-7516/2022/10/089} {\bibfield
  {journal} {\bibinfo  {journal} {JCAP}\ }\textbf {\bibinfo {volume} {10}},\
  \bibinfo {pages} {089} (\bibinfo {year} {2022})},\ \Eprint
  {http://arxiv.org/abs/2207.11041} {arXiv:2207.11041 [astro-ph.CO]}
  \BibitemShut {NoStop}%
\bibitem [{\citenamefont {Dom\`enech}(2024)}]{Domenech:2023jve}%
  \BibitemOpen
  \bibfield  {author} {\bibinfo {author} {\bibfnamefont {G.}~\bibnamefont
  {Dom\`enech}},\ }\href {\doibase 10.1007/s43673-023-00109-z} {\bibfield
  {journal} {\bibinfo  {journal} {AAPPS Bull.}\ }\textbf {\bibinfo {volume}
  {34}},\ \bibinfo {pages} {4} (\bibinfo {year} {2024})},\ \Eprint
  {http://arxiv.org/abs/2311.02065} {arXiv:2311.02065 [gr-qc]} \BibitemShut
  {NoStop}%
\bibitem [{\citenamefont {Kozaczuk}\ \emph {et~al.}(2022)\citenamefont
  {Kozaczuk}, \citenamefont {Lin},\ and\ \citenamefont
  {Villarama}}]{Kozaczuk:2021wcl}%
  \BibitemOpen
  \bibfield  {author} {\bibinfo {author} {\bibfnamefont {J.}~\bibnamefont
  {Kozaczuk}}, \bibinfo {author} {\bibfnamefont {T.}~\bibnamefont {Lin}}, \
  and\ \bibinfo {author} {\bibfnamefont {E.}~\bibnamefont {Villarama}},\ }\href
  {\doibase 10.1103/PhysRevD.105.123023} {\bibfield  {journal} {\bibinfo
  {journal} {Phys. Rev. D}\ }\textbf {\bibinfo {volume} {105}},\ \bibinfo
  {pages} {123023} (\bibinfo {year} {2022})},\ \Eprint
  {http://arxiv.org/abs/2108.12475} {arXiv:2108.12475 [astro-ph.CO]}
  \BibitemShut {NoStop}%
\bibitem [{\citenamefont {Papanikolaou}\ \emph {et~al.}(2022)\citenamefont
  {Papanikolaou}, \citenamefont {Tzerefos}, \citenamefont {Basilakos},\ and\
  \citenamefont {Saridakis}}]{Papanikolaou:2021uhe}%
  \BibitemOpen
  \bibfield  {author} {\bibinfo {author} {\bibfnamefont {T.}~\bibnamefont
  {Papanikolaou}}, \bibinfo {author} {\bibfnamefont {C.}~\bibnamefont
  {Tzerefos}}, \bibinfo {author} {\bibfnamefont {S.}~\bibnamefont {Basilakos}},
  \ and\ \bibinfo {author} {\bibfnamefont {E.~N.}\ \bibnamefont {Saridakis}},\
  }\href {\doibase 10.1088/1475-7516/2022/10/013} {\bibfield  {journal}
  {\bibinfo  {journal} {JCAP}\ }\textbf {\bibinfo {volume} {10}},\ \bibinfo
  {pages} {013} (\bibinfo {year} {2022})},\ \Eprint
  {http://arxiv.org/abs/2112.15059} {arXiv:2112.15059 [astro-ph.CO]}
  \BibitemShut {NoStop}%
\bibitem [{\citenamefont {Tzerefos}\ \emph {et~al.}(2023)\citenamefont
  {Tzerefos}, \citenamefont {Papanikolaou}, \citenamefont {Saridakis},\ and\
  \citenamefont {Basilakos}}]{Tzerefos:2023mpe}%
  \BibitemOpen
  \bibfield  {author} {\bibinfo {author} {\bibfnamefont {C.}~\bibnamefont
  {Tzerefos}}, \bibinfo {author} {\bibfnamefont {T.}~\bibnamefont
  {Papanikolaou}}, \bibinfo {author} {\bibfnamefont {E.~N.}\ \bibnamefont
  {Saridakis}}, \ and\ \bibinfo {author} {\bibfnamefont {S.}~\bibnamefont
  {Basilakos}},\ }\href {\doibase 10.1103/PhysRevD.107.124019} {\bibfield
  {journal} {\bibinfo  {journal} {Phys. Rev. D}\ }\textbf {\bibinfo {volume}
  {107}},\ \bibinfo {pages} {124019} (\bibinfo {year} {2023})},\ \Eprint
  {http://arxiv.org/abs/2303.16695} {arXiv:2303.16695 [gr-qc]} \BibitemShut
  {NoStop}%
\bibitem [{\citenamefont {Basilakos}\ \emph
  {et~al.}(2024{\natexlab{b}})\citenamefont {Basilakos}, \citenamefont
  {Nanopoulos}, \citenamefont {Papanikolaou}, \citenamefont {Saridakis},\ and\
  \citenamefont {Tzerefos}}]{Basilakos:2023xof}%
  \BibitemOpen
  \bibfield  {author} {\bibinfo {author} {\bibfnamefont {S.}~\bibnamefont
  {Basilakos}}, \bibinfo {author} {\bibfnamefont {D.~V.}\ \bibnamefont
  {Nanopoulos}}, \bibinfo {author} {\bibfnamefont {T.}~\bibnamefont
  {Papanikolaou}}, \bibinfo {author} {\bibfnamefont {E.~N.}\ \bibnamefont
  {Saridakis}}, \ and\ \bibinfo {author} {\bibfnamefont {C.}~\bibnamefont
  {Tzerefos}},\ }\href {\doibase 10.1016/j.physletb.2024.138507} {\bibfield
  {journal} {\bibinfo  {journal} {Phys. Lett. B}\ }\textbf {\bibinfo {volume}
  {850}},\ \bibinfo {pages} {138507} (\bibinfo {year} {2024}{\natexlab{b}})},\
  \Eprint {http://arxiv.org/abs/2307.08601} {arXiv:2307.08601 [hep-th]}
  \BibitemShut {NoStop}%
\bibitem [{\citenamefont {Bhaumik}\ \emph {et~al.}(2023)\citenamefont
  {Bhaumik}, \citenamefont {Jain},\ and\ \citenamefont
  {Lewicki}}]{Bhaumik:2023wmw}%
  \BibitemOpen
  \bibfield  {author} {\bibinfo {author} {\bibfnamefont {N.}~\bibnamefont
  {Bhaumik}}, \bibinfo {author} {\bibfnamefont {R.~K.}\ \bibnamefont {Jain}}, \
  and\ \bibinfo {author} {\bibfnamefont {M.}~\bibnamefont {Lewicki}},\ }\href
  {\doibase 10.1103/PhysRevD.108.123532} {\bibfield  {journal} {\bibinfo
  {journal} {Phys. Rev. D}\ }\textbf {\bibinfo {volume} {108}},\ \bibinfo
  {pages} {123532} (\bibinfo {year} {2023})},\ \Eprint
  {http://arxiv.org/abs/2308.07912} {arXiv:2308.07912 [astro-ph.CO]}
  \BibitemShut {NoStop}%
\bibitem [{\citenamefont {Hooper}\ \emph {et~al.}(2019)\citenamefont {Hooper},
  \citenamefont {Krnjaic},\ and\ \citenamefont {McDermott}}]{Hooper:2019gtx}%
  \BibitemOpen
  \bibfield  {author} {\bibinfo {author} {\bibfnamefont {D.}~\bibnamefont
  {Hooper}}, \bibinfo {author} {\bibfnamefont {G.}~\bibnamefont {Krnjaic}}, \
  and\ \bibinfo {author} {\bibfnamefont {S.~D.}\ \bibnamefont {McDermott}},\
  }\href {\doibase 10.1007/JHEP08(2019)001} {\bibfield  {journal} {\bibinfo
  {journal} {JHEP}\ }\textbf {\bibinfo {volume} {08}},\ \bibinfo {pages} {001}
  (\bibinfo {year} {2019})},\ \Eprint {http://arxiv.org/abs/1905.01301}
  {arXiv:1905.01301 [hep-ph]} \BibitemShut {NoStop}%
\bibitem [{\citenamefont {Nesseris}\ \emph {et~al.}(2020)\citenamefont
  {Nesseris}, \citenamefont {Sapone},\ and\ \citenamefont
  {Sypsas}}]{Nesseris:2019fwr}%
  \BibitemOpen
  \bibfield  {author} {\bibinfo {author} {\bibfnamefont {S.}~\bibnamefont
  {Nesseris}}, \bibinfo {author} {\bibfnamefont {D.}~\bibnamefont {Sapone}}, \
  and\ \bibinfo {author} {\bibfnamefont {S.}~\bibnamefont {Sypsas}},\ }\href
  {\doibase 10.1016/j.dark.2019.100413} {\bibfield  {journal} {\bibinfo
  {journal} {Phys. Dark Univ.}\ }\textbf {\bibinfo {volume} {27}},\ \bibinfo
  {pages} {100413} (\bibinfo {year} {2020})},\ \Eprint
  {http://arxiv.org/abs/1907.05608} {arXiv:1907.05608 [astro-ph.CO]}
  \BibitemShut {NoStop}%
\bibitem [{\citenamefont {Lunardini}\ and\ \citenamefont
  {Perez-Gonzalez}(2020)}]{Lunardini:2019zob}%
  \BibitemOpen
  \bibfield  {author} {\bibinfo {author} {\bibfnamefont {C.}~\bibnamefont
  {Lunardini}}\ and\ \bibinfo {author} {\bibfnamefont {Y.~F.}\ \bibnamefont
  {Perez-Gonzalez}},\ }\href {\doibase 10.1088/1475-7516/2020/08/014}
  {\bibfield  {journal} {\bibinfo  {journal} {JCAP}\ }\textbf {\bibinfo
  {volume} {08}},\ \bibinfo {pages} {014} (\bibinfo {year} {2020})},\ \Eprint
  {http://arxiv.org/abs/1910.07864} {arXiv:1910.07864 [hep-ph]} \BibitemShut
  {NoStop}%
\bibitem [{\citenamefont {Papanikolaou}(2023)}]{Papanikolaou:2023oxq}%
  \BibitemOpen
  \bibfield  {author} {\bibinfo {author} {\bibfnamefont {T.}~\bibnamefont
  {Papanikolaou}},\ }\href {\doibase 10.22323/1.436.0265} {\bibfield  {journal}
  {\bibinfo  {journal} {PoS}\ }\textbf {\bibinfo {volume} {CORFU2022}},\
  \bibinfo {pages} {265} (\bibinfo {year} {2023})},\ \Eprint
  {http://arxiv.org/abs/2303.00600} {arXiv:2303.00600 [astro-ph.CO]}
  \BibitemShut {NoStop}%
\bibitem [{\citenamefont {Barrow}\ \emph {et~al.}(1991)\citenamefont {Barrow},
  \citenamefont {Copeland}, \citenamefont {Kolb},\ and\ \citenamefont
  {Liddle}}]{Barrow:1990he}%
  \BibitemOpen
  \bibfield  {author} {\bibinfo {author} {\bibfnamefont {J.~D.}\ \bibnamefont
  {Barrow}}, \bibinfo {author} {\bibfnamefont {E.~J.}\ \bibnamefont
  {Copeland}}, \bibinfo {author} {\bibfnamefont {E.~W.}\ \bibnamefont {Kolb}},
  \ and\ \bibinfo {author} {\bibfnamefont {A.~R.}\ \bibnamefont {Liddle}},\
  }\href {\doibase 10.1103/PhysRevD.43.984} {\bibfield  {journal} {\bibinfo
  {journal} {Phys. Rev. D}\ }\textbf {\bibinfo {volume} {43}},\ \bibinfo
  {pages} {984} (\bibinfo {year} {1991})}\BibitemShut {NoStop}%
\bibitem [{\citenamefont {Hamada}\ and\ \citenamefont
  {Iso}(2017)}]{Hamada:2016jnq}%
  \BibitemOpen
  \bibfield  {author} {\bibinfo {author} {\bibfnamefont {Y.}~\bibnamefont
  {Hamada}}\ and\ \bibinfo {author} {\bibfnamefont {S.}~\bibnamefont {Iso}},\
  }\href {\doibase 10.1093/ptep/ptx011} {\bibfield  {journal} {\bibinfo
  {journal} {PTEP}\ }\textbf {\bibinfo {volume} {2017}},\ \bibinfo {pages}
  {033B02} (\bibinfo {year} {2017})},\ \Eprint
  {http://arxiv.org/abs/1610.02586} {arXiv:1610.02586 [hep-ph]} \BibitemShut
  {NoStop}%
\bibitem [{\citenamefont {Bhaumik}\ \emph {et~al.}(2022)\citenamefont
  {Bhaumik}, \citenamefont {Ghoshal},\ and\ \citenamefont
  {Lewicki}}]{Bhaumik:2022pil}%
  \BibitemOpen
  \bibfield  {author} {\bibinfo {author} {\bibfnamefont {N.}~\bibnamefont
  {Bhaumik}}, \bibinfo {author} {\bibfnamefont {A.}~\bibnamefont {Ghoshal}}, \
  and\ \bibinfo {author} {\bibfnamefont {M.}~\bibnamefont {Lewicki}},\ }\href
  {\doibase 10.1007/JHEP07(2022)130} {\bibfield  {journal} {\bibinfo  {journal}
  {JHEP}\ }\textbf {\bibinfo {volume} {07}},\ \bibinfo {pages} {130} (\bibinfo
  {year} {2022})},\ \Eprint {http://arxiv.org/abs/2205.06260} {arXiv:2205.06260
  [astro-ph.CO]} \BibitemShut {NoStop}%
\bibitem [{\citenamefont {Alexandre}\ \emph {et~al.}(2013)\citenamefont
  {Alexandre}, \citenamefont {Houston},\ and\ \citenamefont
  {Mavromatos}}]{Alexandre:2013iva}%
  \BibitemOpen
  \bibfield  {author} {\bibinfo {author} {\bibfnamefont {J.}~\bibnamefont
  {Alexandre}}, \bibinfo {author} {\bibfnamefont {N.}~\bibnamefont {Houston}},
  \ and\ \bibinfo {author} {\bibfnamefont {N.~E.}\ \bibnamefont {Mavromatos}},\
  }\href {\doibase 10.1103/PhysRevD.88.125017} {\bibfield  {journal} {\bibinfo
  {journal} {Phys. Rev. D}\ }\textbf {\bibinfo {volume} {88}},\ \bibinfo
  {pages} {125017} (\bibinfo {year} {2013})},\ \Eprint
  {http://arxiv.org/abs/1310.4122} {arXiv:1310.4122 [hep-th]} \BibitemShut
  {NoStop}%
\bibitem [{\citenamefont {Alexandre}\ \emph {et~al.}(2014)\citenamefont
  {Alexandre}, \citenamefont {Houston},\ and\ \citenamefont
  {Mavromatos}}]{Alexandre:2013nqa}%
  \BibitemOpen
  \bibfield  {author} {\bibinfo {author} {\bibfnamefont {J.}~\bibnamefont
  {Alexandre}}, \bibinfo {author} {\bibfnamefont {N.}~\bibnamefont {Houston}},
  \ and\ \bibinfo {author} {\bibfnamefont {N.~E.}\ \bibnamefont {Mavromatos}},\
  }\href {\doibase 10.1103/PhysRevD.89.027703} {\bibfield  {journal} {\bibinfo
  {journal} {Phys. Rev. D}\ }\textbf {\bibinfo {volume} {89}},\ \bibinfo
  {pages} {027703} (\bibinfo {year} {2014})},\ \Eprint
  {http://arxiv.org/abs/1312.5197} {arXiv:1312.5197 [gr-qc]} \BibitemShut
  {NoStop}%
\bibitem [{\citenamefont {Alexandre}\ \emph {et~al.}(2015)\citenamefont
  {Alexandre}, \citenamefont {Houston},\ and\ \citenamefont
  {Mavromatos}}]{Alexandre:2014lla}%
  \BibitemOpen
  \bibfield  {author} {\bibinfo {author} {\bibfnamefont {J.}~\bibnamefont
  {Alexandre}}, \bibinfo {author} {\bibfnamefont {N.}~\bibnamefont {Houston}},
  \ and\ \bibinfo {author} {\bibfnamefont {N.~E.}\ \bibnamefont {Mavromatos}},\
  }\href {\doibase 10.1142/S0218271815410047} {\bibfield  {journal} {\bibinfo
  {journal} {Int. J. Mod. Phys. D}\ }\textbf {\bibinfo {volume} {24}},\
  \bibinfo {pages} {1541004} (\bibinfo {year} {2015})},\ \Eprint
  {http://arxiv.org/abs/1409.3183} {arXiv:1409.3183 [gr-qc]} \BibitemShut
  {NoStop}%
\bibitem [{\citenamefont {Fradkin}\ and\ \citenamefont
  {Tseytlin}(1984)}]{Fradkin:1983mq}%
  \BibitemOpen
  \bibfield  {author} {\bibinfo {author} {\bibfnamefont {E.~S.}\ \bibnamefont
  {Fradkin}}\ and\ \bibinfo {author} {\bibfnamefont {A.~A.}\ \bibnamefont
  {Tseytlin}},\ }\href {\doibase 10.1016/0550-3213(84)90074-9} {\bibfield
  {journal} {\bibinfo  {journal} {Nucl. Phys. B}\ }\textbf {\bibinfo {volume}
  {234}},\ \bibinfo {pages} {472} (\bibinfo {year} {1984})}\BibitemShut
  {NoStop}%
\bibitem [{\citenamefont {Akrami}\ \emph
  {et~al.}(2020{\natexlab{a}})\citenamefont {Akrami} \emph
  {et~al.}}]{Planck:2018jri}%
  \BibitemOpen
  \bibfield  {author} {\bibinfo {author} {\bibfnamefont {Y.}~\bibnamefont
  {Akrami}} \emph {et~al.} (\bibinfo {collaboration} {Planck}),\ }\href
  {\doibase 10.1051/0004-6361/201833887} {\bibfield  {journal} {\bibinfo
  {journal} {Astron. Astrophys.}\ }\textbf {\bibinfo {volume} {641}},\ \bibinfo
  {pages} {A10} (\bibinfo {year} {2020}{\natexlab{a}})},\ \Eprint
  {http://arxiv.org/abs/1807.06211} {arXiv:1807.06211 [astro-ph.CO]}
  \BibitemShut {NoStop}%
\bibitem [{\citenamefont {Hellerman}\ \emph {et~al.}(2001)\citenamefont
  {Hellerman}, \citenamefont {Kaloper},\ and\ \citenamefont
  {Susskind}}]{scat1}%
  \BibitemOpen
  \bibfield  {author} {\bibinfo {author} {\bibfnamefont {S.}~\bibnamefont
  {Hellerman}}, \bibinfo {author} {\bibfnamefont {N.}~\bibnamefont {Kaloper}},
  \ and\ \bibinfo {author} {\bibfnamefont {L.}~\bibnamefont {Susskind}},\
  }\href {\doibase 10.1088/1126-6708/2001/06/003} {\bibfield  {journal}
  {\bibinfo  {journal} {JHEP}\ }\textbf {\bibinfo {volume} {06}},\ \bibinfo
  {pages} {003} (\bibinfo {year} {2001})},\ \Eprint
  {http://arxiv.org/abs/hep-th/0104180} {arXiv:hep-th/0104180} \BibitemShut
  {NoStop}%
\bibitem [{\citenamefont {Fischler}\ \emph {et~al.}(2001)\citenamefont
  {Fischler}, \citenamefont {Kashani-Poor}, \citenamefont {McNees},\ and\
  \citenamefont {Paban}}]{scat2}%
  \BibitemOpen
  \bibfield  {author} {\bibinfo {author} {\bibfnamefont {W.}~\bibnamefont
  {Fischler}}, \bibinfo {author} {\bibfnamefont {A.}~\bibnamefont
  {Kashani-Poor}}, \bibinfo {author} {\bibfnamefont {R.}~\bibnamefont
  {McNees}}, \ and\ \bibinfo {author} {\bibfnamefont {S.}~\bibnamefont
  {Paban}},\ }\href {\doibase 10.1088/1126-6708/2001/07/003} {\bibfield
  {journal} {\bibinfo  {journal} {JHEP}\ }\textbf {\bibinfo {volume} {07}},\
  \bibinfo {pages} {003} (\bibinfo {year} {2001})},\ \Eprint
  {http://arxiv.org/abs/hep-th/0104181} {arXiv:hep-th/0104181} \BibitemShut
  {NoStop}%
\bibitem [{\citenamefont {Ooguri}\ and\ \citenamefont {Vafa}(2007)}]{swamp1}%
  \BibitemOpen
  \bibfield  {author} {\bibinfo {author} {\bibfnamefont {H.}~\bibnamefont
  {Ooguri}}\ and\ \bibinfo {author} {\bibfnamefont {C.}~\bibnamefont {Vafa}},\
  }\href {\doibase 10.1016/j.nuclphysb.2006.10.033} {\bibfield  {journal}
  {\bibinfo  {journal} {Nucl. Phys. B}\ }\textbf {\bibinfo {volume} {766}},\
  \bibinfo {pages} {21} (\bibinfo {year} {2007})},\ \Eprint
  {http://arxiv.org/abs/hep-th/0605264} {arXiv:hep-th/0605264} \BibitemShut
  {NoStop}%
\bibitem [{\citenamefont {Ooguri}\ \emph {et~al.}(2019)\citenamefont {Ooguri},
  \citenamefont {Palti}, \citenamefont {Shiu},\ and\ \citenamefont
  {Vafa}}]{swamp2}%
  \BibitemOpen
  \bibfield  {author} {\bibinfo {author} {\bibfnamefont {H.}~\bibnamefont
  {Ooguri}}, \bibinfo {author} {\bibfnamefont {E.}~\bibnamefont {Palti}},
  \bibinfo {author} {\bibfnamefont {G.}~\bibnamefont {Shiu}}, \ and\ \bibinfo
  {author} {\bibfnamefont {C.}~\bibnamefont {Vafa}},\ }\href {\doibase
  10.1016/j.physletb.2018.11.018} {\bibfield  {journal} {\bibinfo  {journal}
  {Phys. Lett. B}\ }\textbf {\bibinfo {volume} {788}},\ \bibinfo {pages} {180}
  (\bibinfo {year} {2019})},\ \Eprint {http://arxiv.org/abs/1810.05506}
  {arXiv:1810.05506 [hep-th]} \BibitemShut {NoStop}%
\bibitem [{\citenamefont {Palti}(2019)}]{swamp3}%
  \BibitemOpen
  \bibfield  {author} {\bibinfo {author} {\bibfnamefont {E.}~\bibnamefont
  {Palti}},\ }\href {\doibase 10.1002/prop.201900037} {\bibfield  {journal}
  {\bibinfo  {journal} {Fortsch. Phys.}\ }\textbf {\bibinfo {volume} {67}},\
  \bibinfo {pages} {1900037} (\bibinfo {year} {2019})},\ \Eprint
  {http://arxiv.org/abs/1903.06239} {arXiv:1903.06239 [hep-th]} \BibitemShut
  {NoStop}%
\bibitem [{\citenamefont {Garg}\ and\ \citenamefont {Krishnan}(2019)}]{swamp5}%
  \BibitemOpen
  \bibfield  {author} {\bibinfo {author} {\bibfnamefont {S.~K.}\ \bibnamefont
  {Garg}}\ and\ \bibinfo {author} {\bibfnamefont {C.}~\bibnamefont
  {Krishnan}},\ }\href {\doibase 10.1007/JHEP11(2019)075} {\bibfield  {journal}
  {\bibinfo  {journal} {JHEP}\ }\textbf {\bibinfo {volume} {11}},\ \bibinfo
  {pages} {075} (\bibinfo {year} {2019})},\ \Eprint
  {http://arxiv.org/abs/1807.05193} {arXiv:1807.05193 [hep-th]} \BibitemShut
  {NoStop}%
\bibitem [{\citenamefont {Agmon}\ \emph {et~al.}(2022)\citenamefont {Agmon},
  \citenamefont {Bedroya}, \citenamefont {Kang},\ and\ \citenamefont
  {Vafa}}]{swamp4}%
  \BibitemOpen
  \bibfield  {author} {\bibinfo {author} {\bibfnamefont {N.~B.}\ \bibnamefont
  {Agmon}}, \bibinfo {author} {\bibfnamefont {A.}~\bibnamefont {Bedroya}},
  \bibinfo {author} {\bibfnamefont {M.~J.}\ \bibnamefont {Kang}}, \ and\
  \bibinfo {author} {\bibfnamefont {C.}~\bibnamefont {Vafa}},\ }\href@noop {}
  {\  (\bibinfo {year} {2022})},\ \Eprint {http://arxiv.org/abs/2212.06187}
  {arXiv:2212.06187 [hep-th]} \BibitemShut {NoStop}%
\bibitem [{\citenamefont {Cunillera}\ \emph {et~al.}(2022)\citenamefont
  {Cunillera}, \citenamefont {Emond}, \citenamefont {Leh\'ebel},\ and\
  \citenamefont {Padilla}}]{Cunillera:2021fbc}%
  \BibitemOpen
  \bibfield  {author} {\bibinfo {author} {\bibfnamefont {F.}~\bibnamefont
  {Cunillera}}, \bibinfo {author} {\bibfnamefont {W.~T.}\ \bibnamefont
  {Emond}}, \bibinfo {author} {\bibfnamefont {A.}~\bibnamefont {Leh\'ebel}}, \
  and\ \bibinfo {author} {\bibfnamefont {A.}~\bibnamefont {Padilla}},\ }\href
  {\doibase 10.1007/JHEP02(2022)012} {\bibfield  {journal} {\bibinfo  {journal}
  {JHEP}\ }\textbf {\bibinfo {volume} {02}},\ \bibinfo {pages} {012} (\bibinfo
  {year} {2022})},\ \Eprint {http://arxiv.org/abs/2112.05771} {arXiv:2112.05771
  [hep-th]} \BibitemShut {NoStop}%
\bibitem [{\citenamefont {Dorlis}\ \emph {et~al.}(2024)\citenamefont {Dorlis},
  \citenamefont {Mavromatos},\ and\ \citenamefont {Vlachos}}]{Dorlis:2024yqw}%
  \BibitemOpen
  \bibfield  {author} {\bibinfo {author} {\bibfnamefont {P.}~\bibnamefont
  {Dorlis}}, \bibinfo {author} {\bibfnamefont {N.~E.}\ \bibnamefont
  {Mavromatos}}, \ and\ \bibinfo {author} {\bibfnamefont {S.-N.}\ \bibnamefont
  {Vlachos}},\ }\href@noop {} {\  (\bibinfo {year} {2024})},\ \Eprint
  {http://arxiv.org/abs/2403.09005} {arXiv:2403.09005 [gr-qc]} \BibitemShut
  {NoStop}%
\bibitem [{\citenamefont {Mavromatos}\ \emph {et~al.}(2024)\citenamefont
  {Mavromatos}, \citenamefont {Dorlis},\ and\ \citenamefont
  {Vlachos}}]{Mavromatos:2024pho}%
  \BibitemOpen
  \bibfield  {author} {\bibinfo {author} {\bibfnamefont {N.~E.}\ \bibnamefont
  {Mavromatos}}, \bibinfo {author} {\bibfnamefont {P.}~\bibnamefont {Dorlis}},
  \ and\ \bibinfo {author} {\bibfnamefont {S.-N.}\ \bibnamefont {Vlachos}},\
  }in\ \href@noop {} {\emph {\bibinfo {booktitle} {{Workshop on the Standard
  Model and Beyond}}}}\ (\bibinfo {year} {2024})\ \Eprint
  {http://arxiv.org/abs/2404.18741} {arXiv:2404.18741 [gr-qc]} \BibitemShut
  {NoStop}%
\bibitem [{\citenamefont {Mavromatos}\ \emph {et~al.}(2022)\citenamefont
  {Mavromatos}, \citenamefont {Spanos},\ and\ \citenamefont
  {Stamou}}]{Mavromatos:2022yql}%
  \BibitemOpen
  \bibfield  {author} {\bibinfo {author} {\bibfnamefont {N.~E.}\ \bibnamefont
  {Mavromatos}}, \bibinfo {author} {\bibfnamefont {V.~C.}\ \bibnamefont
  {Spanos}}, \ and\ \bibinfo {author} {\bibfnamefont {I.~D.}\ \bibnamefont
  {Stamou}},\ }\href {\doibase 10.1103/PhysRevD.106.063532} {\bibfield
  {journal} {\bibinfo  {journal} {Phys. Rev. D}\ }\textbf {\bibinfo {volume}
  {106}},\ \bibinfo {pages} {063532} (\bibinfo {year} {2022})},\ \Eprint
  {http://arxiv.org/abs/2206.07963} {arXiv:2206.07963 [hep-th]} \BibitemShut
  {NoStop}%
\bibitem [{\citenamefont {Germani}\ and\ \citenamefont
  {Musco}(2019)}]{Germani:2018jgr}%
  \BibitemOpen
  \bibfield  {author} {\bibinfo {author} {\bibfnamefont {C.}~\bibnamefont
  {Germani}}\ and\ \bibinfo {author} {\bibfnamefont {I.}~\bibnamefont
  {Musco}},\ }\href {\doibase 10.1103/PhysRevLett.122.141302} {\bibfield
  {journal} {\bibinfo  {journal} {Phys. Rev. Lett.}\ }\textbf {\bibinfo
  {volume} {122}},\ \bibinfo {pages} {141302} (\bibinfo {year} {2019})},\
  \Eprint {http://arxiv.org/abs/1805.04087} {arXiv:1805.04087 [astro-ph.CO]}
  \BibitemShut {NoStop}%
\bibitem [{\citenamefont {Moradinezhad~Dizgah}\ \emph
  {et~al.}(2019)\citenamefont {Moradinezhad~Dizgah}, \citenamefont
  {Franciolini},\ and\ \citenamefont {Riotto}}]{MoradinezhadDizgah:2019wjf}%
  \BibitemOpen
  \bibfield  {author} {\bibinfo {author} {\bibfnamefont {A.}~\bibnamefont
  {Moradinezhad~Dizgah}}, \bibinfo {author} {\bibfnamefont {G.}~\bibnamefont
  {Franciolini}}, \ and\ \bibinfo {author} {\bibfnamefont {A.}~\bibnamefont
  {Riotto}},\ }\href {\doibase 10.1088/1475-7516/2019/11/001} {\bibfield
  {journal} {\bibinfo  {journal} {JCAP}\ }\textbf {\bibinfo {volume} {11}},\
  \bibinfo {pages} {001} (\bibinfo {year} {2019})},\ \Eprint
  {http://arxiv.org/abs/1906.08978} {arXiv:1906.08978 [astro-ph.CO]}
  \BibitemShut {NoStop}%
\bibitem [{\citenamefont {Musco}\ \emph {et~al.}(2009)\citenamefont {Musco},
  \citenamefont {Miller},\ and\ \citenamefont {Polnarev}}]{Musco:2008hv}%
  \BibitemOpen
  \bibfield  {author} {\bibinfo {author} {\bibfnamefont {I.}~\bibnamefont
  {Musco}}, \bibinfo {author} {\bibfnamefont {J.~C.}\ \bibnamefont {Miller}}, \
  and\ \bibinfo {author} {\bibfnamefont {A.~G.}\ \bibnamefont {Polnarev}},\
  }\href {\doibase 10.1088/0264-9381/26/23/235001} {\bibfield  {journal}
  {\bibinfo  {journal} {Class. Quant. Grav.}\ }\textbf {\bibinfo {volume}
  {26}},\ \bibinfo {pages} {235001} (\bibinfo {year} {2009})},\ \Eprint
  {http://arxiv.org/abs/0811.1452} {arXiv:0811.1452 [gr-qc]} \BibitemShut
  {NoStop}%
\bibitem [{\citenamefont {Kolb}\ and\ \citenamefont
  {Turner}(1990)}]{Kolb:1990vq}%
  \BibitemOpen
  \bibfield  {author} {\bibinfo {author} {\bibfnamefont {E.~W.}\ \bibnamefont
  {Kolb}}\ and\ \bibinfo {author} {\bibfnamefont {M.~S.}\ \bibnamefont
  {Turner}},\ }\href@noop {} {\emph {\bibinfo {title} {{The Early
  Universe}}}},\ Vol.~\bibinfo {volume} {69}\ (\bibinfo {year}
  {1990})\BibitemShut {NoStop}%
\bibitem [{\citenamefont {Akrami}\ \emph
  {et~al.}(2020{\natexlab{b}})\citenamefont {Akrami} \emph
  {et~al.}}]{Planck:2019kim}%
  \BibitemOpen
  \bibfield  {author} {\bibinfo {author} {\bibfnamefont {Y.}~\bibnamefont
  {Akrami}} \emph {et~al.} (\bibinfo {collaboration} {Planck}),\ }\href
  {\doibase 10.1051/0004-6361/201935891} {\bibfield  {journal} {\bibinfo
  {journal} {Astron. Astrophys.}\ }\textbf {\bibinfo {volume} {641}},\ \bibinfo
  {pages} {A9} (\bibinfo {year} {2020}{\natexlab{b}})},\ \Eprint
  {http://arxiv.org/abs/1905.05697} {arXiv:1905.05697 [astro-ph.CO]}
  \BibitemShut {NoStop}%
\bibitem [{\citenamefont {Desjacques}\ and\ \citenamefont
  {Riotto}(2018)}]{Desjacques:2018wuu}%
  \BibitemOpen
  \bibfield  {author} {\bibinfo {author} {\bibfnamefont {V.}~\bibnamefont
  {Desjacques}}\ and\ \bibinfo {author} {\bibfnamefont {A.}~\bibnamefont
  {Riotto}},\ }\href {\doibase 10.1103/PhysRevD.98.123533} {\bibfield
  {journal} {\bibinfo  {journal} {Phys. Rev. D}\ }\textbf {\bibinfo {volume}
  {98}},\ \bibinfo {pages} {123533} (\bibinfo {year} {2018})},\ \Eprint
  {http://arxiv.org/abs/1806.10414} {arXiv:1806.10414 [astro-ph.CO]}
  \BibitemShut {NoStop}%
\bibitem [{\citenamefont {Ali-Haimoud}(2018)}]{Ali-Haimoud:2018dau}%
  \BibitemOpen
  \bibfield  {author} {\bibinfo {author} {\bibfnamefont {Y.}~\bibnamefont
  {Ali-Haimoud}},\ }\href {\doibase 10.1103/PhysRevLett.121.081304} {\bibfield
  {journal} {\bibinfo  {journal} {Phys. Rev. Lett.}\ }\textbf {\bibinfo
  {volume} {121}},\ \bibinfo {pages} {081304} (\bibinfo {year} {2018})},\
  \Eprint {http://arxiv.org/abs/1805.05912} {arXiv:1805.05912 [astro-ph.CO]}
  \BibitemShut {NoStop}%
\bibitem [{\citenamefont {Inman}\ and\ \citenamefont
  {Ali-Ha\"\i{}moud}(2019)}]{Inman:2019wvr}%
  \BibitemOpen
  \bibfield  {author} {\bibinfo {author} {\bibfnamefont {D.}~\bibnamefont
  {Inman}}\ and\ \bibinfo {author} {\bibfnamefont {Y.}~\bibnamefont
  {Ali-Ha\"\i{}moud}},\ }\href {\doibase 10.1103/PhysRevD.100.083528}
  {\bibfield  {journal} {\bibinfo  {journal} {Phys. Rev. D}\ }\textbf {\bibinfo
  {volume} {100}},\ \bibinfo {pages} {083528} (\bibinfo {year} {2019})},\
  \Eprint {http://arxiv.org/abs/1907.08129} {arXiv:1907.08129 [astro-ph.CO]}
  \BibitemShut {NoStop}%
\bibitem [{\citenamefont {Inomata}\ \emph {et~al.}(2020)\citenamefont
  {Inomata}, \citenamefont {Kawasaki}, \citenamefont {Mukaida}, \citenamefont
  {Terada},\ and\ \citenamefont {Yanagida}}]{Inomata:2020lmk}%
  \BibitemOpen
  \bibfield  {author} {\bibinfo {author} {\bibfnamefont {K.}~\bibnamefont
  {Inomata}}, \bibinfo {author} {\bibfnamefont {M.}~\bibnamefont {Kawasaki}},
  \bibinfo {author} {\bibfnamefont {K.}~\bibnamefont {Mukaida}}, \bibinfo
  {author} {\bibfnamefont {T.}~\bibnamefont {Terada}}, \ and\ \bibinfo {author}
  {\bibfnamefont {T.~T.}\ \bibnamefont {Yanagida}},\ }\href {\doibase
  10.1103/PhysRevD.101.123533} {\bibfield  {journal} {\bibinfo  {journal}
  {Phys. Rev. D}\ }\textbf {\bibinfo {volume} {101}},\ \bibinfo {pages}
  {123533} (\bibinfo {year} {2020})},\ \Eprint
  {http://arxiv.org/abs/2003.10455} {arXiv:2003.10455 [astro-ph.CO]}
  \BibitemShut {NoStop}%
\bibitem [{\citenamefont {Hawking}(1974)}]{Hawking:1974rv}%
  \BibitemOpen
  \bibfield  {author} {\bibinfo {author} {\bibfnamefont {S.~W.}\ \bibnamefont
  {Hawking}},\ }\href {\doibase 10.1038/248030a0} {\bibfield  {journal}
  {\bibinfo  {journal} {Nature}\ }\textbf {\bibinfo {volume} {248}},\ \bibinfo
  {pages} {30} (\bibinfo {year} {1974})}\BibitemShut {NoStop}%
\bibitem [{\citenamefont {Aghanim}\ \emph {et~al.}(2020)\citenamefont {Aghanim}
  \emph {et~al.}}]{Planck:2018vyg}%
  \BibitemOpen
  \bibfield  {author} {\bibinfo {author} {\bibfnamefont {N.}~\bibnamefont
  {Aghanim}} \emph {et~al.} (\bibinfo {collaboration} {Planck}),\ }\href
  {\doibase 10.1051/0004-6361/201833910} {\bibfield  {journal} {\bibinfo
  {journal} {Astron. Astrophys.}\ }\textbf {\bibinfo {volume} {641}},\ \bibinfo
  {pages} {A6} (\bibinfo {year} {2020})},\ \bibinfo {note} {[Erratum:
  Astron.Astrophys. 652, C4 (2021)]},\ \Eprint
  {http://arxiv.org/abs/1807.06209} {arXiv:1807.06209 [astro-ph.CO]}
  \BibitemShut {NoStop}%
\bibitem [{\citenamefont {Smith}\ \emph {et~al.}(2006)\citenamefont {Smith},
  \citenamefont {Pierpaoli},\ and\ \citenamefont
  {Kamionkowski}}]{Smith:2006nka}%
  \BibitemOpen
  \bibfield  {author} {\bibinfo {author} {\bibfnamefont {T.~L.}\ \bibnamefont
  {Smith}}, \bibinfo {author} {\bibfnamefont {E.}~\bibnamefont {Pierpaoli}}, \
  and\ \bibinfo {author} {\bibfnamefont {M.}~\bibnamefont {Kamionkowski}},\
  }\href {\doibase 10.1103/PhysRevLett.97.021301} {\bibfield  {journal}
  {\bibinfo  {journal} {Phys. Rev. Lett.}\ }\textbf {\bibinfo {volume} {97}},\
  \bibinfo {pages} {021301} (\bibinfo {year} {2006})},\ \Eprint
  {http://arxiv.org/abs/astro-ph/0603144} {arXiv:astro-ph/0603144} \BibitemShut
  {NoStop}%
\bibitem [{\citenamefont {Basilakos}\ \emph
  {et~al.}(2020{\natexlab{c}})\citenamefont {Basilakos}, \citenamefont
  {Mavromatos},\ and\ \citenamefont {Sol\`a~Peracaula}}]{Basilakos:2019acj}%
  \BibitemOpen
  \bibfield  {author} {\bibinfo {author} {\bibfnamefont {S.}~\bibnamefont
  {Basilakos}}, \bibinfo {author} {\bibfnamefont {N.~E.}\ \bibnamefont
  {Mavromatos}}, \ and\ \bibinfo {author} {\bibfnamefont {J.}~\bibnamefont
  {Sol\`a~Peracaula}},\ }\href {\doibase 10.1103/PhysRevD.101.045001}
  {\bibfield  {journal} {\bibinfo  {journal} {Phys. Rev. D}\ }\textbf {\bibinfo
  {volume} {101}},\ \bibinfo {pages} {045001} (\bibinfo {year}
  {2020}{\natexlab{c}})},\ \Eprint {http://arxiv.org/abs/1907.04890}
  {arXiv:1907.04890 [hep-ph]} \BibitemShut {NoStop}%
\bibitem [{\citenamefont {Mavromatos}(2023)}]{Mavromatos:2022xdo}%
  \BibitemOpen
  \bibfield  {author} {\bibinfo {author} {\bibfnamefont {N.~E.}\ \bibnamefont
  {Mavromatos}},\ }\href {\doibase 10.1007/978-3-031-31520-6_1} {\bibfield
  {journal} {\bibinfo  {journal} {Lect. Notes Phys.}\ }\textbf {\bibinfo
  {volume} {1017}},\ \bibinfo {pages} {3} (\bibinfo {year} {2023})},\ \Eprint
  {http://arxiv.org/abs/2205.07044} {arXiv:2205.07044 [hep-th]} \BibitemShut
  {NoStop}%
\bibitem [{\citenamefont {Starobinsky}(1980)}]{Starobinsky:1980te}%
  \BibitemOpen
  \bibfield  {author} {\bibinfo {author} {\bibfnamefont {A.~A.}\ \bibnamefont
  {Starobinsky}},\ }\href {\doibase 10.1016/0370-2693(80)90670-X} {\bibfield
  {journal} {\bibinfo  {journal} {Phys. Lett. B}\ }\textbf {\bibinfo {volume}
  {91}},\ \bibinfo {pages} {99} (\bibinfo {year} {1980})}\BibitemShut {NoStop}%
\bibitem [{\citenamefont {Hwang}\ \emph {et~al.}(2017)\citenamefont {Hwang},
  \citenamefont {Jeong},\ and\ \citenamefont {Noh}}]{Hwang_2017}%
  \BibitemOpen
  \bibfield  {author} {\bibinfo {author} {\bibfnamefont {J.-c.}\ \bibnamefont
  {Hwang}}, \bibinfo {author} {\bibfnamefont {D.}~\bibnamefont {Jeong}}, \ and\
  \bibinfo {author} {\bibfnamefont {H.}~\bibnamefont {Noh}},\ }\href {\doibase
  10.3847/1538-4357/aa74be} {\bibfield  {journal} {\bibinfo  {journal} {The
  Astrophysical Journal}\ }\textbf {\bibinfo {volume} {842}},\ \bibinfo {pages}
  {46} (\bibinfo {year} {2017})}\BibitemShut {NoStop}%
\bibitem [{\citenamefont {Tomikawa}\ and\ \citenamefont
  {Kobayashi}(2020)}]{Tomikawa:2019tvi}%
  \BibitemOpen
  \bibfield  {author} {\bibinfo {author} {\bibfnamefont {K.}~\bibnamefont
  {Tomikawa}}\ and\ \bibinfo {author} {\bibfnamefont {T.}~\bibnamefont
  {Kobayashi}},\ }\href {\doibase 10.1103/PhysRevD.101.083529} {\bibfield
  {journal} {\bibinfo  {journal} {Phys. Rev. D}\ }\textbf {\bibinfo {volume}
  {101}},\ \bibinfo {pages} {083529} (\bibinfo {year} {2020})},\ \Eprint
  {http://arxiv.org/abs/1910.01880} {arXiv:1910.01880 [gr-qc]} \BibitemShut
  {NoStop}%
\bibitem [{\citenamefont {De~Luca}\ \emph {et~al.}(2020)\citenamefont
  {De~Luca}, \citenamefont {Franciolini}, \citenamefont {Kehagias},\ and\
  \citenamefont {Riotto}}]{DeLuca:2019ufz}%
  \BibitemOpen
  \bibfield  {author} {\bibinfo {author} {\bibfnamefont {V.}~\bibnamefont
  {De~Luca}}, \bibinfo {author} {\bibfnamefont {G.}~\bibnamefont
  {Franciolini}}, \bibinfo {author} {\bibfnamefont {A.}~\bibnamefont
  {Kehagias}}, \ and\ \bibinfo {author} {\bibfnamefont {A.}~\bibnamefont
  {Riotto}},\ }\href {\doibase 10.1088/1475-7516/2020/03/014} {\bibfield
  {journal} {\bibinfo  {journal} {JCAP}\ }\textbf {\bibinfo {volume} {03}},\
  \bibinfo {pages} {014} (\bibinfo {year} {2020})},\ \Eprint
  {http://arxiv.org/abs/1911.09689} {arXiv:1911.09689 [gr-qc]} \BibitemShut
  {NoStop}%
\bibitem [{\citenamefont {Yuan}\ \emph {et~al.}(2020)\citenamefont {Yuan},
  \citenamefont {Chen},\ and\ \citenamefont {Huang}}]{Yuan:2019fwv}%
  \BibitemOpen
  \bibfield  {author} {\bibinfo {author} {\bibfnamefont {C.}~\bibnamefont
  {Yuan}}, \bibinfo {author} {\bibfnamefont {Z.-C.}\ \bibnamefont {Chen}}, \
  and\ \bibinfo {author} {\bibfnamefont {Q.-G.}\ \bibnamefont {Huang}},\ }\href
  {\doibase 10.1103/PhysRevD.101.063018} {\bibfield  {journal} {\bibinfo
  {journal} {Phys. Rev. D}\ }\textbf {\bibinfo {volume} {101}},\ \bibinfo
  {pages} {063018} (\bibinfo {year} {2020})},\ \Eprint
  {http://arxiv.org/abs/1912.00885} {arXiv:1912.00885 [astro-ph.CO]}
  \BibitemShut {NoStop}%
\bibitem [{\citenamefont {Inomata}\ and\ \citenamefont
  {Terada}(2020)}]{Inomata:2019yww}%
  \BibitemOpen
  \bibfield  {author} {\bibinfo {author} {\bibfnamefont {K.}~\bibnamefont
  {Inomata}}\ and\ \bibinfo {author} {\bibfnamefont {T.}~\bibnamefont
  {Terada}},\ }\href {\doibase 10.1103/PhysRevD.101.023523} {\bibfield
  {journal} {\bibinfo  {journal} {Phys. Rev. D}\ }\textbf {\bibinfo {volume}
  {101}},\ \bibinfo {pages} {023523} (\bibinfo {year} {2020})},\ \Eprint
  {http://arxiv.org/abs/1912.00785} {arXiv:1912.00785 [gr-qc]} \BibitemShut
  {NoStop}%
\bibitem [{\citenamefont {Dom\`enech}\ and\ \citenamefont
  {Sasaki}(2021)}]{Domenech:2020xin}%
  \BibitemOpen
  \bibfield  {author} {\bibinfo {author} {\bibfnamefont {G.}~\bibnamefont
  {Dom\`enech}}\ and\ \bibinfo {author} {\bibfnamefont {M.}~\bibnamefont
  {Sasaki}},\ }\href {\doibase 10.1103/PhysRevD.103.063531} {\bibfield
  {journal} {\bibinfo  {journal} {Phys. Rev. D}\ }\textbf {\bibinfo {volume}
  {103}},\ \bibinfo {pages} {063531} (\bibinfo {year} {2021})},\ \Eprint
  {http://arxiv.org/abs/2012.14016} {arXiv:2012.14016 [gr-qc]} \BibitemShut
  {NoStop}%
\bibitem [{\citenamefont {Chang}\ \emph {et~al.}(2021)\citenamefont {Chang},
  \citenamefont {Wang},\ and\ \citenamefont {Zhu}}]{Chang:2020tji}%
  \BibitemOpen
  \bibfield  {author} {\bibinfo {author} {\bibfnamefont {Z.}~\bibnamefont
  {Chang}}, \bibinfo {author} {\bibfnamefont {S.}~\bibnamefont {Wang}}, \ and\
  \bibinfo {author} {\bibfnamefont {Q.-H.}\ \bibnamefont {Zhu}},\ }\href
  {\doibase 10.1088/1674-1137/ac0c74} {\bibfield  {journal} {\bibinfo
  {journal} {Chin. Phys. C}\ }\textbf {\bibinfo {volume} {45}},\ \bibinfo
  {pages} {095101} (\bibinfo {year} {2021})},\ \Eprint
  {http://arxiv.org/abs/2009.11025} {arXiv:2009.11025 [astro-ph.CO]}
  \BibitemShut {NoStop}%
\bibitem [{\citenamefont {Arjona}\ \emph {et~al.}(2019)\citenamefont {Arjona},
  \citenamefont {Cardona},\ and\ \citenamefont {Nesseris}}]{Arjona:2018jhh}%
  \BibitemOpen
  \bibfield  {author} {\bibinfo {author} {\bibfnamefont {R.}~\bibnamefont
  {Arjona}}, \bibinfo {author} {\bibfnamefont {W.}~\bibnamefont {Cardona}}, \
  and\ \bibinfo {author} {\bibfnamefont {S.}~\bibnamefont {Nesseris}},\ }\href
  {\doibase 10.1103/PhysRevD.99.043516} {\bibfield  {journal} {\bibinfo
  {journal} {Phys. Rev. D}\ }\textbf {\bibinfo {volume} {99}},\ \bibinfo
  {pages} {043516} (\bibinfo {year} {2019})},\ \Eprint
  {http://arxiv.org/abs/1811.02469} {arXiv:1811.02469 [astro-ph.CO]}
  \BibitemShut {NoStop}%
\bibitem [{\citenamefont {Capozziello}\ and\ \citenamefont
  {De~Laurentis}(2011)}]{Capozziello:2011et}%
  \BibitemOpen
  \bibfield  {author} {\bibinfo {author} {\bibfnamefont {S.}~\bibnamefont
  {Capozziello}}\ and\ \bibinfo {author} {\bibfnamefont {M.}~\bibnamefont
  {De~Laurentis}},\ }\href {\doibase 10.1016/j.physrep.2011.09.003} {\bibfield
  {journal} {\bibinfo  {journal} {Phys. Rept.}\ }\textbf {\bibinfo {volume}
  {509}},\ \bibinfo {pages} {167} (\bibinfo {year} {2011})},\ \Eprint
  {http://arxiv.org/abs/1108.6266} {arXiv:1108.6266 [gr-qc]} \BibitemShut
  {NoStop}%
\bibitem [{\citenamefont {Ananda}\ \emph {et~al.}(2007)\citenamefont {Ananda},
  \citenamefont {Clarkson},\ and\ \citenamefont {Wands}}]{Ananda:2006af}%
  \BibitemOpen
  \bibfield  {author} {\bibinfo {author} {\bibfnamefont {K.~N.}\ \bibnamefont
  {Ananda}}, \bibinfo {author} {\bibfnamefont {C.}~\bibnamefont {Clarkson}}, \
  and\ \bibinfo {author} {\bibfnamefont {D.}~\bibnamefont {Wands}},\ }\href
  {\doibase 10.1103/PhysRevD.75.123518} {\bibfield  {journal} {\bibinfo
  {journal} {Phys. Rev.}\ }\textbf {\bibinfo {volume} {D75}},\ \bibinfo {pages}
  {123518} (\bibinfo {year} {2007})},\ \Eprint
  {http://arxiv.org/abs/gr-qc/0612013} {arXiv:gr-qc/0612013 [gr-qc]}
  \BibitemShut {NoStop}%
\bibitem [{\citenamefont {Baumann}\ \emph
  {et~al.}(2007{\natexlab{b}})\citenamefont {Baumann}, \citenamefont
  {Steinhardt}, \citenamefont {Takahashi},\ and\ \citenamefont
  {Ichiki}}]{Baumann:2007zm}%
  \BibitemOpen
  \bibfield  {author} {\bibinfo {author} {\bibfnamefont {D.}~\bibnamefont
  {Baumann}}, \bibinfo {author} {\bibfnamefont {P.~J.}\ \bibnamefont
  {Steinhardt}}, \bibinfo {author} {\bibfnamefont {K.}~\bibnamefont
  {Takahashi}}, \ and\ \bibinfo {author} {\bibfnamefont {K.}~\bibnamefont
  {Ichiki}},\ }\href {\doibase 10.1103/PhysRevD.76.084019} {\bibfield
  {journal} {\bibinfo  {journal} {Phys. Rev.}\ }\textbf {\bibinfo {volume}
  {D76}},\ \bibinfo {pages} {084019} (\bibinfo {year} {2007}{\natexlab{b}})},\
  \Eprint {http://arxiv.org/abs/hep-th/0703290} {arXiv:hep-th/0703290 [hep-th]}
  \BibitemShut {NoStop}%
\bibitem [{\citenamefont {Kohri}\ and\ \citenamefont
  {Terada}(2018)}]{Kohri:2018awv}%
  \BibitemOpen
  \bibfield  {author} {\bibinfo {author} {\bibfnamefont {K.}~\bibnamefont
  {Kohri}}\ and\ \bibinfo {author} {\bibfnamefont {T.}~\bibnamefont {Terada}},\
  }\href {\doibase 10.1103/PhysRevD.97.123532} {\bibfield  {journal} {\bibinfo
  {journal} {Phys. Rev.}\ }\textbf {\bibinfo {volume} {D97}},\ \bibinfo {pages}
  {123532} (\bibinfo {year} {2018})},\ \Eprint
  {http://arxiv.org/abs/1804.08577} {arXiv:1804.08577 [gr-qc]} \BibitemShut
  {NoStop}%
\bibitem [{\citenamefont {Espinosa}\ \emph {et~al.}(2018)\citenamefont
  {Espinosa}, \citenamefont {Racco},\ and\ \citenamefont
  {Riotto}}]{Espinosa:2018eve}%
  \BibitemOpen
  \bibfield  {author} {\bibinfo {author} {\bibfnamefont {J.~R.}\ \bibnamefont
  {Espinosa}}, \bibinfo {author} {\bibfnamefont {D.}~\bibnamefont {Racco}}, \
  and\ \bibinfo {author} {\bibfnamefont {A.}~\bibnamefont {Riotto}},\ }\href
  {\doibase 10.1088/1475-7516/2018/09/012} {\bibfield  {journal} {\bibinfo
  {journal} {JCAP}\ }\textbf {\bibinfo {volume} {1809}},\ \bibinfo {pages}
  {012} (\bibinfo {year} {2018})},\ \Eprint {http://arxiv.org/abs/1804.07732}
  {arXiv:1804.07732 [hep-ph]} \BibitemShut {NoStop}%
\bibitem [{\citenamefont {Ellis}\ and\ \citenamefont
  {Mavromatos}(2013)}]{Ellis:2013zsa}%
  \BibitemOpen
  \bibfield  {author} {\bibinfo {author} {\bibfnamefont {J.}~\bibnamefont
  {Ellis}}\ and\ \bibinfo {author} {\bibfnamefont {N.~E.}\ \bibnamefont
  {Mavromatos}},\ }\href {\doibase 10.1103/PhysRevD.88.085029} {\bibfield
  {journal} {\bibinfo  {journal} {Phys. Rev. D}\ }\textbf {\bibinfo {volume}
  {88}},\ \bibinfo {pages} {085029} (\bibinfo {year} {2013})},\ \Eprint
  {http://arxiv.org/abs/1308.1906} {arXiv:1308.1906 [hep-th]} \BibitemShut
  {NoStop}%
\bibitem [{\citenamefont {Maggiore}(2000)}]{Maggiore_2000}%
  \BibitemOpen
  \bibfield  {author} {\bibinfo {author} {\bibfnamefont {M.}~\bibnamefont
  {Maggiore}},\ }\href {\doibase 10.1016/s0370-1573(99)00102-7} {\bibfield
  {journal} {\bibinfo  {journal} {Physics Reports}\ }\textbf {\bibinfo {volume}
  {331}},\ \bibinfo {pages} {283–367} (\bibinfo {year} {2000})}\BibitemShut
  {NoStop}%
\bibitem [{\citenamefont {Dimopoulos}(2006)}]{Dimopoulos:2006ms}%
  \BibitemOpen
  \bibfield  {author} {\bibinfo {author} {\bibfnamefont {K.}~\bibnamefont
  {Dimopoulos}},\ }\href {\doibase 10.1103/PhysRevD.74.083502} {\bibfield
  {journal} {\bibinfo  {journal} {Phys. Rev. D}\ }\textbf {\bibinfo {volume}
  {74}},\ \bibinfo {pages} {083502} (\bibinfo {year} {2006})},\ \Eprint
  {http://arxiv.org/abs/hep-ph/0607229} {arXiv:hep-ph/0607229} \BibitemShut
  {NoStop}%
\bibitem [{\citenamefont {Kawasaki}\ and\ \citenamefont
  {Murai}(2024)}]{Kawasaki:2023rfx}%
  \BibitemOpen
  \bibfield  {author} {\bibinfo {author} {\bibfnamefont {M.}~\bibnamefont
  {Kawasaki}}\ and\ \bibinfo {author} {\bibfnamefont {K.}~\bibnamefont
  {Murai}},\ }\href {\doibase 10.1088/1475-7516/2024/01/050} {\bibfield
  {journal} {\bibinfo  {journal} {JCAP}\ }\textbf {\bibinfo {volume} {01}},\
  \bibinfo {pages} {050} (\bibinfo {year} {2024})},\ \Eprint
  {http://arxiv.org/abs/2308.13134} {arXiv:2308.13134 [astro-ph.CO]}
  \BibitemShut {NoStop}%
\bibitem [{\citenamefont {Alexandre}\ \emph {et~al.}(2024)\citenamefont
  {Alexandre}, \citenamefont {Dvali},\ and\ \citenamefont
  {Koutsangelas}}]{Alexandre:2024nuo}%
  \BibitemOpen
  \bibfield  {author} {\bibinfo {author} {\bibfnamefont {A.}~\bibnamefont
  {Alexandre}}, \bibinfo {author} {\bibfnamefont {G.}~\bibnamefont {Dvali}}, \
  and\ \bibinfo {author} {\bibfnamefont {E.}~\bibnamefont {Koutsangelas}},\
  }\href@noop {} {\  (\bibinfo {year} {2024})},\ \Eprint
  {http://arxiv.org/abs/2402.14069} {arXiv:2402.14069 [hep-ph]} \BibitemShut
  {NoStop}%
\bibitem [{\citenamefont {Thoss}\ \emph {et~al.}(2024)\citenamefont {Thoss},
  \citenamefont {Burkert},\ and\ \citenamefont {Kohri}}]{Thoss:2024hsr}%
  \BibitemOpen
  \bibfield  {author} {\bibinfo {author} {\bibfnamefont {V.}~\bibnamefont
  {Thoss}}, \bibinfo {author} {\bibfnamefont {A.}~\bibnamefont {Burkert}}, \
  and\ \bibinfo {author} {\bibfnamefont {K.}~\bibnamefont {Kohri}},\
  }\href@noop {} {\  (\bibinfo {year} {2024})},\ \Eprint
  {http://arxiv.org/abs/2402.17823} {arXiv:2402.17823 [astro-ph.CO]}
  \BibitemShut {NoStop}%
\bibitem [{\citenamefont {Dvali}\ \emph {et~al.}(2024)\citenamefont {Dvali},
  \citenamefont {Valbuena-Berm\'udez},\ and\ \citenamefont
  {Zantedeschi}}]{Dvali:2024hsb}%
  \BibitemOpen
  \bibfield  {author} {\bibinfo {author} {\bibfnamefont {G.}~\bibnamefont
  {Dvali}}, \bibinfo {author} {\bibfnamefont {J.~S.}\ \bibnamefont
  {Valbuena-Berm\'udez}}, \ and\ \bibinfo {author} {\bibfnamefont
  {M.}~\bibnamefont {Zantedeschi}},\ }\href@noop {} {\  (\bibinfo {year}
  {2024})},\ \Eprint {http://arxiv.org/abs/2405.13117} {arXiv:2405.13117
  [hep-th]} \BibitemShut {NoStop}%
\bibitem [{\citenamefont {Dvali}\ \emph {et~al.}(2020)\citenamefont {Dvali},
  \citenamefont {Eisemann}, \citenamefont {Michel},\ and\ \citenamefont
  {Zell}}]{Dvali:2020wft}%
  \BibitemOpen
  \bibfield  {author} {\bibinfo {author} {\bibfnamefont {G.}~\bibnamefont
  {Dvali}}, \bibinfo {author} {\bibfnamefont {L.}~\bibnamefont {Eisemann}},
  \bibinfo {author} {\bibfnamefont {M.}~\bibnamefont {Michel}}, \ and\ \bibinfo
  {author} {\bibfnamefont {S.}~\bibnamefont {Zell}},\ }\href {\doibase
  10.1103/PhysRevD.102.103523} {\bibfield  {journal} {\bibinfo  {journal}
  {Phys. Rev. D}\ }\textbf {\bibinfo {volume} {102}},\ \bibinfo {pages}
  {103523} (\bibinfo {year} {2020})},\ \Eprint
  {http://arxiv.org/abs/2006.00011} {arXiv:2006.00011 [hep-th]} \BibitemShut
  {NoStop}%
\bibitem [{\citenamefont {Balaji}\ \emph {et~al.}(2024)\citenamefont {Balaji},
  \citenamefont {Dom\`enech}, \citenamefont {Franciolini}, \citenamefont
  {Ganz},\ and\ \citenamefont {Tr\"ankle}}]{Balaji:2024hpu}%
  \BibitemOpen
  \bibfield  {author} {\bibinfo {author} {\bibfnamefont {S.}~\bibnamefont
  {Balaji}}, \bibinfo {author} {\bibfnamefont {G.}~\bibnamefont {Dom\`enech}},
  \bibinfo {author} {\bibfnamefont {G.}~\bibnamefont {Franciolini}}, \bibinfo
  {author} {\bibfnamefont {A.}~\bibnamefont {Ganz}}, \ and\ \bibinfo {author}
  {\bibfnamefont {J.}~\bibnamefont {Tr\"ankle}},\ }\href@noop {} {\  (\bibinfo
  {year} {2024})},\ \Eprint {http://arxiv.org/abs/2403.14309} {arXiv:2403.14309
  [gr-qc]} \BibitemShut {NoStop}%
\bibitem [{\citenamefont {Tsujikawa}(2007)}]{Tsujikawa:2007gd}%
  \BibitemOpen
  \bibfield  {author} {\bibinfo {author} {\bibfnamefont {S.}~\bibnamefont
  {Tsujikawa}},\ }\href {\doibase 10.1103/PhysRevD.76.023514} {\bibfield
  {journal} {\bibinfo  {journal} {Phys. Rev. D}\ }\textbf {\bibinfo {volume}
  {76}},\ \bibinfo {pages} {023514} (\bibinfo {year} {2007})},\ \Eprint
  {http://arxiv.org/abs/0705.1032} {arXiv:0705.1032 [astro-ph]} \BibitemShut
  {NoStop}%
\end{thebibliography}%

\end{document}